\documentclass[prb,twocolumn,superscriptaddress,float,aps]{revtex4-2}
\usepackage{bm,color,amsmath,amssymb,mathrsfs,latexsym,graphicx,psfrag,tabularx}

\usepackage[hidelinks,colorlinks,linkcolor=blue,
citecolor=blue,urlcolor=blue]{hyperref}

\newcommand{\braket}[2]{\left\langle #1 | #2 \right\rangle}
\newcommand{\bra}[1]{\left\langle#1\right|}
\newcommand{\ket}[1]{\left|#1\right\rangle}

\newcommand{\beq}{\begin{equation}}
\newcommand{\eneq}{\end{equation}}

\newcommand{\eqnref}[1]{Eq.\,\eqref{#1}}

\usepackage[dvipsnames]{xcolor}

\def\dd{\mathbf{d}}

\begin{document}

\title{Giant and Helical Exciton Dipole from Berry Curvature in Flat Chern Bands}

\author{Kaijie Yang}
    \affiliation{Department of Materials Science and Engineering, University of Washington, Seattle, Washington 98195, USA}
\author{Huiyuan Zheng}
    \affiliation{Department of Materials Science and Engineering, University of Washington, Seattle, Washington 98195, USA}
\author{Xiaodong Xu}
    \affiliation{Department of Physics, University of Washington, Seattle, Washington 98195, USA}
    \affiliation{Department of Materials Science and Engineering, University of Washington, Seattle, Washington 98195, USA}
\author{Di Xiao}
    \affiliation{Department of Materials Science and Engineering, University of Washington, Seattle, Washington 98195, USA}
    \affiliation{Department of Physics, University of Washington, Seattle, Washington 98195, USA}
\author{Ting Cao}
    \affiliation{Department of Materials Science and Engineering, University of Washington, Seattle, Washington 98195, USA}

\begin{abstract} 
We show that excitons forming between moir\'e flat Chern bands possess a substantial electric dipole moment comparable to the moir\'e lattice parameter times the elementary charge ($\sim10^2$ Debye).
At a hole filling factor of one in twisted MoTe$_2$, the dipole moment of the lowest-energy exciton branch develops in-plane helical texture in momentum space from the intrinsic Berry curvature of electron and hole.
By solving the Bethe-Salpeter equations, we demonstrate that an out-of-plane displacement field induces a Frenkel-to-Wannier exciton transition, accompanied by a reversal of the dipole texture helicity.
The resulting attractive exciton dipole-dipole interactions lead to quadrupolar biexcitons that can be probed via two-photon spectroscopy. 
Our findings establish band topology as a tunable knob to engineer exciton dipole moments and pave the way to manipulate many-body interactions in the terahertz regime.
\end{abstract}

\date{\today}

\maketitle

{\it Introduction -} 
Excitons are charge-neutral quasiparticles, but their internal charge separation can endow them with electric dipoles that strongly influence transport, optical response, and many-body interactions \cite{knox_theory_1963, haug2009quantum, wang2018colloquium}. 
A prominent realization is the interlayer exciton in van der Waals heterostructures and coupled quantum wells, where  the spatial separation between the electron and hole layers produces a permanent out-of-plane dipole and significantly modifies optical properties and interactions \cite{rivera2018interlayer, rivera2015observation, fang2014strong, chiu2014spectroscopic, charbonneau1988transformation, chen1987effect, kuo2005strong, miller1984band, klein2016stark, kim2014ultrafast, sie2017valley, withers2015wse2}. 
By contrast, engineering a substantial in-plane exciton dipole is far more challenging.
Unlike a fixed out-of-plane polarization, an in-plane dipole could in principle be locked to the exciton momentum, generating dipolar textures and anisotropic interactions.
The canonical example is the magneto-exciton, in which a perpendicular magnetic field displaces the electron and hole in opposite directions through the Lorentz force and produces a dipole perpendicular to the center-of-mass momentum \cite{kallin1984excitations, cao2021quantum}.
At zero magnetic field, Berry curvature offers an analogous geometric mechanism \cite{xiao2010berry}. 
However, in many topological or near-inverted band structures, the conduction and valence bands carry large Berry curvature of opposite signs, so the electron and hole contributions to the exciton dipole tend to cancel.

Flat Chern bands in moir\'e materials provide a natural route to overcome this obstacle. 
When the electron and hole bands carry the Berry curvature of the same sign and are flattened in the moir\'e Brillouin zone (MBZ), the exciton envelope function can coherently sample the quantum geometry of both bands, resulting in a large in-plane dipole moment.
This regime is distinct from most moir\'e excitons studied so far in the visible and near-infrared ranges, where the moir\'e potential primarily modifies optical excitations inherited from the constituent monolayers or heterostructures \cite{naik2022intralayer, lagoin2022extended, xiong2023correlated, zeng2023exciton, park2023dipole}.
Here, instead, the excitons arise across a low-energy moir\'e miniband gap at integer filling, in the same flat-topological-band regime that hosts integer and fractional quantum anomalous Hall physics, and therefore naturally falls within the THz energy scale  \cite{li2021quantum, serlin2020intrinsic, cai2023signatures, park2023observation, lu2024fractional, xu2023observation, xu2025signatures}. 
This raises a basic question: can flat-band topology be used as a practical knob to engineer a giant, tunable in-plane exciton dipole, and what new interactions and optical phenomena would follow?

\begin{figure}
\centering
\includegraphics[width=0.6\columnwidth]{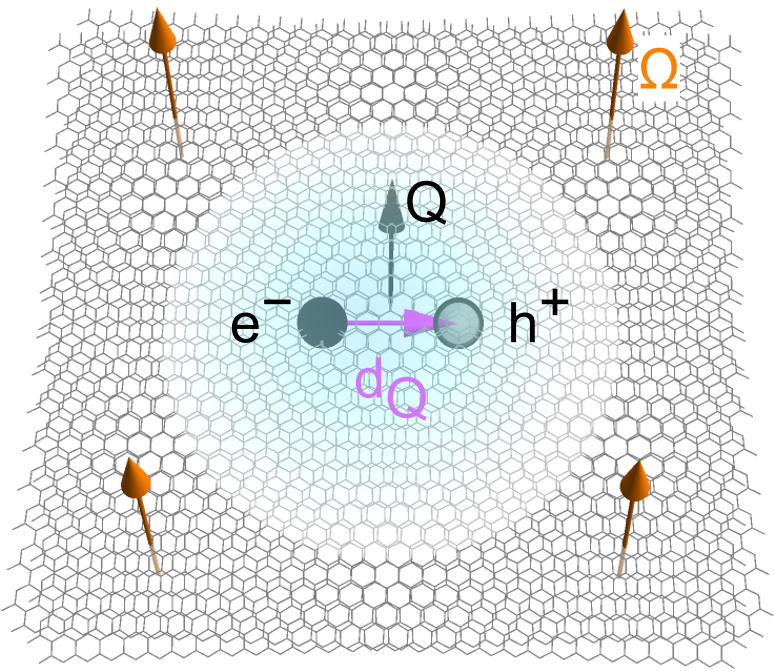}
\caption{
A schematic view of the exciton dipole in flat Chern band systems.
The electron ($e^-$) and hole ($h^+$) are bound with an envelope function (cyan), move with a center-of-mass momentum $\mathbf Q$, and are separated by a displacement vector $\mathbf d_\mathbf Q$ (purple) originated from their Berry curvature $\Omega$.
}
\label{fig: schematic} 
\end{figure}

\begin{figure*}
   \centering
\includegraphics[width=\textwidth]{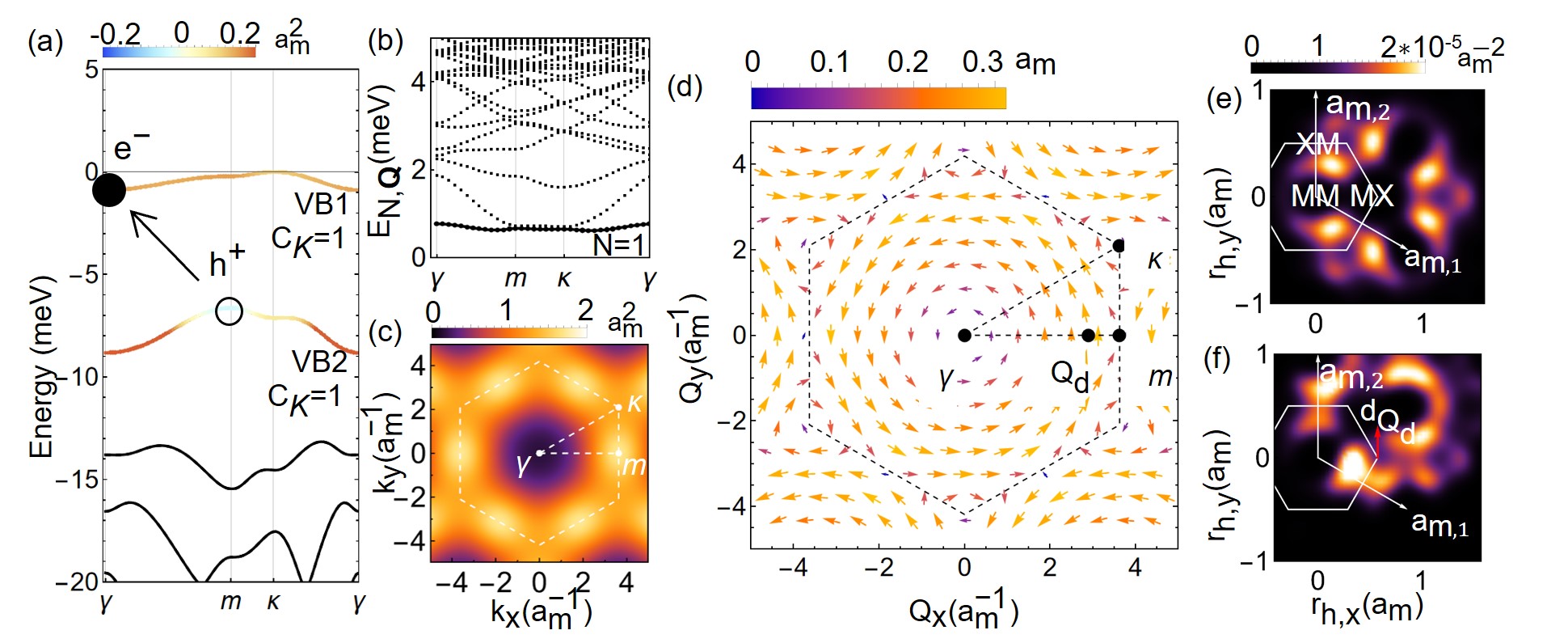}
\caption{
(a) The single-particle valence moir\'e minibands of tMoTe$_2$ at $2.1^\circ$ in valley K.
VB1 and VB2 are the top two valence moir\'e minibands. 
$C_K$ labels the valley Chern number of a miniband.
Color represents the Berry curvature $\Omega_n(\mathbf k)$.
(b) The exciton excitation spectrum $E_{N,\mathbf Q}$. The solid line is the lowest energy exciton band labeled by $N=1$.
(c) Modular square of the exciton envelope function $\vert \Psi_{\mathbf Q=\gamma}(\mathbf k)\vert^2$ for $N=1$ in momentum space.
(d) The exciton dipole textures of the $N=1$ exciton branch in MBZ.
The direction and length of arrows represent the direction and magnitude of the displacement vectors $\mathbf d_{\mathbf Q}$ of the exciton dipole, respectively.
The magnitude is also color coded in the unit of the moir\'e lattice constant $a_\text{m}$.
(e)(f) Modular square of exciton wavefunction in real space with fixed electron position at MX, i.e., $\vert \Psi_{\mathbf Q}(\mathbf r_e = \mathbf{a}_{\text m, 1} 2/3+\mathbf{a}_{\text m, 2}/3, \mathbf r_h)\vert^2$ for $\mathbf Q = \gamma$ (e) and $\mathbf Q = \mathbf Q_d$ (f). 
$\mathbf{a}_{\text m, 1}, \mathbf{a}_{\text m, 2}$ are moir\'e lattice vectors.
MX, XM, MM are high symmetry stacking positions.
Red arrow labels $\mathbf d_{\mathbf Q_d}$.} 
\label{fig:flatChernband} 
\end{figure*}

In this work, we investigate the in-plane exciton dipoles formed between two flat Chern bands with the same Chern number.
As schematically illustrated in Fig.~\ref{fig: schematic}, when an exciton moves in real space with a center-of-mass momentum $\mathbf Q$, the Berry curvature of Chern bands induces opposite anomalous velocities for the electron and hole perpendicular to $\mathbf Q$, resulting in an electric dipole moment $\mathbf d_\mathbf Q |e|$ stabilized by  the Coulomb attraction.
$\mathbf d_{\mathbf Q}$, the displacement vector, is perpendicular to $\mathbf{Q}$ and forms a helical texture in momentum space.
Taking $2.1^\circ$ twisted MoTe$_2$ (tMoTe$_2$) at a hole filling factor of one as an example, we demonstrate $\mathbf d_\mathbf Q = \bar{\Omega} \hat z \times \mathbf Q$ near high symmetry points
in MBZ, where $\bar\Omega$ is the Berry curvature weighted by the exciton envelope function.
Remarkably, the concentration of Berry curvature within the MBZ leads to exciton dipole moments reaching 150 D (Debye).
Increasing the displacement field induces the exciton transition from Frenkel to Wannier types, rendering both the dipole magnitude and helicity tunable.
Furthermore, the giant exciton dipole gives rise to anisotropic, attractive dipole-dipole interactions that exceed the exciton bandwidth, leading to the formation of optically bright biexcitons with in-plane quadrupole moments.
Our work suggests moir\'e systems as compelling platforms for exploring tunable exciton dipole physics in the terahertz regime.

{\it Frenkel excitons in flat Chern bands -}
tMoTe$_2$ bilayers exhibit multiple flat valley Chern bands at the twist angle near $2^\circ$ \cite{cao2025fractional, wang2025higher, deng2025nonmonotonic}.
We start from the single-particle electronic structure at $2.1^\circ$, corresponding to a moir\'e lattice constant $a_\text m = 9.6$ nm \cite{zhang2024polarization, jia2024moire, wang2024fractional, ahn2024non}.
Due to the spontaneous valley polarization at a filling factor of one charge per moir\'e unit cell \cite{li2025universal, kang2024evidence}, we focus on excitons formed within the flat Chern bands of the K valley.
For the same filling factor in the K' valley, the exciton dipole moments flip due to the time reversal symmetry. 

Fig.~\ref{fig:flatChernband}(a) shows frontier valence moir\'e minibands at $\Delta_D=0$ meV, where $\Delta_D$ is the layer potential difference induced by the displacement field (See details in supplementary material (SM) Sec.~\ref{sec: single particle hamiltonian}).
The top two minibands VB1 and VB2 share the same valley Chern number $C_K=1$ with nearly uniform and positive Berry curvature across MBZ.
The flatness of VB1 and VB2 is reflected in their bandwidth of $W_1=0.9$ meV and $W_2=2.2$ meV, respectively.
These bandwidths are smaller than both the electron-hole binding energy $U_0 \approx 6.0$ meV and the average band gap $E_g \approx 10$ meV (See SM Sec.~\ref{sec: bse}).
Given that $U_0 < E_g$, the Tamm–Dancoff approximation is used in the exciton calculations, in which VB1 can be taken as the electron band and VB2 the hole band \cite{fetter2012quantum}. 

Figure.~\ref{fig:flatChernband}(b) shows the excitation spectrum $E_{N, \mathbf Q}$ as a function of the center-of-mass momentum $\mathbf Q$, obtained by solving the Bethe-Salpeter equation (See SM Sec.~\ref{sec: bse} for details).
We focus on the lowest energy exciton band labeled by $N=1$ (the solid line in Fig.~\ref{fig:flatChernband}(b)) and omit the $N$ index hereafter for brevity.
Due to the indirect gap between VB1 and VB2, the exciton band maximum is located at $\gamma$ with an exciton bandwidth of $0.15$ meV.
This bandwidth is even smaller than those of both the flat electron and hole bands due to the strong electron-hole Coulomb interactions, revealing localization characteristic of a Frenkel-type exciton \cite{mattis1984mass}. 
Second, the Frenkel exciton is tightly bound, with a binding energy calculated to be $U_0 \approx 6.0$ meV, making the exciton excitation energy of approximately 1 meV (0.2 THz).
We note the exciton excitation energy is underestimated due to the omission of quasiparticle self-energy corrections in the electron and hole bands. 
These corrections would mainly lead to a rigid blue shift of the miniband gaps and excitation energies, while preserving the Frenkel character of the exciton.
Third, Fig.~\ref{fig:flatChernband}(c) shows the exciton envelope function $\Psi_\mathbf Q(\mathbf k)$ at $\mathbf Q = \gamma$ is nearly uniformly distributed in MBZ except for the region near $\gamma$, where the amplitude approaches zero.
The real space wavefunction of exciton at $\mathbf Q =\gamma$ is plotted in Fig.~\ref{fig:flatChernband}(e), defined as
$
    \Psi_\mathbf Q(\mathbf r_e, \mathbf r_h) = \sum_\mathbf k \Psi_{\mathbf Q}(\mathbf k) \psi_{e,\mathbf k+\mathbf{Q/2}}(\mathbf r_e)  \psi_{h,\mathbf k-\mathbf Q/2}(\mathbf r_h)^*,
$
where the $\psi_{e,\mathbf k+\mathbf{Q}/2}(\mathbf r_e)$, $\psi_{h,\mathbf k-\mathbf Q/2}(\mathbf r_h)$ denote Bloch states of the electron in VB1 and hole in VB2, respectively.
By fixing the electron's position at MX high-symmetry stacking location, where the metal (M) atom is vertically aligned with chalcogen (X) atom of the adjacent layer, we find the hole density is distributed at the XM stacking region across adjacent moir\'e unit cells, confirming the Frenkel nature of the exciton \cite{frenkel1931transformation}.

{\it Giant and helical exciton dipole moment -}
We proceed to study the dipole moment as a function of exciton center-of-mass momentum (Fig.~\ref{fig:flatChernband}(d)) using a gauge-invariant formulation. 
The exciton dipole displacement vector, pointing from the electron to the hole, can be written in a two-band system as
\begin{equation}
\begin{split}
    &\mathbf d_{\mathbf Q}= \int \dd^2 \mathbf r_e \dd^2 \mathbf r_h \vert\Psi_\mathbf Q(\mathbf r_e, \mathbf r_h)\vert^2 (\mathbf r_h-\mathbf r_e)    \\
    & =\sum_\mathbf k \Psi_{\mathbf Q}(\mathbf k)^*(-i \partial_\mathbf k)  \Psi_{\mathbf Q}(\mathbf k)\\ 
    & +\Psi_{\mathbf Q}(\mathbf k)^*(\mathbf A_e (\mathbf k+\mathbf Q/2) - \mathbf A_h (\mathbf k-\mathbf Q/2))\Psi_{\mathbf Q}(\mathbf k)
\end{split}
\label{eq: dipole definition}
\end{equation}
where $\mathbf A_{e/h} (\mathbf k)$ is the Berry connection of the electron/hole band, respectively \cite{cao2021quantum, paiva2024shift, davenport2025exciton} (See details in SM Sec.~\ref{sec: exciton dipole}).
The term involving $-i \partial_\mathbf k$ comes solely from the variation of the exciton envelope function, denoted as the envelope contribution $\mathbf d^\text E_{\mathbf Q}$.
The terms involving Berry connections are primarily inherited from the intrinsic electron/hole Bloch state.
While $\mathbf d_{\mathbf Q}$ is gauge invariant, each term is respectively gauge dependent, and is evaluated under the optimal Chern gauge (See SM Sec.~\ref{sec: optimal Chern gauge}) \cite{xie2024chern}, which can be continuously tuned to the hydrogen gauge previously employed for excitons in gapped Dirac fermion systems \cite{cao2018unifying, zhang2018optical}.
Generalization of Eq. \ref{eq: dipole definition} to multi-band systems is straightforward.

To visualize the exciton dipole in real space, we examine the exciton wavefunction $\Psi_{\mathbf Q}(\mathbf r_e, \mathbf r_h)$ at $\mathbf Q= 0$ (Fig. \ref{fig:flatChernband}(e)) and $\mathbf Q_d$ (Fig. \ref{fig:flatChernband}(f)).
At $\mathbf Q= 0$ the wavefunction amplitude shows a threefold rotational symmetry, suggesting a vanishing dipole.
At $\mathbf Q_d$ where the dipole moment is maximized, the overall hole density is shifted up perpendicular to  $\mathbf Q_d$ on the scale of $a_\text m$, yielding a giant exciton dipole moment of $0.31 |e| a_\text m \sim $ $150$ D ($3.0 |e|\cdot$nm).
This magnitude is comparable to the largest exciton dipole moment observed in multilayer van der Waals heterostructures to date \cite {zhu2025observation}.
However, in contrast to the charge-transfer origin characteristic of existing out-of-plane dipoles, the in-plane dipole here arises from Berry phase effects. 
The dipole moment can in principle increase under larger Berry curvatures by varying twist angle and engineering higher Chern bands.

Near $\gamma$, the exciton dipole moment exhibits a helical texture.
Fig.~\ref{fig:flatChernband}(d) shows that the dipole moment is oriented perpendicular to the $\mathbf Q$ and its magnitude scales linearly with $\vert \mathbf Q\vert$ in magnitude, i.e.,
\begin{equation}
    \mathbf d_{\mathbf Q} \simeq  v  \hat z \times \mathbf Q / 2
    \label{eq: helical dipole}
\end{equation}
with the sign of $v$ defining the helicity.
The helical texture is constrained by the symmetries $\mathcal C_{3z}, \mathcal C_{2y} \mathcal T$ and permitted by the absence of the in-plane mirror and $\mathcal T$ in one valley, where $\mathcal C_{3z}, \mathcal C_{2y}, \mathcal T$ are threefold rotation  about z axis, twofold rotation about y axis, and time reversal symmetry, respectively (See details in SM Sec.~\ref{sec: exciton dipole}).

\begin{figure}
    \centering
\includegraphics[width=\columnwidth]{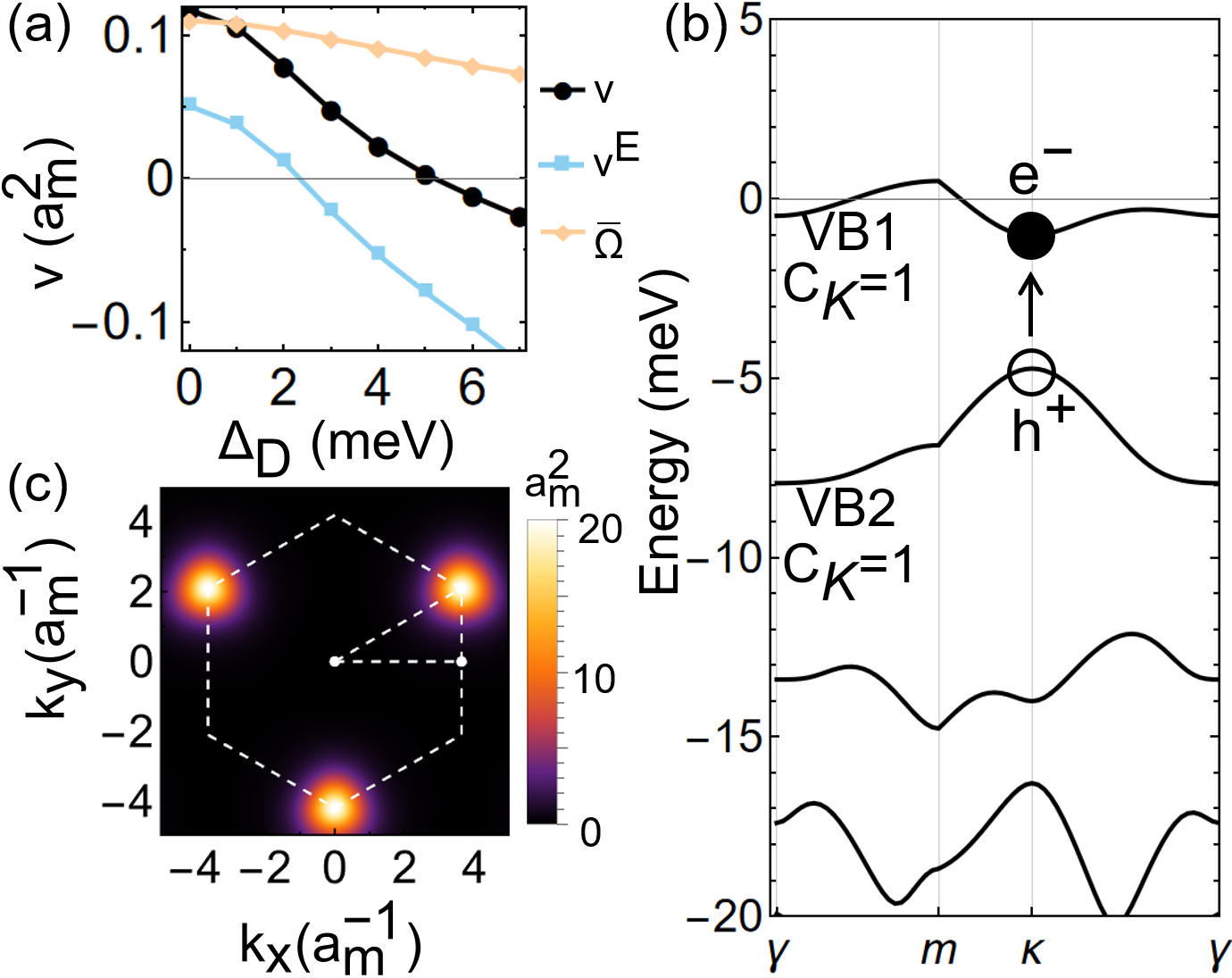}
\caption{
(a) Helicity coefficient $v$ of the exciton dipole texture near $\gamma$ versus displacement fields $\Delta_D$.
$v^\text E$ and $ \bar \Omega$ represent contributions to $v$ from the envelope function gradient and the averaged Berry curvature, respectively. 
(b) The single particle valence moir\'e bands of tMoTe$_2$ for the displacement field of $\Delta_D=6$ meV in valley K.
(c) Modular square of lowest energy exciton envelope function in momentum space $\vert \Psi_{\mathbf Q=\gamma}(\mathbf k)\vert^2$ at $\Delta_D=6$ meV.}
\label{fig: gappedDiraccone} 
\end{figure}

Beyond the symmetry constraints, the helicity of the exciton dipole moment is directly governed by the Berry curvature of the electron and hole bands. 
When neglecting the contributions from the envelope functions, we approximate $ v= \nabla_\mathbf Q \times \mathbf d_\mathbf Q \vert_{\mathbf Q=\gamma} \approx \sum _\mathbf k \vert \Psi_{\mathbf Q=\gamma}(\mathbf k)\vert^2(\Omega_e(\mathbf k)+\Omega_h(\mathbf k)) /2 = \bar \Omega$, where $\Omega_{e/h}(\mathbf k) = \nabla_\mathbf k \times \mathbf A_{e/h}(\mathbf k) $ is the Berry curvature of the electron/hole band and $\bar \Omega$ is the envelope-function weighted Berry curvature averaged over bands.
Fig.~\ref{fig: gappedDiraccone}(a) provides a numerical comparison between $v$ and $\bar \Omega$.
Notably, they share the same sign and match quantitatively at $\Delta_D = 0$ meV, and the contribution of the envelope function is negligible because the function is smooth and relatively uniform across MBZ as shown in Fig.~\ref{fig:flatChernband}(c).
From $\gamma$ to $\kappa$, the dipole moment initially grows linearly in magnitude before decreasing to zero at $\kappa$ due to the constraint of threefold rotational symmetry.

{\it Gate-induced Frenkel to Wannier exciton transition and tunable helicity - }
Both the magnitude and the helicity of the exciton dipole moment are tunable via an out-of-plane displacement field.
As shown in Fig.~\ref{fig: gappedDiraccone}(a), when $\Delta_D$ is increased, the helicity coefficient $v$ initially decreases and subsequently undergoes a sign reversal at $\Delta_D\approx5$ meV.
The variation of $v$ is primarily driven by the envelope function contribution $v^\text E$ where $v^\text E$ is the linear coefficient of $\mathbf d^\text E_\mathbf Q$ near $\gamma$.

The dominant variation of $v^\text E$ originates from the transformation of the exciton envelope function from the Frenkel type in the flat Chern bands at zero $\Delta_D$ to the Wannier type in the gapped Dirac cone at large $\Delta_D$.
Taking $\Delta_D=6$ meV as an example, the displacement field reduces the band gap at $\kappa$, rendering the gap between VB1 and VB2 direct (Fig.~\ref{fig: gappedDiraccone}(b)).
Fig.~\ref{fig: gappedDiraccone}(c) shows the envelope function $\Psi_{\mathbf Q=\gamma}(\mathbf k)$ of the lowest energy exciton band becomes highly localized near $\kappa$.
The momentum space localization indicates the transition of exciton to the Wannier type and enhances the gradient of the exciton envelope function, giving large $v^\text E$ that dominates over the band Berry curvature contribution.
We have also developed a perturbative theory to understand why $v^\text E$ is opposite to $\bar \Omega$ at large $\Delta_D$ (i.e., approaching the limit of a gapped Dirac cone, see SM Sec.~\ref{sec: gapped Dirac cone}).

\begin{figure}
    \centering
\includegraphics[width=\columnwidth]{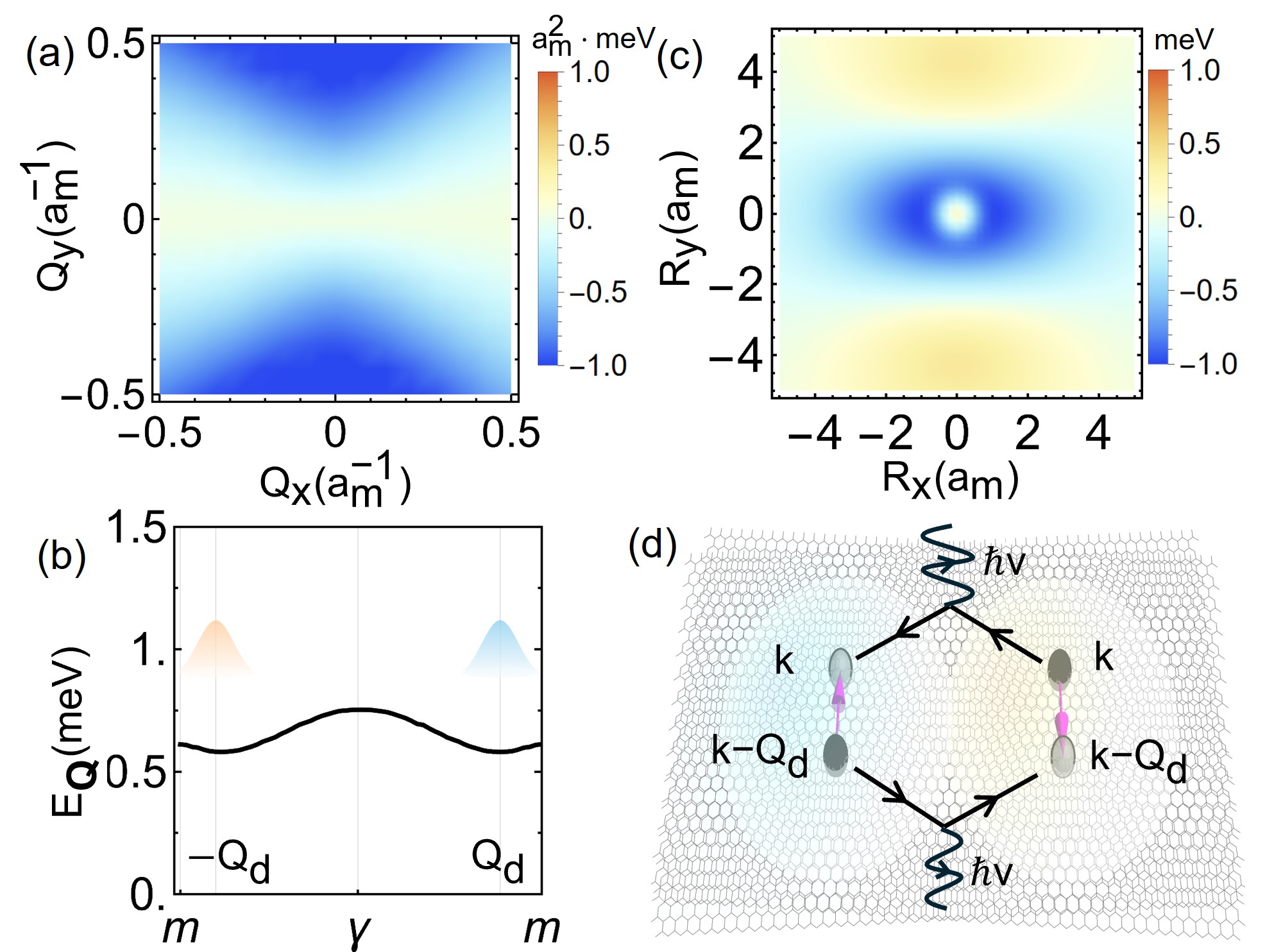}
\caption{
(a) The two-body exciton interaction potential $\mathcal V^X_{\mathbf Q_d,-\mathbf Q_d}(\mathbf Q) $ in momentum space, evaluated for excitons at $\pm \mathbf Q_d$ with antiparallel dipole moments along y at $\Delta_D = 0$ meV.
(b) Schematic representation of exciton wavepackets centered at $\pm \mathbf Q_d$. The black line is the lowest-energy exciton branch.
(c) The real-space interaction potential $\tilde {\mathcal V}^X_{\mathbf Q_d,-\mathbf Q_d}(\mathbf R)$ between the two exciton wavepackets with antiparallel dipole moments along y.
(d) Schematic two-photon process of the biexciton. The wavepackets correspond to those color-coded in (b). The radiative emission of two photons, each with energy $\hbar \nu$, is represented by the black arrows in Feynman diagrams.
}
\label{fig: dipoledipoleinteraction} 
\end{figure}

{\it Exciton dipole-dipole interaction - }
The giant exciton dipole of $150$ D opens a range of interaction-driven optical phenomena. 
At filling factor $\nu = 1$, charged excitations such as trions or localized defects can electrostatically polarize the excitons, producing radial dipole alignment and, through the helical dipole texture, winding momentum-space patterns. 
We focus here on exciton-exciton dipole interactions, which dominate the low-energy optical response.
We first analyze this interaction in the exciton momentum space. 
Considering two excitons in the dilute limit with creation operators $X^\dagger_{\mathbf Q_1}$ and $X^\dagger_{\mathbf Q_2}$, the two-body exciton interaction matrix element is defined as 
$
\mathcal V^X_{\mathbf Q_1,\mathbf Q_2}(\mathbf Q) = \langle X_{\mathbf Q_1 + \mathbf Q} X_{\mathbf Q_2-\mathbf Q} \mathcal V X^\dagger_{\mathbf Q_2} X^\dagger_{\mathbf Q_1}\rangle
$,
where $\mathcal V$ is the screened Coulomb interaction and $\langle\cdots\rangle$ is taken over the integer-filling ground state.
Here the index $\mathbf Q$ denotes the momentum transfer of the exciton-exciton interaction.
In the long-wavelength limit, the direct component of the exciton interaction has the leading term
\begin{equation}
    \mathcal V^X_{\mathbf Q_1,\mathbf Q_2}(\mathbf Q) \approx V(\mathbf Q) (\mathbf Q \cdot \mathbf d_{\mathbf Q_1})(\mathbf Q \cdot \mathbf d_{\mathbf Q_2}),
\label{eq：biexciton_linear_in_q}
\end{equation}
which recovers the classical in-plane dipole-dipole interaction, and scales linearly with $\mathbf Q$ due to $ 1/ \vert \mathbf Q\vert$ dependence of the 2D screened Coulomb potential $V(\mathbf Q)$ (See SM Sec.~\ref{sec: dipole-dipole interaction}).

The interaction potential $\mathcal V^X_{\mathbf Q_1,\mathbf Q_2}$ is anisotropic not only in the dipole orientations $\mathbf d_{\mathbf Q_1}$ and $\mathbf d_{\mathbf Q_2}$ as expected from the dipole-dipole interactions, but also in the momentum transfer $\mathbf Q$.
Setting $\Delta_D = 0$ meV and $\mathbf Q_1 = -\mathbf Q_2 = \mathbf Q_d$ (maximum dipole), Fig.~\ref{fig: dipoledipoleinteraction}(a) shows $\mathcal V^X_{\mathbf Q_d,-\mathbf Q_d}(\mathbf Q)$ for $\mathbf d_{\mathbf Q_d}= -\mathbf d_{ -\mathbf Q_d}$ oriented along the $y$-axis.
The exciton interaction is attractive when $\mathbf Q$ is parallel to dipole moments, but repulsive for $\mathbf Q$ along the  $x$-axis.
For $\mathbf Q$ along the $y$-axis, $\mathcal V^X_{\mathbf Q_d,-\mathbf Q_d}(\mathbf Q)$ is linear in $|\mathbf{Q}|$, consistent with Eq. \ref{eq：biexciton_linear_in_q}.
For $\mathbf Q$ along the $x$-axis, $\mathcal V^X_{\mathbf Q_d,-\mathbf Q_d}(\mathbf Q)$ is quadratic due to higher-order repulsive contributions, and its overall magnitude remains an order of magnitude smaller.

The large, linear in $|\mathbf{Q}|$ interaction in momentum space translates to strong attraction between excitons in real space. 
We evaluate the exciton-exciton interaction potential as a function of their center-of-mass separation $\mathbf{R}$ using a semiclassical approach, in which we construct two exciton wavepackets at momenta $\pm \mathbf Q_d$ as schematically illustrated in Fig.~\ref{fig: dipoledipoleinteraction}(b) (See SM Sec.~\ref{sec: Exciton interaction in real space}).
Fig.~\ref{fig: dipoledipoleinteraction}(c) shows that the interaction potential $\tilde{\mathcal V}_{\mathbf Q_d,-\mathbf Q_d}(\mathbf R)$ is attractive for $\vert \mathbf R \vert \lesssim 2 a_m$.
For excitons occupying adjacent moir\'e cells, the interaction potential reaches -1.03 meV, an order of magnitude larger than the exciton bandwidth of 0.15 meV shown in Fig.~\ref{fig: dipoledipoleinteraction}(b).

{\it Biexcitons in flat Chern bands - }
Strong exciton–exciton interactions can produce interaction-driven optical phenomena such as bound biexcitons, strong optical nonlinearity, and multipolar excitonic complexes, whose observability is maximized in the zero total momentum channel that directly couples to light \cite{wang2018colloquium, wilson2021excitons, combescot1992semiconductors}.
We focus here on the optically active zero-total-momentum biexciton channel.
In the flat exciton band, the kinetic energy is quenched.
So the biexciton binding is set by the interaction energy.
For a pair of excitons at $\mathbf Q_1$ and $-\mathbf Q_1$ with zero total momentum, we approximate the biexciton enery as
\begin{equation}
    E^\text{BX} \simeq E_{\mathbf Q_1} + E_{-\mathbf Q_1} + \min_\mathbf R \tilde {\mathcal V}^X_{\mathbf Q_1, -\mathbf Q_1}(\mathbf R).
\label{eq：biexciton_energy}
\end{equation}
The potential $\tilde {\mathcal V}^X_{\mathbf Q_1, -\mathbf Q_1}(\mathbf R)$ is most attractive at $\mathbf{Q_1} \approx \mathbf Q_d$ with maximum dipole moments.
Given the short range repulsion, the two excitons maintain a minimal separation of $|\mathbf R| \sim a_m$. 
Thus, the biexciton, comprising exciton pairs bound by opposite dipoles and distributed in two adjacent moir\'e cells shown in Fig.~\ref{fig: dipoledipoleinteraction}(d), has an estimated energy of 0.11 meV. 
This is significantly smaller than $2E_{\mathbf{Q_1}=0} = 1.51$ meV of two noninteracting excitons at the direct optical gap.
In the meanwhile, $E^\text{BX}$ is tunable by gating through the single exciton energy and the exciton dipole moment.

The bound biexciton develops a sizable quadrupole moment due to the spatial separation of exciton wavepackets and the antiparallel alignment of their dipole moments.
At $|\mathbf R| \sim a_m$, the $xy$ quadrupole moment is $\sim 30 \vert e \vert \cdot nm^2$, two orders of magnitude larger than the $zz$ quadrupole moment observed in other quadrupolar exciton species in TMD \cite{jasinski2025quadrupolar, lian2023quadrupolar, li2023quadrupolar, wang2023intercell, slobodkin2020quantum}. The giant quadrupole moment also enables dispersive optical responses \cite{gao2019nonreciprocal}.

The biexciton structures facilitate two-photon adsorption and recombination processes involving two dark excitons at finite and opposite momenta $\pm\mathbf Q_d$.
As schematically shown in Fig.~\ref{fig: dipoledipoleinteraction}(d), the electron and hole within each exciton have momenta differed by finite $\mathbf Q_d$, preventing the direct radiative recombination.
As a result, the sequential two-step optical transition via an intermediate single exciton state is suppressed \cite{klingshirn2007semiconductor}.
Nevertheless, the photon emission can occur via the cross annihilation between the electron in one exciton and the hole in the other, which satisfies the momentum conservation and opens a two-photon emission channel.

{\it Conclusion and discussion - }
In summary, we have demonstrated the existence of giant and helical exciton dipole arising from Berry curvature of electron and hole Chern bands with gate tunable helicity, magnitude on the order of $10^2$ D in tMoTe$_2$, leading to long-range and attractive dipole-dipole interactions and the optically bright biexciton.

The exciton proposed here possesses an excitation energy on the order of meV, positioning tMoTe$_2$ naturally in the THz frequency range and establishing it as a promising system for exploring excitonic physics beyond the visible regime.
Unlike visible-light techniques, THz spectroscopy enables phase-sensitive, time-domain access to the dynamical response of excitations \cite{yoon2024terahertz, li2024coherent, pashkin2003phase}. 
The intrinsic exciton dipole moment gives rise to a variety of dynamical phenomena absent in conventional optical regimes, including center-of-mass drift motion under an external electric field at the single exciton level (See SM Sec.~\ref{sec: Exciton drift velocity}) \cite{cao2021quantum,tang2024inheritance}, as well as inelastic scattering mediated by the dipole-dipole interactions \cite{klingshirn2007semiconductor}.
In addition, the accessibility of the biexciton via two-photon processes opens a nonlinear response channel in the THz regime, potentially serving as a source of entangled photon pairs for applications in quantum information science \cite{klingshirn2007semiconductor, utzat2019coherent, stevenson2006semiconductor, akopian2006entangled}.

{{\it Acknowledgment} - The authors thank Zhenglu Li and Xiaoyang Zhu for helpful discussions. This work is primarily supported by the U.S. Department of Energy, Office of Basic Energy Sciences, under Contract No. DE-SC0025327. 
This research used resources of the National Energy Research Scientific Computing Center, a DOE Office of Science User Facility supported by the Office of Science of the U.S. Department of Energy under Contract No. DE-AC02-05CH11231 using NERSC award BES-ERCAP0032546.
This work was facilitated through the use of advanced computational, storage, and networking infrastructure provided by the AI-core as well as the Hyak supercomputer system funded by the University of Washington Molecular Engineering Materials Center at the University of Washington (DMR-2308979).}

\bibliographystyle{apsrev4-2}
\bibliography{ref}

%========================================
% Supplemental materials
%========================================

\clearpage
\pagebreak
\widetext
\begin{center}
	{\large{\bf Supplementary Materials for "Giant and helical exciton dipole in flat Chern bands" }}
\end{center}

\setcounter{equation}{0}
\setcounter{figure}{0}
\setcounter{secnumdepth}{2}
\setcounter{page}{1}
\setcounter{section}{0}
\renewcommand{\theequation}{S\arabic{equation}}
\renewcommand{\thefigure}{S\arabic{figure}}
\renewcommand{\thetable}{S\arabic{table}}

In this Supplementary Material, we write down the details of the formalism for in-plane exciton dipole and perturbative analysis of exciton dipole for the gapped Dirac cones.

\section{Single Particle Hamiltonian}
\label{sec: single particle hamiltonian}
\subsection{Continuum model}

In this section, we construct the continuum model for tMoTe$_2$ from symmetries following Ref.~\cite{jia2024moire}.

\begin{figure}
    \centering
\includegraphics[width=0.5\columnwidth]{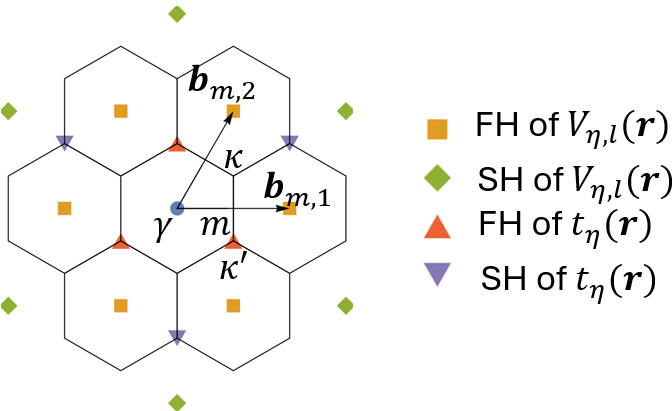}
\caption{
The moir\'e BZ. $\mathbf b_{\text{m},1}, \mathbf b_{\text{m},2}$ are primitive moir\'e reciprocal lattice vectors. 
FH, SH stand for the first, second harmonic, respectively.
Black honeycombs stand for the extended moir\'e Brillouin zone.}
\label{fig: moireBZ} 
\end{figure}

The lattice structure of tMoTe$_2$ is composed of two layers with the top layer rotated clockwise for an angle $-\theta/2$ and the bottom layer counterclockwise with the opposite angle $\theta/2$.
The moir\'e primitive lattice vectors are defined as 
\begin{equation}
    \mathbf a_{\text m,1}=a_{\text m, 0}\left(\cos \pi/6, -\sin \pi/6 \right) \quad 
    \mathbf a_{\text m,2}=a_{\text m, 0}\left(\cos \pi/6, \sin \pi/6 \right).
\end{equation}
$a_{\text m, 0} = a_0 / (2 \sin \theta / 2)$ is the moir\'e lattice constant and $a_0 = 3.52$ \AA \ is the lattice constant for the monolayer MoTe$_2$.
We take $\theta = 2.1^\circ$, which has a series of flat valence Chern bands mimic Landau level in terms of the quantum geometry \cite{wang2025higher}.
The reciprocal lattice vectors as shown in Fig.~\ref{fig: moireBZ} are
\begin{equation}
    \mathbf b_{\text m,1}=\frac{4\pi}{\sqrt 3 a_{\text m, 0}}\left(1,0\right) \quad 
    \mathbf b_{\text m,2}=\frac{4\pi}{\sqrt 3 a_{\text m, 0}}\left(\cos \pi/3, \sin \pi/3 \right).
\end{equation}
The moir\'e BZ corner is $\kappa = \mathbf b_{\text m,1} / 3 + \mathbf b_{\text m,2} / 3$ and $\kappa' = \mathbf b_{\text m,1} 2 / 3 - \mathbf b_{\text m,2} / 3$.
$\kappa (\kappa')$ corresponds to the atomic BZ corner $K$ of the bottom (top) layer in the moir\'e BZ.
We set $K_b =\kappa, K_t =\kappa', \tilde q = \kappa - \kappa'$.

The space group for tMoTe$_2$ is SG 150 generated by three-fold rotation about z axis $\mathcal C_{3z}$, two-fold rotation about y axis $\mathcal C_{2y}$, time reversal symmetry $\mathcal T$, and the moir\'e translation symmetry $T_{\mathbf R_\text m}$.
The creation operator in valley $\eta$ and layer $l$ at position $\mathbf r$ is denoted as $c_{\eta, l,\mathbf r} ^\dagger$.
$\eta=+1 (-1)$ for the valley $K (K')$ and $l=-1 (+1)$ for the bottom (top) layer.
The creation operation in the real space is the Fourier transform of the creation operator of plane waves in momentum space with the origin set as $\eta K_l$ or $\eta l \tilde q$.
The symmetry operators on the $c_{\eta, l,\mathbf r} ^\dagger$ are 
\begin{equation}
\begin{split}
\mathcal C_{3 z} c_{\eta ,l,\mathbf r}^{\dagger } \mathcal C_{3 z}^{-1} &= e^{i \eta \pi /3} c_{\eta ,l, \mathbf r}^{\dagger } \\
\mathcal C_{2 y} \mathcal T c_{\eta, l, \mathbf r}^{\dagger} \left(\mathcal C_{2 y} \mathcal T \right)^{-1} &= c_{\eta, -l, \mathbf r}^{\dagger} \\
T_{\mathbf R_\text m} c_{\eta ,l, \mathbf r}^{\dagger} T_{\mathbf R_\text m}^{-1} & =e^{-i \eta K_l\cdot \mathbf R_\text m} c_{\eta ,l, \mathbf r + \mathbf R_\text m}^{\dagger } \\
\mathcal T c_{\eta ,l, \mathbf r}^{\dagger } \mathcal T^{-1} &= c_{-\eta ,l, \mathbf r}^{\dagger }(-\eta ).
\end{split}
\end{equation}

Next, we apply the symmetry operator on the continuum model in the real space.
The continuum model is 
\begin{equation}
    h_{\eta }^0=
\int d^2r\left(c_{\eta ,b, \mathbf r}^{\dagger },c_{\eta ,t, \mathbf r}^{\dagger }\right)\left(
\begin{array}{cc}
 h_{\eta ,b}(\mathbf r) &  t_{\eta }(\mathbf r) \\
 t_{\eta }(\mathbf r)^* & h_{\eta ,t}(\mathbf r) \\
\end{array}\right)
\left(
\begin{array}{c}
 c_{\eta ,b,\mathbf r} \\
 c_{\eta ,t, \mathbf r} \\
\end{array}
\right).
\end{equation}
$h_{\eta ,l}(\mathbf r)=\frac{-\hbar ^2\partial_\mathbf r^2}{2 m_{\eta ,l}} + V_{\eta ,l}(\mathbf r)+ \frac{l \Delta_D}{2}$ with $V_{\eta ,l}(\mathbf r) $ as the intralayer potential for the valley $\eta$ and layer $l$ and $m_{\eta ,l}$ is the effective mass.
$ \Delta_D$ is the potential difference between two layers induced by the displacement field.
$t_{\eta }(\mathbf r)$ is the interlayer potential in the valley $\eta$.
The moir\'e translation symmetry determines the periodicity of the intra and inter layer moir\'e potential by
\begin{equation}
h_{\eta, \mathbf r}^0=T_{\mathbf R_\text{m}} h_{\eta ,\mathbf r}^0 T_{\mathbf R_\text{m}}^{-1}
=\int d^2r \left(c_{\eta ,b,r}^{\dagger },c_{\eta ,t,r}^{\dagger }\right)
\left(
\begin{array}{cc}
  h_{\eta ,b} \left(\mathbf r-\mathbf R_\text{m}\right) & t_{\eta } \left(\mathbf r-\mathbf R_\text{m}\right) e^{-i \eta  \left(K_b-K_t\right)\cdot \mathbf R_\text{m}} \\
  t_{\eta } \left(\mathbf r-\mathbf R_\text{m}\right)^* e^{i \eta \left(K_b-K_t\right)\cdot \mathbf R_\text{m}} & h_{\eta ,t}\left(\mathbf r-\mathbf R_\text{m}\right) \\
\end{array}
\right)\left(
\begin{array}{c}
 c_{\eta ,b,\mathbf r} \\
 c_{\eta ,t,\mathbf r} \\
\end{array}
\right).
\end{equation}
$V_{\eta ,l}(\mathbf r-\mathbf R_\text{m})=V_{\eta ,l}(\mathbf r)$ leads to  the moir\'e lattice periodicity of $V_{\eta ,l}(\mathbf r)$, which is Fourier transformed as
\begin{equation}
    V_{\eta ,l}(\mathbf r)=\sum _{\mathbf G} V_{\eta ,l,\mathbf G} e^{ -i\eta  \mathbf G\cdot \mathbf r}.
\end{equation}
$\mathbf G$ is the moir\'e reciprocal lattice vectors.
$t_{\eta }\left(\mathbf r-\mathbf R_\text{m}\right) e^{- i\eta \tilde q \cdot( \mathbf r-\mathbf R_\text{m})}=t_{\eta }(\mathbf r)$ results in the moir\'e periodicity of $t_{\eta }(\mathbf r) e^{i \eta \mathbf  q_1\cdot \mathbf r}$, which can be Fourier transformed as
\begin{equation}
    t_{\eta }(\mathbf r)=\sum _{\mathbf G} t_{\eta ,\tilde q + \mathbf G} e^{-i \eta \left(\mathbf G+\tilde q\right)\cdot \mathbf r}.
\end{equation}
$\mathcal C_{3z}$ relates the Fourier transform of moir\'e potentials by
\begin{equation}
     h_{\eta,\mathbf r }^0=\mathcal C_{3z} h_{\eta,\mathbf r }^0\mathcal C_{3z}^{-1}=
     \int d^2r \left(c_{\eta ,b,\mathbf r}^{\dagger },c_{\eta ,t,\mathbf r}^{\dagger }\right)\left(
\begin{array}{cc}
 h_{\eta ,b}\left(\mathcal C_{3z}^{-1}\mathbf r\right) & t_{\eta }(\mathcal C_{3z}^{-1}\mathbf r) \\
 t_{\eta}(\mathcal C_{3z}^{-1}\mathbf r)^* & h_{\eta ,t}\left(\mathcal C_{3z}^{-1}\mathbf r\right) \\
\end{array}
\right)\left(
\begin{array}{c}
 c_{\eta ,b,\mathbf r} \\
 c_{\eta ,t,\mathbf r} \\
\end{array}
\right),
\end{equation}
and constrains the moir\'e potentials by
\begin{equation}
\begin{split}
    V_{\eta,l}\left(\mathcal C_{3z}^{-1}\mathbf r\right) = V_{\eta,l}\left(\mathbf r\right) & \rightarrow V_{\eta ,l, \mathbf G}=V_{\eta ,l,\mathcal C_{3z}^{-1} \mathbf G} \\
    t_{\eta}\left(\mathcal C_{3z}^{-1}\mathbf r\right) = t_{\eta}\left(\mathbf r\right) & \rightarrow 
    t_{\eta, \tilde q+\mathbf G}=t_{\eta, \mathcal C_{3z}^{-1}(\tilde q+\mathbf G)}.
\end{split}
\label{eq: c3}
\end{equation}
$\mathcal C_{2y} \mathcal T$ relates layer degree of freedom by
\begin{equation}
    h_{\eta, \mathbf r}^0= (\mathcal C_{2y} \mathcal T) h_{\eta, \mathbf r}^0 (\mathcal C_{2y} \mathcal T)^{-1}
    =\int d^2 r \left(c_{\eta ,b,\mathbf r}^{\dagger },c_{\eta ,t,\mathbf r}^{\dagger }\right)\left(
\begin{array}{cc}
 h_{\eta ,t}(\mathcal C_{2y}^{-1}\mathbf r)^* & t_{\eta }(\mathcal C_{2y}^{-1}\mathbf r) \\
 t_{\eta }(\mathcal C_{2y}^{-1}\mathbf r)^* & h_{\eta ,b}(\mathcal C_{2y}^{-1}\mathbf r)^* \\
\end{array}
\right)\left(
\begin{array}{c}
 c_{\eta,b,\mathbf r} \\
 c_{\eta,t,\mathbf r} \\
\end{array}
\right),
\end{equation}
and constrains the Hamiltonian by
\begin{equation}
\begin{split}
    m_{\eta ,-l} & = m_{\eta ,l} \\
    V_{\eta ,-l}(\mathcal C_{2y}^{-1}\mathbf r)^*=V_{\eta ,l}(\mathbf r) & \rightarrow V_{\eta ,-l,-\mathcal C_{2y}^{-1}\mathbf G}^*=V_{\eta ,l, \mathbf G} \\
    t_{\eta}\left(\mathcal C_{2y}^{-1}\mathbf r\right) = t_{\eta}\left(\mathbf r\right) & \rightarrow t_{\eta, \mathcal C_{2y}^{-1}(\tilde q + \mathbf G)}=t_{\eta,(\tilde q + \mathbf G)}.
\end{split}
\label{eq: c2yT}
\end{equation}
$\mathcal T$ relates the Hamiltonian in two valleys by
\begin{equation}
    h_{\eta, \mathbf r}^0= \mathcal T h_{\eta, \mathbf r}^0 \mathcal T^{-1}
    =\int d^2r\left(c_{\eta ,b,\mathbf r}^{\dagger },c_{\eta ,t,\mathbf r}^{\dagger }\right)\left(
\begin{array}{cc}
 h_{-\eta ,b}(\mathbf r)^* & t_{-\eta }(\mathbf r)^* \\
 t_{-\eta }(\mathbf r) & h_{-\eta ,t}(\mathbf r)^* \\
\end{array}
\right)\left(
\begin{array}{c}
 c_{\eta ,b,\mathbf r} \\
 c_{\eta ,t,\mathbf  r} \\
\end{array}
\right),
\end{equation}
and constrains the Hamiltonian by
\begin{equation}
\begin{split}
    m_{\eta ,l} & =m_{-\eta ,l} \\
    V_{-\eta ,l}(\mathbf r)^* = V_{\eta ,l}(\mathbf r) & \rightarrow V_{-\eta ,l, \mathbf G}^*=V_{\eta ,l, \mathbf G} \\
    t_{-\eta }(\mathbf r)^* = t_{\eta }(\mathbf r) & \rightarrow t_{-\eta , \tilde q + \mathbf G}^* = t_{\eta , \tilde q + \mathbf G}.
\end{split}
\label{eq: T}
\end{equation}

Next, we transform the Hamiltonian into the momentum space.
The momentum space creation operator is defined as
\begin{equation}
    c_{\eta ,l, \mathbf r}^{\dagger }=\frac{1}{\sqrt{V}}\sum _{\mathbf k\in \text{MBZ}, \mathbf q \in \mathbf q_{\eta ,l}}e^{-i (\mathbf k-\mathbf q)\cdot \mathbf r} c_{\eta ,l,\mathbf k-\mathbf q}^{\dagger }\ ,
\end{equation}
where $\mathbf q_{\eta ,l}=\eta  l \tilde q + \mathbf G_{\text{m}}$ and MBZ is the first moir\'e BZ.
Each term in the Hamiltonian is transformed as
\begin{equation}
\begin{split}
    \int d^2  c_{\eta ,l,\mathbf r}^\dagger V_{\eta,l}(\mathbf r) c_{\eta ,l, \mathbf r} & = \sum _{\mathbf k_1, \mathbf k_2\in \text{MBZ},\mathbf q_1, \mathbf q_2\in \mathbf q_{\eta ,l}}
    c_{\eta ,l,\mathbf k_1- \mathbf q_1}^{\dagger } \delta _{\mathbf k_1, \mathbf k_2} V_{\eta ,l,\eta  \left(\mathbf q_1- \mathbf q_2\right)} c_{\eta,l,\mathbf k_2-\mathbf q_2} \\
    \int d^2  c_{\eta ,l,\mathbf r}^\dagger (\frac{-\hbar ^2\partial_\mathbf r^2}{2 m_{\eta ,l}} + \frac{l \Delta_D}{2})  c_{\eta ,l, \mathbf r} & =
    \sum _{\mathbf k_1, \mathbf k_2\in \text{MBZ},\mathbf q_1, \mathbf q_2\in \mathbf q_{\eta ,l}}
    c_{\eta ,l,\mathbf k_1- \mathbf q_1}^{\dagger } \delta _{\mathbf k_1, \mathbf k_2} \delta _{\mathbf q_1, \mathbf q_2} (\frac{\hbar^2 \vert \mathbf q_1 \vert^2}{2 m_{\eta,l}} + \frac{l \Delta_D}{2}) c_{\eta,l,\mathbf k_2-\mathbf q_2} \\
    \int d^2 c_{\eta ,l, \mathbf r}^{\dagger } t_{\eta }(\mathbf r) c_{\eta ,-l,\mathbf r} 
    &=\sum _{\mathbf k_1, \mathbf k_2\in \text{MBZ},\mathbf q_1\in \mathbf q_{\eta ,l},\mathbf q_2\in \mathbf q_{\eta ,-l}} c_{\eta ,l, \mathbf k_1-\mathbf q_1}^{\dagger } \delta _{\mathbf k_1, \mathbf k_2} t_{\eta ,\eta  \left(\mathbf q_1-\mathbf q_2\right)} c_{\eta, -l, \mathbf k_2-\mathbf q_2}.
\end{split}
\end{equation}

Finally, We consider the symmetry constraints on the first and second harmonic Fourier coefficients of the moir\'e potentials.
For the intralayer potential $V_{\eta ,l}(\mathbf r)$, the first harmonic of $\mathbf G$ is  $ \left\{\pm\mathbf b_{\text{m},1}, \pm\mathcal C_{3z}\mathbf b_{\text{m},1},\pm\mathcal C_{3z}^2\mathbf b_{\text{m},1}\right\}$, shown yellow square in Fig.~\ref{fig: moireBZ}.
For the interlayer potential $t_{\eta }(\mathbf r)$, the first harmonic of $\tilde q +\mathbf  G$  is $\left\{\tilde q,\mathcal C_{3z} \tilde q,\mathcal C_{3z}^2 \tilde q\right\}$, shown as red upper triangle in Fig.~\ref{fig: moireBZ}.
$\mathcal C_{3z}$ in \eqnref{eq: c3} and the Hermitian condition of the Hamiltonian tells there is only one independent complex parameter $V_{\eta ,l,\mathbf b_{\text{m},1}}$ for the intralayer moir\'e potential in one layer and one valley and $t_{\eta ,\tilde q}$ for the interlayer moir\'e potential.
$t_{\eta ,\tilde q}$ can be chosen as a real number by choosing the relative phase between $c_{\eta ,t, \mathbf r}^{\dagger }$ and $c_{\eta ,b, \mathbf r}$.
$\mathcal C_{2y} \mathcal T$ in \eqnref{eq: c2yT} tells 
\begin{equation}
    V_{\eta ,l, \mathbf b_{\text{m},1}}=V_{\eta ,-l, \mathbf b_{\text{m},1}}^*=V_{\eta ,-l,-\mathbf b_{\text{m},1}}.
\end{equation}
$\mathcal T$ in \eqnref{eq: T} tells 
\begin{equation}
\begin{split}
    V_{\eta ,l,\mathbf b_{\text{m},1}} &= V_{-\eta ,l,\mathbf b_{\text{m},1}}^*=V_{-\eta ,l,-\mathbf b_{\text{m},1}} \\
    t_{\eta ,\tilde q} &= t_{-\eta ,\tilde q}^*.
\end{split}
\end{equation}
An effective intravalley inversion symmetry emerges when all these constraints for the first harmonic coefficients are satisfied, which is defined as
\begin{equation}
    \mathcal{I} c_{\eta ,l,\mathbf r}^{\dagger }\mathcal{I}^{-1}=c_{\eta ,-l,-\mathbf r}^{\dagger }.
\end{equation}

Summarizing all symmetry constraints, there is only one independent complex parameter $V_{\eta ,l,\mathbf b_{\text{m},1}}$ for the intralayer potentials in both layers and valleys, one real parameter $t_{\eta ,\tilde q}$ for the interlayer potentials in both valleys, and the same mass $m$ for both layers and valleys.
We can analyze the symmetry constraints of the second harmonic expansion of the intralayer and interlayer moir\'e potentials similarly.
The second harmonics  of $ G_\text{m}$ for the $V_{\eta ,l}(\mathbf r)$ is $
\left\{\pm \left(\mathbf b_{\text{m},1}+\mathbf b_{\text{m},2}\right),\pm \left(2 \mathbf b_{\text{m},2}-\mathbf b_{\text{m},1},\right),\pm \left(\mathbf b_{\text{m},2}-2 \mathbf b_{\text{m},1}\right)\right\}$, shown as green diamond in Fig.~\ref{fig: moireBZ}, and there exist one independent complex parameter $V_{\eta, l, \mathbf b_{\text{m},1}+\mathbf b_{\text{m},2}}$.
The second harmonics  of $ G_\text{m}$ for the $t_{\eta}(\mathbf r)$ is $\tilde q +\mathbf  G_\text{m}$  is $\left\{\tilde q +\mathbf b_{\text{m},1},\mathcal C_{3z} \tilde q - \mathbf b_{\text{m},1}+\mathbf b_{\text{m},2},\mathcal C_{3z}^2 \tilde q-\mathbf b_{\text{m},2}\right\}$, shown as purple downward triangle in Fig.~\ref{fig: moireBZ}, and there exist one independent complex parameter $t_{\eta ,\tilde q +\mathbf b_{\text{m},1}}$.
If we keep the intravalley inversion symmetry, $V_{\eta, l, \mathbf b_{\text{m},1}+\mathbf b_{\text{m},2}}$ and $t_{\eta ,\tilde q +\mathbf b_{\text{m},1}}$ are chosen as real.
We take the parameters \cite{jia2024moire, ahn2024non, wang2024fractional} as
\begin{equation}
\begin{split}
    m &= 0.6m_e, \\
    \quad V_{\eta = K ,l =b,\mathbf b_{\text{m},1}}&=20.51 e^{-i 61.49\pi/180 } \text{meV}, 
    \quad t_{\eta=K, \tilde q} = -7.01 \text{meV}, \\
    \quad V_{\eta, l, \mathbf b_{\text{m},1}+\mathbf b_{\text{m},2}} &= -9.08 \text{meV},
    \quad t_{\eta ,\tilde q +\mathbf b_{\text{m},1}} = 11.08 \text{meV}.
\end{split}
\end{equation}
$m_e$ is the bare electron mass.

The eigen equation in the valley K reads
\begin{equation}
    h^0_{\eta = K}(\mathbf k) \ket{ u_{n, \mathbf k} }= \epsilon_{n, \mathbf k} \ket{ u_{n, \mathbf k} }
\label{eq: single particle state}
\end{equation}
$h^0_{\eta}(\mathbf k)$ is the single particle Hamiltonian expanded in the basis $c_{\eta ,l=b,\mathbf k -(-\eta \tilde q + \mathbf G)}^{\dagger }, c_{\eta ,l=b,\mathbf k -(\eta \tilde q + \mathbf G)}^{\dagger }$ with an cut off of $\mathbf G$.
The coefficient of $\ket{ u_{n, \mathbf k}}$ in the basis is denoted as $u_{n, \mathbf k, \mathbf G, l}$. 
In the following sections, we focus on the valley K and drop the $\eta$ index.
We also treat the layer $l$ as pseudo spin.

\subsection{Optimal Chern gauge}
\label{sec: optimal Chern gauge}

\begin{figure}
    \centering
\includegraphics[width=\columnwidth]{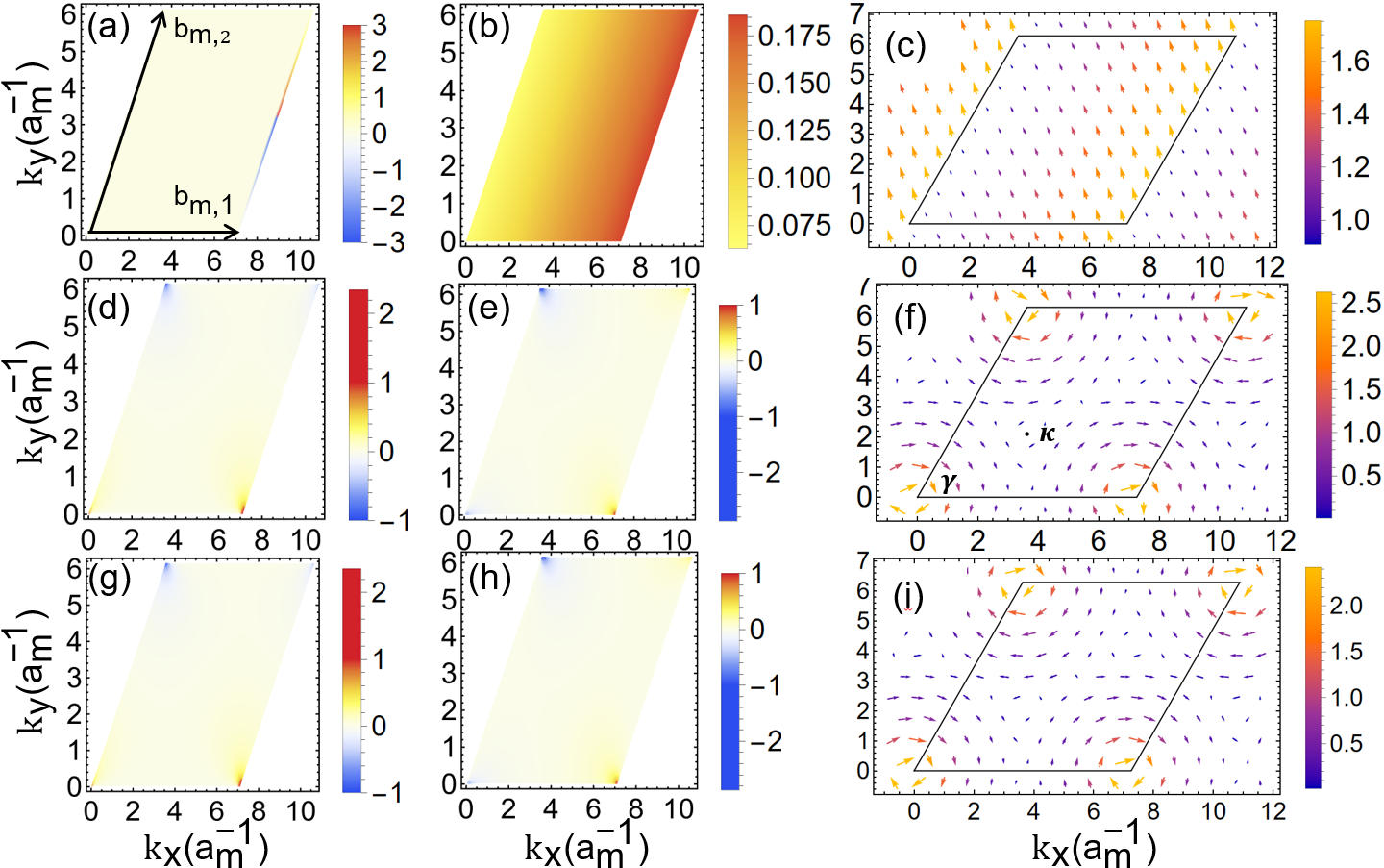}
\caption{
(a)(b)(c) The Wilson line $w_1(\mathbf k)$, Wilson line $w_2(\mathbf k)$, Berry connection $A^s(\mathbf k)$ after applying the smooth gauge with a line discontinuity in the first step, respectively.
The arrows in (c) is the direction of $(A^s_x(\mathbf k),A^s_y(\mathbf k))$ and its length and color represent the magnitude of $A^s(\mathbf k)$.
Here, $N_1 = N_2 =50$.
(d)(e)(f) The Wilson line $w_1(\mathbf k)$, Wilson line $w_2(\mathbf k)$, Berry connection $A^s(\mathbf k)$ after applying the gauge $\chi^v(\mathbf k)$ with a singular vortex at $\gamma$ in the second step, respectively. 
(g)(h)(i) The Wilson line $w_1(\mathbf k)$, Wilson line $w_2(\mathbf k)$, Berry connection $A^s(\mathbf k)$ after applying the gauge $\chi^c(\mathbf k)$ in the third step, respectively. }
\label{fig: optimal chern gauge} 
\end{figure}

In this section, we specify the gauge chosen for the states in VB1 and VB2 with the valley Chern number $C_K=1$ following Ref.~\cite{xie2024chern}, known as the optimal Chern gauge.
The optimal Chern gauge is defined as $\nabla_\mathbf k \cdot A_n (\mathbf k) = 0$ with $A_n (\mathbf k)$ as the Berry connection.

We start from a random gauge of single particle states and construct the Wilson line on a momentum mesh.
The momentum mesh is defined as 
\begin{equation}
    \mathbf k = \frac{k_1}{N_1} \mathbf b_{\text{m},1} + \frac{k_2}{N_2} \mathbf b_{\text{m},2}
\end{equation}
with $N_1, N_2$ as integers, $k_1 = 0, 1, \dots, N_1-1$, and $k_2 = 0, 1, \dots, N_2-1$.
We denote the differential length in the direction $\mathbf b_{\text{m},1}$ as $\delta_1 =\vert\mathbf b_{\text{m},1}\vert /N_1$, the differential length in the direction $\mathbf b_{\text{m},2}$ as $\delta_2 = \vert\mathbf b_{\text{m},2}\vert /N_2$, and the differential MBZ area as $\delta A_\text{BZ} = \vert \mathbf b_{\text{m},1}\times \mathbf b_{\text{m},2} \vert / N_1 N_2$.
After the eigenstates $\ket{u_{n,k_1,k_2}}$ are obtained, we construct the inner product of the adjacent states by
\begin{equation}
\begin{split}
    e^{i w_1\left(k_1,k_2\right)}=\frac{\left\langle u_{n,k_1,k_2}|u_{n,k_1+1 ,k_2}\right\rangle }{\left| \left\langle u_{n,k_1,k_2}|u_{n,k_1+1,k_2}\right\rangle \right| } \\
    e^{i w_2\left(k_1,k_2\right)}=\frac{\left\langle u_{n,k_1,k_2}|u_{n,k_1,k_2+1}\right\rangle }{\left| \left\langle u_{n,k_1,k_2}|u_{n,k_1,k_2+1}\right\rangle \right| }.
\end{split}
\end{equation}
At the MBZ boundary, $\ket{u_{n,k_1 = N_1,k_2}} = U_{\mathbf b_{\text{m},1}}\ket{u_{n,k_1 =0,k_2}}$ and $\ket{u_{n,k_1,k_2 = N_2}} = U_{\mathbf b_{\text{m},2}}\ket{u_{n,k_1,k_2=0}}$ where $U_{\mathbf G_{\text{m}}} h^0_{\eta }(\mathbf k) U_{\mathbf G_{\text{m}}}^\dagger = h^0_{\eta }(\mathbf k + \mathbf G_{\text{m}})$.
We drop the $n$ index in this section and we take VB1 in the valley K as an example.
The Berry connection $A_n(\mathbf k) = - i \left\langle u_{n,\mathbf k}|\partial_\mathbf k u_{n,\mathbf k}\right\rangle$ is connected to the Wilson line by 
\begin{equation}
     w_1\left(k_1, k_2 \right) = \delta_1 A_1(k_1 +1/2, k_2), \quad
     w_2\left(k_1, k_2 \right) = \delta_2 A_2(k_1, k_2+1/2).
\end{equation}
The Berry curvature can be calculated by
\begin{equation}
    e^{i \tilde \Omega(k_1,k_2)}=e^{i w_1\left(k_1,k_2\right)} \left(e^{i w_2\left(k_1,k_2\right)}\right)^* e^{i w_2\left(k_1+1,k_2\right)}\left(e^{i w_1\left(k_1,k_2+1\right)}\right)^*,
    \label{eq: berrycurvature}
\end{equation}
with $\tilde \Omega(k_1,k_2) = \Omega _{n}(k_1,k_2) \delta A_\text{BZ}$ \cite{fukui2005chern}.

The first step is constructing a smooth gauge over MBZ except a line discontinuity at the MBZ boundary as shown in Fig.~\ref{fig: optimal chern gauge}(a)-(c).
We start from the Wilson loops defined as
\begin{equation}
    e^{i W_1\left(k_2\right)}=\prod _{k_1} e^{i w_1\left(k_1,k_2\right)}, \quad e^{i W_2\left(k_1\right)}=\prod _{k_2} e^{i w_2\left(k_1,k_2\right)}.
\end{equation}
$W_1\left(k_2 = N_2\right)- W_1\left(k_2 = 0\right) = -2 \pi C$ and $W_2\left(k_1 = N_1\right)- W_2\left(k_1 = 0\right) = 2 \pi C$ with $C$ as the Chern number of the band.
A smooth gauge can be chosen by fixing the Wilson line as
\begin{equation}
    w_1\left(k_1,0\right)=\frac{W_1(0)}{N_1}, \quad w_2\left(k_1,k_2\right)=\frac{W_2\left(k_1\right)}{N_2}.
\label{eq: wilson line 2}
\end{equation}
\eqnref{eq: berrycurvature} generates the remaining Wilson line by
\begin{equation}
\begin{split}
    w_1\left(k_1,k_2 +1\right) & =w_1\left(k_1,k_2\right)-\tilde\Omega\left(k_1,k_2\right)+\frac{W_2\left(k_1+1\right)}{N_2}-\frac{W_2\left(k_1\right)}{N_2} \\
    &=w_1\left(k_1,k_2 =0\right) - \sum_{k_2'=0}^{k_2} \tilde\Omega\left(k_1,k_2\right) +k_2 (\frac{W_2\left(k_1+1\right)}{N_2}-\frac{W_2\left(k_1\right)}{N_2})
\end{split}
\label{eq: wilson line 1}
\end{equation}
At the boundary $k_1 = N_1-1$,
\begin{equation}
\begin{split}
    w_1\left(k_1 = N_1 -1,k_2 +1\right)
    &=w_1\left(k_1,k_2 =0\right) - \sum_{k_2'=0}^{k_2} \tilde\Omega\left(k_1,k_2\right) +k_2 (\frac{W_2\left(N_1\right)}{N_2}-\frac{W_2\left(N_1 - 1 \right)}{N_2}) \\
    & = w_1\left(k_1,k_2 =0\right) - \sum_{k_2'=0}^{k_2} \tilde\Omega\left(k_1,k_2\right) +k_2 (\frac{W_2\left(0\right)}{N_2}-\frac{W_2\left(N_1 - 1 \right)}{N_2})  + 2\pi C k_2 / N_2
\end{split}
\end{equation}
% \begin{equation}
% \begin{split}
%         & w_1\left(k_1 = N_1-1, k_2 +1 \right) \\
%         &=w_1\left(k_1 = N_1-1, k_2 \right)+\frac{W_2(k_1 = N_1)}{N_2}-\frac{W_2 \left(k_1 = N_1 -1\right)}{ N_2}-\tilde \Omega \left(k_1 = N_1-1, k_2\right) \\ 
%         &=w_1\left(k_1 = N_1-1, k_2 \right)+\frac{W_2(k_1 = 0)}{N_2}-\frac{W_2 \left(k_1 = N_1 -1\right)}{ N_2}-\tilde \Omega \left(k_1 = N_1-1, k_2 \right) + 2\pi C / N_2,
% \end{split}
% \end{equation}
with a line discontinuity proportional to $2\pi C k_2 / N_2$ when $C$ is nonzero as shown in Fig.~\ref{fig: optimal chern gauge}(a).
Fig.~\ref{fig: optimal chern gauge}(b) shows the Wilson line $w_2\left(k_1,k_2\right)$ has a constant jump at the MBZ boundary by $w_2\left(k_1 = N_1 ,k_2\right) - w_2\left(k_1 = 0 ,k_2\right) = 2 \pi C / N_2$.
The Berry connection after the gauge choice is denoted as $A^s(\mathbf k)$ as shown in Fig.~\ref{fig: optimal chern gauge}(c), which has a line discontinuity at one side of the MBZ boundary with $k_1 = N_1$.

The second step transforms the line discontinuity to a vortex by the Weierstrass sigma function as shown in Fig.~\ref{fig: optimal chern gauge}(d)-(f).
The gauge transformation, denoted as $e^{i \chi_v(\mathbf k)}$ is chosen to satisfy
\begin{equation}
    \chi_v (\mathbf k + \mathbf b_{\text{m},1}) - \chi_v (\mathbf k) = - 2 \pi C k_2 /N_2, \quad
    \chi_v (\mathbf k + \mathbf b_{\text{m},2}) - \chi_v (\mathbf k) = 0.
\label{eq: singular gauge boundary condition}
\end{equation}
The $\chi_v(\mathbf k)$ is chosen as 
\begin{equation}
    \chi _v(k)=-\mathcal{C} \Im\left(\log \left(\sigma \left(z_\mathbf k-z_v, \left\{z_{\text{1}},z_{\text{2}}\right\}\right)\right)+A z_\mathbf k^2+B z_\mathbf k^* z_k+C z_\mathbf k\right),
    \label{eq: singular gauge}
\end{equation}
where $z_\mathbf k=k_x+i k_y, z_{1}=b_{\text{m},1,x}+i b_{\text{m},1,y}, z_{2}=b_{\text{m},2,x}+i b_{\text{m},2,y}$, and $\sigma$ is the Weierstrass sigma function with the double periodicity $\left\{z_{\text{1}},z_{\text{2}}\right\}$.
The constants $A,B,C$ are determined by matching the boundary condition in \eqnref{eq: singular gauge boundary condition} and given by
\begin{equation}
\begin{split}
    B &=\frac{\pi  i \left(z_{\text{1}}^* z_{\text{2}}+z_{\text{1}} z_{\text{2}}^*\right)}{\left(z_\text{1}^* z_{\text{2}}-z_{\text{1}} z_{\text{2}}^*\right)^2} \\
    A & =-\frac{2 B z_{\text{2}}^*+\eta _2}{2 z_{\text{2}}} \\
    C&=2 i \frac{z_{\text{2}}^*\left(\pi -\Im\left(A z_{\text{1}}^2+B z_{\text{1}}^* z_{\text{1}}+\frac{\eta _1 z_{\text{1}}}{2}-\eta_1 z_v\right)\right)-z_{\text{1}}^*\left(\pi -\Im\left(A z_{\text{2}}^2+B z_{\text{2}}^* z_{\text{2}}+\frac{\eta_2 z_{\text{2}}}{2}-\eta _2 z_v\right)\right)}{z_{\text{1}} z_{\text{2}}^*-z_{\text{1}}^* z_{\text{2}}}.
\end{split}
\end{equation}
$\eta_{1,2}$ is the Weierstrass invariants defined as  
\begin{equation}
\begin{split}
    \log\left(\sigma \left(z_\mathbf k +z_{\text{1}} ,\left\{z_{\text{1}},z_{\text{2}}\right\}\right)\right)& =\log \left(\sigma \left(z_\mathbf k,\left\{z_{\text{1}},z_{\text{2}}\right\}\right)\right)+\eta _1\left(\frac{z_{\text{1}}}{2}+z_\mathbf k\right)+i \pi \\ 
    \log\left(\sigma \left(z_\mathbf k +z_{\text{2}} ,\left\{z_{\text{1}},z_{\text{2}}\right\}\right)\right)& =\log \left(\sigma \left(z_\mathbf k,\left\{z_{\text{1}},z_{\text{2}}\right\}\right)\right)+\eta _2\left(\frac{z_{\text{2}}}{2}+z_\mathbf k\right)+i \pi
\end{split}
\end{equation}
$z_v$ is the position of the vortex and we choose $z_v$ as the reciprocal lattice vectors to locate the vortex at $\gamma$ and the specific reciprocal lattice vectors is determined by setting $\mathbf A_n (\kappa)=0$.
For VB1, we take the $z_v = -b_{\text{m}, 2, x} - i b_{\text{m}, 2, y}$.
The Berry connection after the gauge transformation of $\chi_v(\mathbf k)$ is shown in Fig.~\ref{fig: optimal chern gauge}(f), denoted as $A^v(\mathbf k)$ with $A^v(\mathbf k)=  A^s(\mathbf k)+\nabla_\mathbf k \chi^v(\mathbf k)$.
The maximum of $A^v(\mathbf k)$ is located at $\gamma$ with a vortex with divergent magnitude while the minimum of $A^v(\mathbf k)$ is located at $\kappa$ with zero magnitude.
Except the vortex, the Berry connection and the Wilson lines in Fig.~\ref{fig: optimal chern gauge}(d)(e) are continuous across both the MBZ boundaries. 

\begin{figure}
    \centering
\includegraphics[width=\columnwidth]{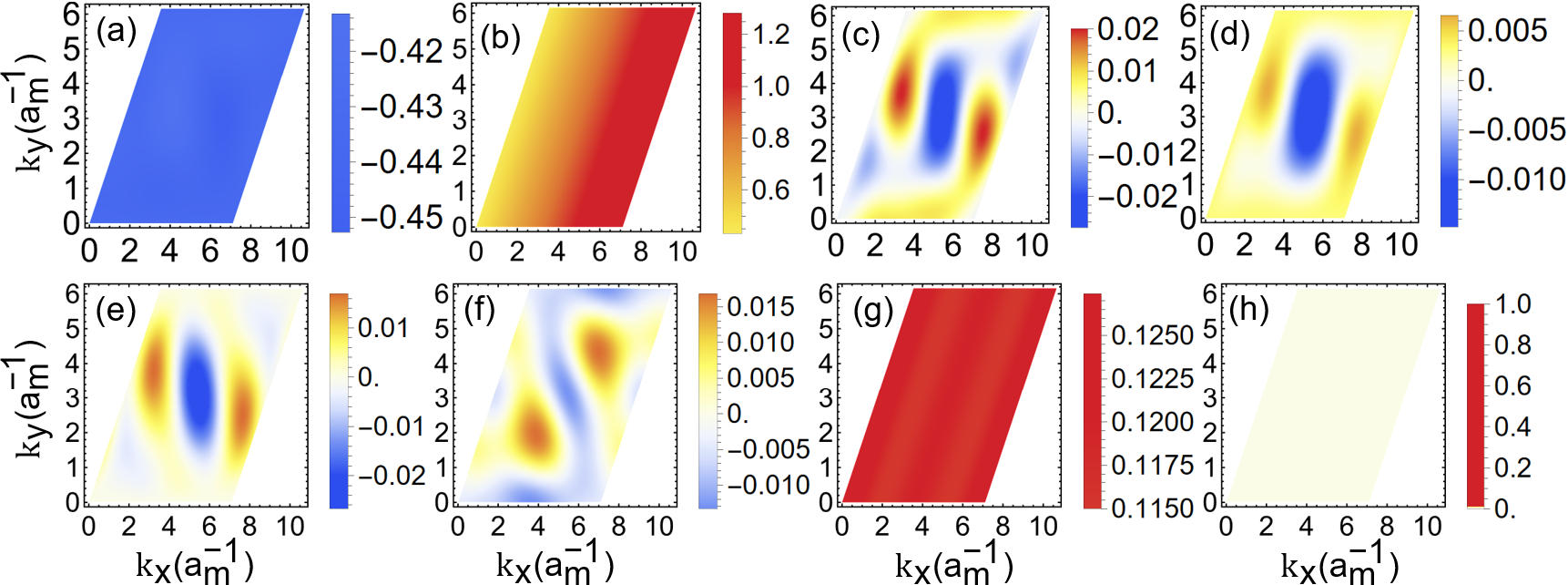}
\caption{
(a)(b) The Berry connection without the line discontinuity $\tilde A^s_1(\mathbf k), \tilde A^s_2(\mathbf k)$, respectively.
(c) The divergence of the Berry connection $\nabla_\mathbf k\cdot A^v(\mathbf k)$.
(d) The gauge transformation $\chi^c(\mathbf k)$ in the third step.
(e)-(h) The derivative of the Berry connection $\partial_1 \tilde A^s_1(\mathbf k), \partial_2 \tilde A^s_1(\mathbf k), \partial_1 \tilde A^s_2(\mathbf k), \partial_2 \tilde A^s_2(\mathbf k)$, respectively.
}
\label{fig: derivative of A} 
\end{figure}

The third step clears the remaining divergence of the Berry connection by solving the Poisson equation $\nabla_\mathbf k^2 \chi_c(\mathbf k) = - \nabla_\mathbf k\cdot A^v(\mathbf k)$.
We first derive $\nabla_\mathbf k\cdot A^v(\mathbf k)$.
We analytically drop the line continuity in  $A^s(\mathbf k)$, which is canceled in $A^v(\mathbf k)$, and compute the divergence of $\chi_v(\mathbf k)$ to avoid the numerical differentiation of the line discontinuity and the singular gauge vortex.
We remove the line discontinuity in the Wilson line in \eqnref{eq: wilson line 1} and \eqnref{eq: wilson line 2} as shown in Fig.~\ref{fig: derivative of A}(a)(b) by 
\begin{equation}
\begin{split}
    \tilde A^s_1 (k_1 +1/2, k_2) &= \frac{1}{\delta_1} (w_1(k_1, k_2)- \delta_{k_1, N_1-1} 2 \pi C k_2 / N_2 ) \\
    \tilde A^s_2 (k_1, k_2+1/2) &= \frac{1}{\delta_2} (w_2(k_1, k_2) + \delta_{k_1, N_1} 2 \pi C / N_2),
\end{split}
\label{eq: smooth gauge without line discontinuity}
\end{equation}
where $\delta_{k_1, N_1-1}, \delta_{k_1, N_1}$ are Kronecker delta functions. Their derivatives can be numerically computed as shown in Fig.~\ref{fig: derivative of A}(e)-(h), which are periodic in MBZ.
The divergence $\nabla_\mathbf k\cdot \tilde A^s(\mathbf k)$ on a non square mesh can be computed as
\begin{equation}
    \nabla_\mathbf k \cdot \tilde A^s(\mathbf k) = 
    \frac{\partial_1 \tilde A^s_1+\partial_2 \tilde A^s_2-\frac{\mathbf b_{\text{m},1}\cdot\mathbf b_{\text{m},2}}{\vert \mathbf b_{\text{m},1}\vert \vert\mathbf b_{\text{m},2}\vert} \left(\partial_2 \tilde A^s_1+\partial_1\tilde A^s_2\right)}{1-\left(\frac{\mathbf b_{\text{m},1}\cdot\mathbf b_{\text{m},2}}{\vert \mathbf b_{\text{m},1}\vert \vert\mathbf b_{\text{m},2}\vert}\right)^2}.
\end{equation}
The divergence of $\nabla_\mathbf k \chi^v(\mathbf k)$ from \eqnref{eq: singular gauge} is
\begin{equation}
    \nabla_\mathbf k^2 \chi^v(\mathbf k) =  4 \partial_z \partial_{z^*}  \chi^v(z_\mathbf k) = - 4 C \Im B,
\label{eq: div singular gauge}
\end{equation}
as the Weierstrass sigma function is a holomorphic function.
Combining \eqnref{eq: smooth gauge without line discontinuity} and \eqnref{eq: div singular gauge} results in 
\begin{equation}
    \nabla_\mathbf k \cdot A^v(\mathbf k) = \nabla_\mathbf k \cdot \tilde A^s(\mathbf k) - 4 C \Im B,
\end{equation}
which is shown in Fig.~\ref{fig: derivative of A}(c) and periodic over MBZ.
Then, $\chi^c(\mathbf k)$ shown in Fig.~\ref{fig: derivative of A}(d) is solved by Fourier transform of the Poisson equation
\begin{equation}
\begin{split}
    \mathcal F(-\nabla_\mathbf k \cdot A^v)_{\mathbf R_{\text{m}} } & = \frac{1}{N_1 N_2} \sum_{\mathbf k} e^{-i \mathbf k \cdot \mathbf R_{\text{m}}}(-\nabla_\mathbf k \cdot A^v(\mathbf k)) \\
    \chi^c(\mathbf k) & = \sum_{\mathbf R_{\text{m}} \neq 0} \frac{-1}{\vert \mathbf R_{\text{m}} \vert^2}e^{i \mathbf k \cdot \mathbf R_{\text{m}}} \mathcal F(-\nabla_\mathbf k \cdot A^v)_{\mathbf R_{\text{m}} },
\end{split}
\end{equation}
where $\mathbf R_{\text{m}}$ is the real space moir\'e lattice vectors.
By $\mathbf R_{\text{m}} \neq 0$, we remove the constant term in $\chi^c(\mathbf k)$ that does not contribute to $A^c(\mathbf k)$.
The resulting Berry connection, denoted as $A^c(\mathbf k)$, and Wilson lines are  shown in Fig.~\ref{fig: optimal chern gauge}(g)-(i), which qualitatively agrees with the $A^v(\mathbf k)$ and removes the numerical redundancy in  $A^v(\mathbf k)$.

\subsection{Quantum geometry of the top two valence minibands}
\label{sec: quantum geometry of the top two valence minibands}
In this section, we discuss the Berry connection, Berry curvature, and quantum metric for VB1 and VB2 at $\Delta_D = 0$ and $6$meV induced by the displacement field.
The optimal Chern gauge for VB1 and VB2 has the following three properties:
1) A singular vortex located at $\gamma$.
2) $A_n (\mathbf k) \approx \Omega_n(\kappa)\hat z\times (\mathbf k-\kappa) / 2$ near $\kappa$ with $\Omega_n(\kappa)$ as the Berry curvature at $\kappa$.
3) The continuous variation of the Berry connection mainly at $\kappa, \kappa'$ determined by the Berry curvature when the displacement field is tuned.

We briefly mention the numerical evaluation of the quantum geometry quantities.
The Berry connection $A_n(\mathbf k)$ is chosen as discussed in Sec.~\ref{sec: optimal Chern gauge}.
The Berry curvature $\Omega_n(\mathbf k)$ is calculated as Eq.~\ref{eq: berrycurvature}.
The quantum metric \cite{provost1980riemannian,souza2000polarization} is defined as
\begin{equation}
    g_{n,ij}(\mathbf k) = \frac{1}{2} \text{Tr} (\partial_i \ket{u_{n,\mathbf k}}\bra{u_{n,\mathbf k}})(\partial_j \ket{u_{n,\mathbf k}}\bra{u_{n,\mathbf k}}).
\end{equation}
The Berry curvature and quantum metric are invariant of the gauge transformation of the Bloch states.

\begin{figure}
    \centering
\includegraphics[width=\columnwidth]{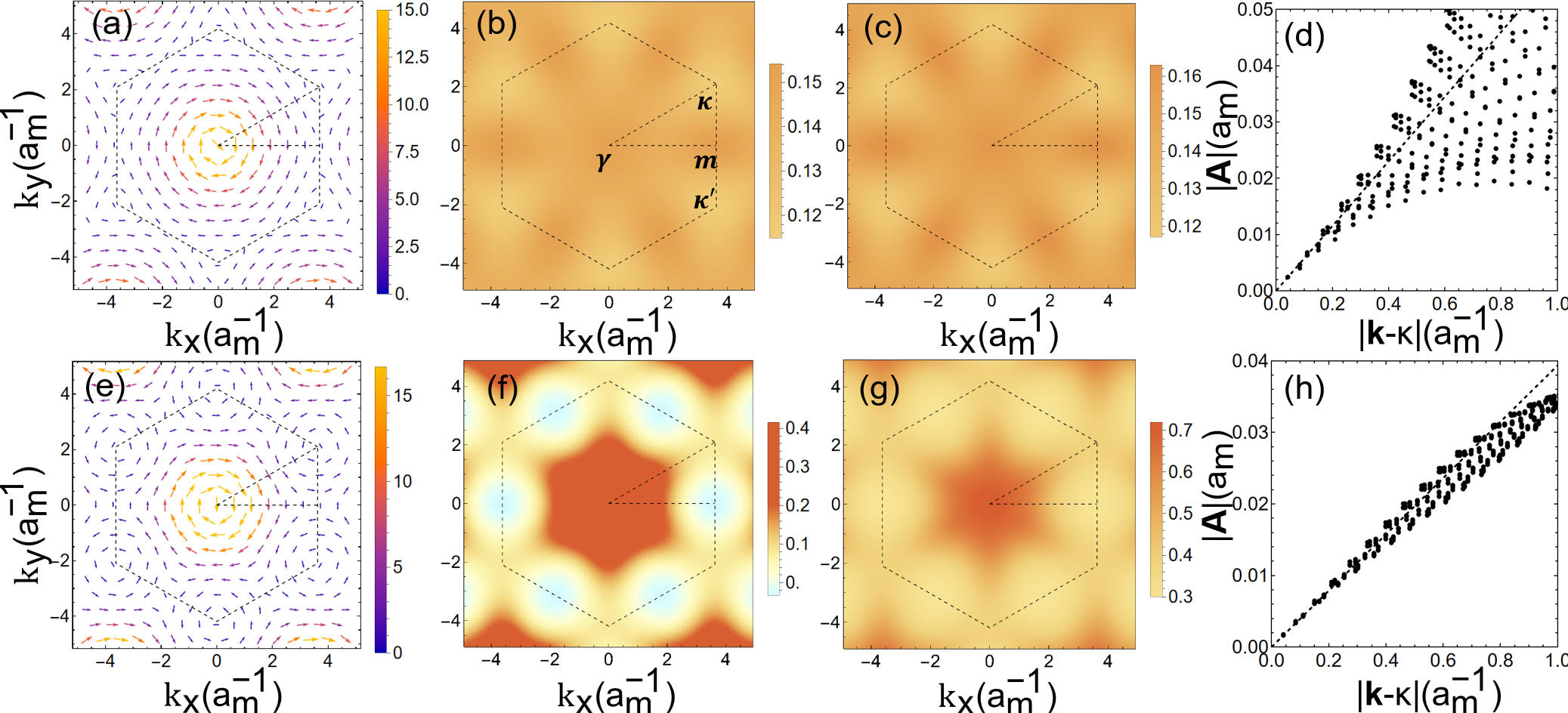}
\caption{
(a)(b)(c) The Berry connection $A_n(\mathbf k)$, the Berry curvature $\Omega_n(\mathbf k)$, and the trace of the quantum metric tensor $\sum_i g_{n,ii}(\mathbf k)$, for VB1 under $\Delta_D=0$ meV, respectively.
Dashed lines are MBZ and the high symmetry lines $\gamma-\kappa$ and $\gamma-m$.
(d)(h) The magnitude of the Berry connection near $\kappa$ for VB1 (VB2). The dashed lines are $\vert \Omega_n(\kappa) \times (\mathbf k - \kappa)/2 \vert$.
(e)(f)(g) $A_n(\mathbf k)$, $\Omega_n(\mathbf k)$, and $\sum_i g_{n,ii}(\mathbf k)$ under $\Delta_D=0$ meV.
}
\label{fig: quantum geometry D0} 
\end{figure}

We first discuss the quantum geometry for VB1 and VB2 under $\Delta_D=0$meV as shown in Fig.~\ref{fig: quantum geometry D0}, which mimics the lowest and first Landau level, respectively.
Fig.~\ref{fig: quantum geometry D0} (a)(d) shows the resulting Berry connection for VB1 and VB2, respectively.
The vortex position for VB1 is $z_v = -b_{\text{m}, 2, x} - i b_{\text{m}, 2, y}$ and that for VB2 is $z_v = 0$.
The singular vortex with the largest magnitude is located at $\gamma$ while the minimum is located at $\kappa$ and $\kappa'$.
Fig.~\ref{fig: quantum geometry D0}(d)(h) shows the magnitude of the Berrry connection near $\kappa$.
For $\vert \mathbf k-\kappa\vert < 0.2 a_\text{m}^{-1}$, $\vert A_n(\mathbf k) \vert$ matches well with $\vert \Omega_n(\kappa) \times (\mathbf k - \kappa)/2 \vert$.
The direction of $A_n(\mathbf k)$ follows the direction of $\Omega_n(\kappa) \times (\mathbf k - \kappa)/2$ for a positive Berry curvature $\Omega_n(\kappa)$ at $\kappa$  shown in Fig.~\ref{fig: quantum geometry D0}(b)(f).
The Berry connection at $\kappa'$ is the same as those at $\kappa$ by the $\mathcal C_{2y} \mathcal T$ symmetry.
The quantum metric is shown in Fig.~\ref{fig: quantum geometry D0}(c)(g) for VB1 and VB2.
The total quantum metric is defined as
\begin{equation}
    \text{Tr}g_n = \frac{1}{2\pi}\sum_{\mathbf k,i}\delta A_\text{BZ} \ g_{n,ii}(\mathbf k).
\end{equation}
$\text{Tr}g_{n=\text{VB1}} = 1.05, \text{Tr}g_{n=\text{VB2}} = 3.21$.
A series of Chern bands with $C=+1$ and the total quantum metric near $1,3,\cdots$ mimics the $0,1,\cdots$ Landau levels \cite{wang2025higher}, respectively.

\begin{figure}
    \centering
\includegraphics[width=\columnwidth]{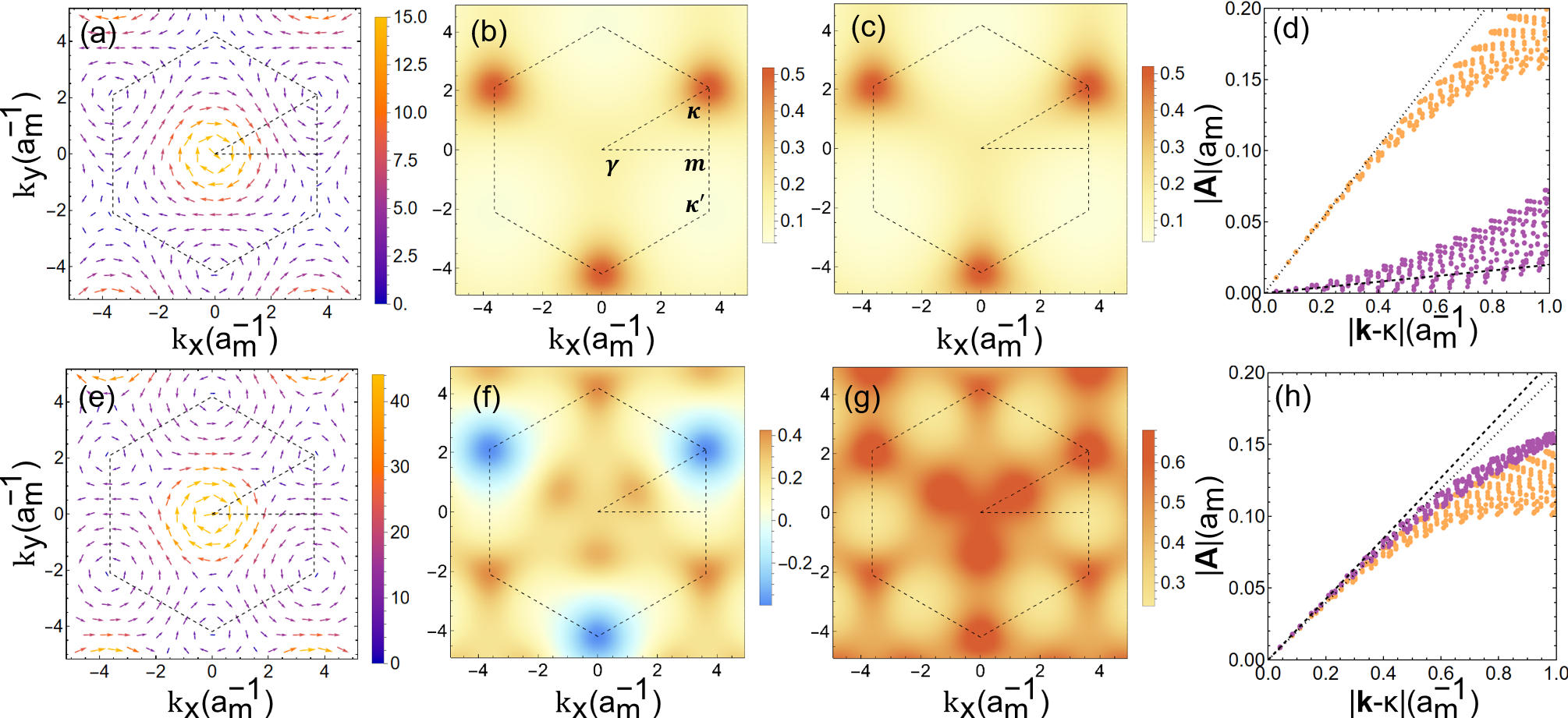}
\caption{
(a)(b)(c) The Berry connection $A_n(\mathbf k)$, the Berry curvature $\Omega_n(\mathbf k)$, and the trace of the quantum metric tensor $\sum_i g_{n,ii}(\mathbf k)$, for VB1 under $\Delta_D=6$ meV, respectively.
Dashed lines are MBZ and the high symmetry lines $\gamma-\kappa$ and $\gamma-m$.
(d)(h) The magnitude of the Berry connection near $\kappa$ (orange) and $\kappa'$ (purple) for VB1 and VB2, respectively. The dotted line is $\vert \Omega_n(\kappa) \times (\mathbf k - \kappa)/2 \vert$. The dashed line is $\vert \Omega_n(\kappa') \times (\mathbf k - \kappa')/2 \vert$.
(e)(f)(g) $A_n(\mathbf k)$, $\Omega_n(\mathbf k)$, and $\sum_i g_{n,ii}(\mathbf k)$ under $\Delta_D=0$ meV.
}
\label{fig: quantum geometry D6} 
\end{figure}

The quantum geometry of VB1 and VB2 under $\Delta_D=6$meV are shown in Fig.~\ref{fig: quantum geometry D6}, similar to those in gapped Dirac cones.
The singular vortices for VB1 and VB2 in Fig.~\ref{fig: quantum geometry D6}(a)(d) are still located at $\gamma$.
The vortex position for VB1 is $z_v = -b_{\text{m}, 2, x} - i b_{\text{m}, 2, y}$ and that for VB2 is $z_v = b_{\text{m}, 1, x} + i b_{\text{m}, 1, y}$.
The Berry connection near $\kappa, \kappa'$ follows the variation of the Berry curvature.
Its magnitude still follows the $\vert \Omega_n(\kappa) \times (\mathbf k - \kappa)/2 \vert$ as shown in orange dots and dotted lines in Fig.~\ref{fig: quantum geometry D6}(d)(h).
For VB1, $\Omega_{n=\text{VB1}}(\kappa)$ is larger than $\Omega_{n=\text{VB1}}(\kappa')$ due to the break of the $\mathcal C_{2y} \mathcal T$ symmetry by the displacement field.
The region near $\kappa$ that matches $\vert \Omega_n(\kappa) \times (\mathbf k - \kappa)/2 \vert$ is $\vert \mathbf k -\kappa\vert < 0.35$, larger than that near $\kappa'$ with $\vert \mathbf k -\kappa\vert < 0.15$.
For VB2, $\Omega_{n=\text{VB2}}(\kappa)$ shown in Fig.~\ref{fig: quantum geometry D6}(f) reverses its sign compared to that in Fig~\ref{fig: quantum geometry D0}(f).
So is the $A_{n=\text{VB2}}(\mathbf k)$ Fig.~\ref{fig: quantum geometry D6}(e) near $\kappa$ with its direction reversed compared to that in Fig.~\ref{fig: quantum geometry D0}(e).
While $\Omega_{n=\text{VB2}}(\kappa')$ in Fig.~\ref{fig: quantum geometry D6}(f) has the same sign as that in in Fig.~\ref{fig: quantum geometry D6}(f), the direction of the Berry connection is kept.
Fig.~\ref{fig: quantum geometry D6}(c)(g) shows an increase of the quantum metric near $\kappa$ compared to those in Fig.~\ref{fig: quantum geometry D0}(c)(g).
The opposite sign of the Berry curvature near $\kappa$ for VB1 and VB2 and the concentration of the quantum metric, together with the band dispersion shown in Fig.~\ref{fig: gappedDiraccone}(b), matches the gapped Dirac cone model in Sec.~\ref{sec: gapped Dirac cone}.

The singular vortex at $\gamma$ in property 1) and the Berry curvature near $\kappa, \kappa'$ in property 2) qualitatively capture the continuous variation of the Berry connection over MBZ as property 3), whose quantitative redundancy is removed by $\nabla_\mathbf k\cdot A_n(\mathbf k)$ in the optimal Chern gauge.
Comparing the Berry connection for $\Delta_D=0$ meV in Fig.~\ref{fig: quantum geometry D0}(a)(e) and that for for $\Delta_D=6$ meV in Fig.~\ref{fig: quantum geometry D6}(a)(e), the singular vortex is fixed at $\gamma$ and the main variation comes from the Berry connection near $\kappa, \kappa'$.
The Berry connection near $\kappa, \kappa'$ follows the variation of the Berry curvature.
When the Berry curvature increases, the range of Berry connection following the property 2) increases.
When the Berry curvature reverses the direction, the Berry connection also reverses the direction.
Because Berry curvature is gauge invariant and continuously varies when the displacement field is tuned, the Berry connection near $\kappa, \kappa'$ is also continuously varied for displacement fields.
From $\gamma$ to $\kappa, \kappa'$, $A_n(\mathbf k)$ smoothly connects and quantitatively fixed by $\nabla_\mathbf k A_n(\mathbf k)=0$.
%The clear picture of the variation of $A_n(\mathbf k)$ over the displacement field is crucial for understanding the Bloch contribution to exciton dipoles in Sec.XXX.

\section{Formalism of Exciton Dipole}
\subsection{Bethe-Saplter Equation in the band basis}
\label{sec: bse}
In this section, we derive the Bethe-Saplter Equation (BSE) \cite{qiu2015nonanalyticity} for exciton from the electron-hole excitations following Ref.~\cite{wu2015exciton}.

We first define the band basis and project the interacting Hamiltonian into the band basis.
The transformation between the band basis and the plane wave basis for the continuum model is
\begin{equation}
\begin{split}
    c_{n, \mathbf k}^{\dagger }=\sum _{\mathbf G, l} c_{l, \mathbf k- l \tilde q - \mathbf G}^{\dagger } u_{n, \mathbf k,\mathbf G,l}, & \quad 
    c_{n,\mathbf k}=\sum _{\mathbf G, l} c_{l, \mathbf k - l \tilde q - \mathbf G} u_{n, \mathbf k,\mathbf G,l}^* \\
    c_{l, \mathbf k - l \tilde q - \mathbf G}^{\dagger} =\sum _{n} c_{n, \mathbf k}^{\dagger } u_{n, \mathbf k,\mathbf G,l}^*, & \quad 
    c_{l, \mathbf k - l \tilde q - \mathbf G}=\sum _{n} c_{n,\mathbf k} u_{n, \mathbf k,\mathbf G,l},
\end{split}
\end{equation}
where $u_{n, \mathbf k,\mathbf G,l}$ is the eigenstate solved by \eqnref{eq: single particle state} with the basis $c^\dagger_{\eta=K, l, \mathbf k - \eta l \tilde q-\mathbf G}$.
Here $\mathbf k\in \text{MBZ}$.
We consider exciton in one valley $K$ and drop the valley index here.
The $h^0_{\eta=K}$ in the band basis is
\begin{equation}
    h^0 = \sum_{n, \mathbf k \in \text{MBZ}} \epsilon_n(\mathbf k) c_{n,\mathbf k}^{\dagger }c_{n,\mathbf k}.
\end{equation}
The transformation of the Coulomb interaction into the band basis reads
\begin{equation}
\begin{split}
    \mathcal V & = \sum _{l_1, l_2, \mathbf k_1, \mathbf k_2, \mathbf q \in \text{MBZ},\mathbf G_1, \mathbf G_2, \mathbf G}\frac{1}{2} V(\mathbf q+\mathbf G)  c_{l_1,\mathbf k_1 - \mathbf G_1 -l_1 \tilde q + \mathbf q +\mathbf G}^{\dagger } 
    c_{l_2,\mathbf k_2 - \mathbf G_2 -l_2 \tilde q - \mathbf q -\mathbf G}^{\dagger }c_{l_2,\mathbf k_2 - \mathbf G_2 -l_2 \tilde q} c_{l_1,\mathbf k_1 - \mathbf G_1 -l_1 \tilde q} \\
    & = \sum _{n_1,n_2,n_3,n_4, \mathbf k_1, \mathbf k_2, \mathbf q \in \text{MBZ}, \mathbf G}\frac{1}{2} V(\mathbf q + \mathbf G)  c_{n_1,\mathbf k_1 + \mathbf q}^{\dagger } 
    c_{n_2,\mathbf k_2 - \mathbf q }^{\dagger }c_{n_4,\mathbf k_2} c_{n_3, \mathbf k_1} \braket{u_{\mathbf n_1, \mathbf k_1 + \mathbf q +\mathbf G}}{u_{\mathbf n_3, \mathbf k_1}} \braket{u_{\mathbf n_2, \mathbf k_2 - \mathbf q - \mathbf G}}{u_{\mathbf n_4, \mathbf k_2}},
\end{split}
\label{eq: coulomb interaction}
\end{equation}
making use of $u_{n, \mathbf k - \mathbf G_0,\mathbf G,l} = u_{n, \mathbf k,\mathbf G +\mathbf G_0,l}$ from
% {\color{blue} check the boundary condition of Chern band}
\begin{equation}
\begin{split}
    c_{n, \mathbf k}^{\dagger } &=\sum _{\mathbf G, l} c_{l, \mathbf k  - l \tilde q - \mathbf G - \mathbf G_0}^{\dagger } u_{n, \mathbf k,\mathbf G + \mathbf G_0,l} \\
    = c_{n, \mathbf k-\mathbf G_0}^{\dagger } & = \sum _{\mathbf G, l} c_{l, \mathbf k -\mathbf G_0- l \tilde q - \mathbf G}^{\dagger } u_{n, \mathbf k - \mathbf G_0,\mathbf G,l}.
\end{split}
\end{equation}
We neglect the layer dependence of the Coulomb interaction $V(\mathbf q)$ due to the interlayer distance an order smaller than the moir\'e periodicity.
We take the Coulomb interaction as the Keyldsh potential
\begin{equation}
    V(\mathbf q) = \frac{2 \pi e^2}{\epsilon_0 \epsilon_r S^\text{uc}} \frac{1}{\vert \mathbf q\vert (1+ \vert \mathbf q \vert r_0 )},
\end{equation}
where $e$ is the electron charge, $\epsilon_0$ is the vacuum permittivity, $\epsilon_r = 30$ is the relative permittivity, $S^\text{uc}=a_{\text{m},0}^2 \sin{\pi/3}$ is the moir\'e unit cell area \cite{chernikov2014exciton, keldysh2024coulomb}.
$r_0 = 74 $ \AA is the screening length \cite{kumar2012tunable, berkelbach2013theory, szyniszewski2017binding}.
% \Or{We take the dual-gated Coulomb interaction as 
% \begin{equation}
%     V(\mathbf q) = \frac{2 \pi e^2}{\epsilon_0 \epsilon_r a_{\text{m},0} \sin{\pi/3}} \frac{\tanh \vert \mathbf q\vert a_{\text{m},0}  d/a_{\text{m},0}}{\vert \mathbf q\vert a_{\text{m},0}},
% \end{equation}
% where $e$ is the electron charge, $\epsilon_0$ is the vacuum permittivity, $\epsilon_r = 100$ is the relative permittivity, $S^\text{uc}=a_{\text{m},0}^2 \sin{\pi/3}$ is the moir\'e unit cell area and $d \approx 10$nm is the gate distance.
% $\epsilon_r\approx 40$ comes from both the substrate and screening of doped holes (See SM Sec.XXX).}
%$\epsilon_r$ is taken an order larger that in bilayer TMD  to make $U_0 = \frac{2 \pi e^2}{\epsilon_0 \epsilon_r a_{\text{m},0}} \approx 9.42$ meV similar to the single particle gap.
%$d$ is taken to regularize the $V(\mathbf q)$ and affect its value at $\vert\mathbf q \vert a_{\text{m},0} < 0.01$, which is nearly only $\mathbf q=0$ on a finite momentum mesh.
%Here the unit of $V(\mathbf q)$ is meV by absorbing $a_{\text{m},0}^{-2}$ from $\sum_\mathbf q$ in \eqnref{eq: coulomb interaction}.

Next, we write the BSE equations in terms of the electron and hole pair basis.
We denote the electron band as $n=e=\text{VB1}$ and the hole band as $n=h=\text{VB2}$.
The inclusion of higher valence minibands does not affect the numerical convergence of exciton formed between VB1 and VB2 because the interaction energy $U_0$ is comparable to the gap between VB1 and VB2.
The basis of the electron hole pair is
\begin{equation}
    c^\dagger_{e, \mathbf k + \mathbf Q} c_{h, \mathbf k} \ket{\text{FS}}.
\end{equation}
$\ket{\text{FS}}$ stands for the Fermi sea of the filled VB2 and lower valence minibands. 
$\mathbf Q$ is the center-of-mass momentum of the electron hole pair.
Define the exciton creation operator as
\begin{equation}
    X^\dagger_{N,\mathbf Q} \ket{\text{FS}} = \sum_{\mathbf k \in \text{MBZ}} \Psi_N(e, \mathbf k +\mathbf Q, h, \mathbf k ) c^\dagger_{e, \mathbf k + \mathbf Q} c_{h, \mathbf k} \ket{\text{FS}}, 
\end{equation}
$\Psi_N(e, \mathbf k +\mathbf Q; h, \mathbf k)$ is the envelope function of the exciton.
BSE for the envelope function is the exact diagonalization in the electron hole pair basis \cite{rohlfing2000electron} as
\begin{equation}
\begin{split}
    & (\epsilon_{e,\mathbf k_1+\mathbf Q} - \epsilon_{h,\mathbf k_1}) \Psi_N(e, \mathbf k_1 +\mathbf Q, h, \mathbf k_1 )  \\
    & - \sum_{\mathbf G} V\left(\mathbf k_1- \mathbf k_2 + \mathbf G\right)  \left\langle u_{e,\mathbf k_1+\mathbf Q +\mathbf G}|u_{e,\mathbf k_2+\mathbf Q}\right\rangle \left\langle u_{h,\mathbf k_2}|u_{h,\mathbf k_1 +\mathbf G}\right\rangle \Psi_N(e, \mathbf k_1 +\mathbf Q, h, \mathbf k_1 ) \\
   &+ \sum _{\mathbf G}  V^\text{EX}\left(\mathbf Q + \mathbf G\right) \left\langle u_{ e,\mathbf k_1+\mathbf Q+\mathbf G}|u_{h,\mathbf k_1}\right\rangle  \left\langle u_{h,\mathbf k_2}|u_{e,\mathbf k_2+\mathbf Q+ \mathbf G}\right\rangle \Psi_N(e, \mathbf k_1 +\mathbf Q, h, \mathbf k_1 ) \\ 
   &= E_{N,\mathbf Q} \Psi_N(e, \mathbf k_1 +\mathbf Q, h, \mathbf k_1 ) ,
\end{split}
\label{eq: BSE V matrix element}
\end{equation}
We denote the first line in \eqnref{eq: BSE V matrix element} as the kinetic term $\mathcal H^\text{K}(e,\mathbf k_1+\mathbf Q_1, h, \mathbf k_1)$, the second line in \eqnref{eq: BSE V matrix element} with the minus sign as the direct term $-\mathcal H^\text{D}(e,\mathbf k_1+\mathbf Q_1, h, \mathbf k_1; e,\mathbf k_2+\mathbf Q_2, h, \mathbf k_2)$, and the third line with the plus sign as the exchange term $\mathcal H^\text{EX}(e,\mathbf k_1+\mathbf Q_1, h, \mathbf k_1; e,\mathbf k_2+\mathbf Q_2, h, \mathbf k_2)$.
$V^\text{EX}(\mathbf q)$ is the unscreened Coulomb interaction as $V^\text{EX} (\mathbf q) = 2\pi e^2 / \epsilon_0 \vert \mathbf q\vert S^\text{uc}$.
Rewriting all terms, BSE reads
\begin{equation}
\begin{split}
    &\sum_{\mathbf k_2 \in \text{MBZ}} (\mathcal H^\text{K}(e,\mathbf k_1+\mathbf Q, h, \mathbf k_1) \tilde \delta_{\mathbf k_1, \mathbf k_2} -
    \mathcal H^\text{D}(e,\mathbf k_1+\mathbf Q, h, \mathbf k_1; e,\mathbf k_2+\mathbf Q, h, \mathbf k_2) \\
    & +\mathcal H^\text{EX}(e,\mathbf k_1+\mathbf Q, h, \mathbf k_1; e,\mathbf k_2+\mathbf Q, h, \mathbf k_2)) \Psi_N(e, \mathbf k_2 +\mathbf Q, h, \mathbf k_2) \\
    &= E_{N,\mathbf Q} \Psi_N(e, \mathbf k_1 +\mathbf Q, h, \mathbf k_1).
\end{split}
\label{eq: BSE}
\end{equation}
We consider the exciton near $\mathbf Q = \gamma$ and the exchange term is negligible due to the overlap $\left\langle u_{ e,\mathbf k_1}|u_{h,\mathbf k_1}\right\rangle = 0$ between the electron and hole band at the same momentum.
The binding energy of the $N=1$ exciton changes by an order of magnitude less upon inclusion of the exchange term.
The exciton envelope function is periodic in $\mathbf k$ with 
\begin{equation}
    \Psi_N(e, \mathbf k_1 +\mathbf Q + \mathbf G, h, \mathbf k_1 + \mathbf G) = \Psi_N(e, \mathbf k_1 +\mathbf Q, h, \mathbf k_1)
\end{equation}
from the same BSE equations under $\braket{u_{n_1, \mathbf k_1+ \mathbf G}}{u_{n_2, \mathbf k_2+ \mathbf G}} = \braket{u_{n_1, \mathbf k_1}}{u_{n_2, \mathbf k_2}}$.

\subsection{In-plane exciton dipole}
\label{sec: exciton dipole}
In this section, we discuss the in-plane exciton dipole definition, the gauge transformation of Bloch states, the choice of the electron-hole basis, and the symmetry constraints on the exciton dipole.

We start from the intuitive definition of the exciton dipole displacement vector as the in-plane real space separation of its constituent electron and hole.
We start from the real space exciton wavefunction as
\begin{equation}
    \Psi_{N,\mathbf Q, l_e, l_h}(\mathbf r_e, \mathbf r_h) = \sum_{\mathbf k \in \text{MBZ}} \Psi_{N}(e, \mathbf k + \mathbf Q, h, \mathbf k) \psi_{e, \mathbf k + \mathbf Q, l_e}(\mathbf r_e) \psi_{h, \mathbf k, l_h}(\mathbf r_h)^*,
\label{eq: exciton real space wavefunction}
\end{equation}
where $\psi_{n, \mathbf k}(\mathbf r)$ is the Bloch wavefunction in the real space defined from its periodic part
\begin{equation}
    \psi_{n, \mathbf k, l}(\mathbf r) = e^{i \mathbf k \cdot \mathbf r} u_{n,\mathbf k,l} (\mathbf r) = e^{i \mathbf k \cdot \mathbf r}  \sum_{\mathbf G} u_{n, \mathbf k, \mathbf G, l}  e^{- i ( l \tilde q + \mathbf G)\cdot \mathbf r}.
\label{eq: Bloch wavefunction}
\end{equation}
The intuitive definition of the dipole displacement vector is
\begin{equation}
\begin{split}
    \mathbf d_{N,\mathbf Q} &= \int \dd^2 r_e \dd ^2 r_h \sum_{l_e, l_h} \vert \Psi_{N,\mathbf Q, l_e, l_h}(\mathbf r_e, \mathbf r_h)\vert^2 (\mathbf r_h -\mathbf r_e) \\
    & = \sum_{\mathbf k \in \text{MBZ}} \vert \Psi_{N}(e, \mathbf k + \mathbf Q, h, \mathbf k)\vert^2 (-\bra{\psi_{e,\mathbf k+ \mathbf Q}} \hat{\mathbf r} \ket{\psi_{e,\mathbf k+ \mathbf Q}} +\bra{\psi_{h,\mathbf k}} \hat{\mathbf r} \ket{\psi_{h,\mathbf k}}),
\label{eq: dipole definition SM}
\end{split}
\end{equation}
where $\bra{\psi_{n,\mathbf k}} \hat{\mathbf r} \ket{\psi_{n,\mathbf k}}$ with the in-plane position operator $\hat{\mathbf r}$ is
\begin{equation}
    \bra{\psi_{n,\mathbf k}} \hat{\mathbf r} \ket{\psi_{n,\mathbf k}} = 
    \int \dd^2 r\sum_{l} \vert \psi_{n, \mathbf k, l}(\mathbf r) \vert^2 \mathbf r.
\end{equation}
Next, we replace $\mathbf r$ with $-i \partial_\mathbf k e^{i \mathbf k \cdot \mathbf r} $ in the the Bloch wavefunctions \eqnref{eq: Bloch wavefunction} and integrate by parts as
% In the first line of \eqnref{eq: dipole definition SM}, one can make use of the periodicity of Bloch wavefunctions and exciton envelope function in $\mathbf k$ as
\begin{equation}
\begin{split}
    & 0 = \sum_{\mathbf k} i\partial_{\mathbf k} (\Psi_{N}(e, \mathbf k + \mathbf Q, h, \mathbf k) \psi_{e, \mathbf k + \mathbf Q, l_e}(\mathbf r_e) \psi_{h, \mathbf k, l_h}(\mathbf r_h)^*) \\
    & = \sum_{\mathbf k} e^{i \mathbf k \cdot (\mathbf r_e-\mathbf r_h)} e^{i \mathbf Q \cdot \mathbf r_e} (i\partial_{\mathbf k} \Psi_{N}(e, \mathbf k + \mathbf Q, h, \mathbf k) u_{e, \mathbf k + \mathbf Q, l_e}(\mathbf r_e) u_{h, \mathbf k, l_h}(\mathbf r_h)^*) \\
    & +  (\mathbf r_h -\mathbf r_e)\Psi_{N}(e, \mathbf k + \mathbf Q, h, \mathbf k) \psi_{e, \mathbf k + \mathbf Q, l_e}(\mathbf r_e) \psi_{h, \mathbf k, l_h}(\mathbf r_h)^*.
\end{split}
\end{equation}
The resulting displacement vector
is 
\begin{equation}
\begin{split}
    \mathbf d_{N,\mathbf Q} & = \sum_{\mathbf k} \text{Tr} \left(i  \partial_{\mathbf k} (|u_{h, \mathbf k}\rangle  \Psi_N\left(e, \mathbf k+ \mathbf Q; h, \mathbf k\right)^*\langle u_{e,\mathbf k + \mathbf Q}|)\right)  |u_{e, \mathbf k+ \mathbf Q}\rangle  \Psi _N\left(e, \mathbf k+\mathbf Q, h, \mathbf k\right) \langle u_{h,\mathbf k}| \\
    & = \sum _\mathbf k \Psi _N(e,\mathbf k+\mathbf Q,h,\mathbf k)^* \left(-i \partial_\mathbf k +A_e(\mathbf k+\mathbf Q)-A_h(\mathbf k) \right) \Psi _N(e,\mathbf k+\mathbf Q,h,\mathbf k).
\end{split}
\label{eq: dipole definition SM momentum}
\end{equation}
Another way is to derive the Eq.~\ref{eq: dipole definition SM momentum} is making use of \cite{blount1962formalisms}
\begin{equation}
    \bra{\psi_{n_1,\mathbf k_1}} \hat{\mathbf r} \ket{\psi_{n_2,\mathbf k_2}} = \delta _{n_1,n_2}\left(-i \partial_{\mathbf k_2}  \braket{\psi_{n_1,\mathbf k_1}}{\psi_{n_2,\mathbf k_2}} \right)+\tilde \delta_{\mathbf k_1, k_2} \left\langle u_{n_1,\mathbf k_1}|i\partial_{\mathbf k_2} u_{n_2,\mathbf k_2}\right\rangle
\end{equation}
in the second line of \eqnref{eq: dipole definition SM} with the singularity in the position operator of the Bloch wavefunctions regularized by the exciton envelope function.

The exciton dipole is invariant under the gauge transformation of the Bloch wavefunctions  \cite{davenport2025exciton} and the separation of envelop and Bloch contribution depends on a particular gauge choice.
A gauge transformation of the Bloch states is
\begin{equation}
    \ket{\tilde u_{n,\mathbf k}} = e^{i \phi_n(\mathbf k)} \ket{u_{n,\mathbf k}}.
\end{equation}
The exciton envelope function after the gauge transformation is
\begin{equation}
    \tilde \Psi _N(e,\mathbf k+\mathbf Q,h,\mathbf k) = \Psi _N(e,\mathbf k+\mathbf Q,h,\mathbf k) e^{-i \phi_{e, \mathbf k+ \mathbf Q}} e^{i \phi_{h, \mathbf k}}.
\end{equation}
The exciton dipole in \eqnref{eq: dipole definition SM momentum} is gauge invariant by
\begin{equation}
    |\tilde u_{e, \mathbf k+ \mathbf Q}\rangle  \tilde \Psi _N\left(e, \mathbf k+\mathbf Q, h, \mathbf k\right) \langle \tilde u_{h,\mathbf k}| = |u_{e, \mathbf k+ \mathbf Q}\rangle  \Psi _N\left(e, \mathbf k+\mathbf Q, h, \mathbf k\right) \langle u_{h,\mathbf k}|.
\end{equation}
Under the choice of the optimal Chern gauge, one may separate the dipole displacement vector into the envelope function contribution and the contribution from the Berry connection of the Bloch states
\begin{equation}
\begin{split}
    \mathbf d_{N,\mathbf Q}^\text{E} & = \sum _\mathbf k \Psi _N(e,\mathbf k+\mathbf Q,h,\mathbf k)^* (-i) \partial_\mathbf k \Psi _N(e,\mathbf k+\mathbf Q,h,\mathbf k). \\
    \mathbf d_{N,\mathbf Q}^\text{B} & = \sum _\mathbf k \Psi _N(e,\mathbf k+\mathbf Q,h,\mathbf k)^* (A_e(\mathbf k+\mathbf Q)-A_h(\mathbf k)) \Psi _N(e,\mathbf k+\mathbf Q,h,\mathbf k).    
\end{split}
\end{equation}
$\mathbf d_{N,\mathbf Q}^\text{E}, \mathbf d_{N,\mathbf Q}^\text{B}$ individually depends on the gauge.
The optimal Chern gauge locates the vortex at $\gamma$ outside the envelope function shown in Fig.~\ref{fig:flatChernband}(e) and Fig.~\ref{fig: gappedDiraccone}(c) to avoid numerical differential of singular points.
Then main variation of the Berry connection under the envelope function at $\kappa, \kappa'$ is smooth and controlled by the Berry curvature at $\kappa, \kappa'$, making the $\mathbf d_{N,\mathbf Q}^\text{B}$ continuous over the displacement field as shown in Fig.~\ref{fig: gappedDiraccone}(a).

The exciton dipole is also invariant under the choice of the electron hole pair basis.
The choice of the electron hole pair basis can be picked as
\begin{equation}
    X^\alpha_{N,\mathbf Q}{}^\dagger \ket{\text{FS}} = \sum_{\mathbf k \in \text{MBZ}} \Psi_N(e, \mathbf k + \alpha \mathbf Q, h, \mathbf k -(1-\alpha)\mathbf Q) c^\dagger_{e, \mathbf k + \alpha\mathbf Q} c_{h, \mathbf k-(1-\alpha)\mathbf Q} \ket{\text{FS}}
\end{equation}
for different $\alpha$.
The exciton dipole displacement vector becomes
\begin{equation}
    \mathbf d_{N,\mathbf Q}^\alpha =
     \sum_\mathbf k \Psi_N(e,\mathbf k+\alpha \mathbf Q,h,\mathbf k-(1-\alpha) \mathbf Q)^* \left(-i \partial_\mathbf k + A_e(\mathbf k+\alpha \mathbf Q)-A_h(\mathbf k-(1-\alpha)\mathbf Q) \right) \Psi _N(e,\mathbf k+\alpha \mathbf Q,h,\mathbf k-(1-\alpha)\mathbf Q).
\end{equation}
The invariant of the exciton dipole displacement vector is equivalent to 
\begin{equation}
\partial_\alpha\mathbf d_{N,\mathbf Q}^\alpha = 0.
\end{equation}
$\partial_\alpha$ acting on components in the dipole displacement vector gives 
\begin{equation}
\begin{split}
    \partial_\alpha \Psi_N(e,\mathbf k+\alpha \mathbf Q,h,\mathbf k-(1-\alpha) \mathbf Q) &= \partial_\alpha \Psi_N(e,\mathbf k+\alpha \mathbf Q, h,\mathbf k- \mathbf Q + \alpha \mathbf Q) = \mathbf Q \cdot \partial_\mathbf k \Psi_N(e,\mathbf k+\alpha \mathbf Q, h,\mathbf k- \mathbf Q + \alpha \mathbf Q) \\
    \partial_\alpha A_e(\mathbf k+\alpha \mathbf Q) &= \mathbf Q\cdot\partial_\mathbf k A_e(\mathbf k+\alpha \mathbf Q) \\
    \partial_\alpha A_h(\mathbf k - (1-\alpha) \mathbf Q) &= \mathbf Q\cdot\partial_\mathbf k A_h(\mathbf k - (1-\alpha) \mathbf Q).
\end{split}
\end{equation}
Thus, $\partial_\alpha = \mathbf Q\cdot \partial_\mathbf k$ is a total derivative inside the sum of $\mathbf k$ in the dipole displacement vector. 
Specifically,
\begin{equation}
\begin{split}
    &\partial_\alpha\mathbf d_{N,\mathbf Q}^\alpha = \sum_{\mathbf k} \mathbf Q\cdot \partial_\mathbf k \Big( \\
    &\Psi_N(e,\mathbf k+\alpha \mathbf Q,h,\mathbf k-(1-\alpha) \mathbf Q)^* \left(-i \partial_\mathbf k + A_e(\mathbf k+\alpha \mathbf Q)-A_h(\mathbf k-(1-\alpha)\mathbf Q) \right) \Psi _N(e,\mathbf k+\alpha \mathbf Q,h,\mathbf k-(1-\alpha)\mathbf Q) \Big),
\end{split}
\end{equation}
which is zero by the periodicity of the Berry connection and the exciton envelope function over MBZ.
So the dipole is the same for the choice of electron hole pair basis in the \eqnref{eq: dipole definition SM} and \eqnref{eq: dipole definition} in the main text.
Similarly, $\mathbf d_{N,\mathbf Q}^\text{E}, \mathbf d_{N,\mathbf Q}^\text{B}$ are also invariant for $\alpha$.

The symmetry of tMoTe$_2$ constrains the exciton dipole displacement vector near $\gamma$ with the form $\mathbf d_{N,\mathbf Q} = v_N \hat z\times \mathbf Q/2$.
As symmetry analysis apply to all exciton states, we briefly drop $N$ index in this paragraph.
First, we treat the $\mathbf d$ as a real space vector and $\mathbf Q$ as a momentum space vector and discuss its transformation under $\mathcal C_{3z}$, $\mathcal C_{2y} \mathcal T$, $\mathcal T$, and the in-plane mirror symmetry $\mathcal M_y$ with the mirror plane perpendicular to the y-direction.
The transformation of the dipole displacement vector under the rotation symmetry is
\begin{equation}
    \mathcal C_{3z} \mathbf d_\mathbf Q = (\mathcal C_{3z} \mathbf d)_\mathbf Q =\mathbf d_{(\mathcal C_{3z} \mathbf Q)}.
\end{equation}
To linear order, there are two basis terms obeying this rotation symmetry $\mathbf Q$ and $\hat z \times \mathbf Q$, which are two orthogonal basis vectors in the 2D polar coordinate.
Any constant linear combination of the two terms are rotation invariant.
For the remaining symmetries,
\begin{equation}
\begin{split}
    \mathcal C_{2y}\mathcal T \mathbf d_\mathbf Q &= (-d_{x}, d_{y})_{(Q_x,Q_y)} =  (d_{x}, d_{y})_{(Q_x,-Q_y)} \\
    \mathcal T \mathbf d_\mathbf Q & = (d_{x}, d_{y})_{(Q_x,Q_y)}= (d_{x}, d_{y})_{(-Q_x,-Q_y)} \\
    \mathcal M_y \mathbf d_\mathbf Q & = (d_{x}, -d_{y})_{(Q_x,Q_y)}= (d_{x}, d_{y})_{(Q_x, - Q_y)},
\end{split}
\end{equation}
where the first equal applies symmetries on $\mathbf d$ and the second equal applies symmetries on $\mathbf Q$.
$\mathcal C_{2y} \mathcal T$ is present in tMoTe$_2$, selects the $\hat z\times \mathbf Q$ and forbids the $\mathbf Q$ term.
$\mathcal T$ forbids both linear terms and $\mathcal M_y$ forbids the $\hat z \times \mathbf Q$, which are absent in tMoTe$_2$.
Combining the presence of the $\mathcal C_{3z}$ and $\mathcal C_{2y}\mathcal T$ and the absence of $\mathcal T$ and in-plane mirror symmetries in one valley, the exciton dipole displacement vector takes the form $\mathbf d_{N,\mathbf Q} = v_N \hat z\times \mathbf Q/2$ near $\mathbf Q=\gamma$.
One can separate the envelope function contribution $v^\text{E}_N$ and the Bloch contribution $v^\text{B}_N$ to $v_N$.

We make clear the distinction between $\mathbf d_{N,\mathbf Q}^\text{B}$ and the average of the Berry curvature $\bar \Omega$ over the envelope function and the electron hole bands.
The curl of the dipole displacement vector from the Bloch contribution from \eqnref{eq: dipole definition} in the main text is
\begin{equation}
\begin{split}
    v^\text{B}_N & =\nabla_\mathbf Q \times \mathbf d_{N,\mathbf Q}^\text{B} \vert_{\mathbf Q=\gamma} \\
    & = \sum_{\mathbf k} (\nabla_\mathbf Q \vert\Psi _N(e,\mathbf k+\mathbf Q/2,h,\mathbf k-\mathbf Q/2)\vert^2 ) \times (A_e(\mathbf k+\mathbf Q/2)-A_h(\mathbf k-\mathbf Q/2)) \vert_{\mathbf Q=\gamma} \\
    & + \sum_{\mathbf k}  \vert \Psi _N(e,\mathbf k,h,\mathbf k)\vert^2  (\Omega_e(\mathbf k)/2+\Omega_h(\mathbf k)/2),
\end{split}
\end{equation}
where the second line is $\bar \Omega$.

\subsection{Dipole-dipole interaction}
\label{sec: dipole-dipole interaction}
In this section, we discuss about the direct interaction between exciton of the same exciton band in the dilute limit of the exciton density following Ref.~\cite{perea2025exciton, erkensten2021exciton}.

We consider two exciton interacted through $\mathcal V$ in \eqnref{eq: coulomb interaction}, which is
\begin{equation}
    \left\langle \text{FS}\left| X_{\mathbf Q+\mathbf Q_1} X_{ \mathbf Q_2-\mathbf Q} \mathcal{V} X_{\mathbf Q_2}^{\dagger } X_{\mathbf Q_1}^{\dagger }\right|\text{FS}\right\rangle.
\end{equation}
Here, we limit the exciton to one band and drop the $N$ index here.
We consider first the momentum without MBZ for $\mathbf k\cdot \mathbf p$ models and then the momentum in MBZ for the continuum model.
Write out $\mathcal V$ for electron holes normal order to the Fermi Sea as
\begin{equation}
\begin{split}
    \mathcal V & = \mathcal V^{ee} + \mathcal V^{hh} + \mathcal V^{eh} \\
    & = \sum_{\mathbf k_1, \mathbf k_2,\mathbf q \in \text{MBZ}} 
    \frac{1}{2} V(\mathbf q) \left\langle u_{e, \mathbf k_1+ \mathbf q } |u_{e,\mathbf k_1}\right\rangle  \left\langle u_{e,\mathbf k_2-\mathbf q}|u_{e,\mathbf  k_2} \right\rangle 
    c_{e,\mathbf k_1+\mathbf q}^{\dagger } c_{e,\mathbf k_2-\mathbf q}^{\dagger }c_{e,\mathbf k_2} c_{e,\mathbf k_1} \\
    & + \frac{1}{2} V(\mathbf q) \left\langle u_{h, \mathbf k_1+ \mathbf q} |u_{h,\mathbf k_1}\right\rangle  \left\langle u_{h,\mathbf k_2-\mathbf q}|u_{h,\mathbf  k_2} \right\rangle c_{h,\mathbf k_2} c_{h,\mathbf k_1}    c_{h,\mathbf k_1+\mathbf q}^{\dagger } c_{h,\mathbf k_2-\mathbf q}^{\dagger } \\
    & - V(\mathbf q) \left\langle u_{e, \mathbf k_1+ \mathbf q} |u_{e,\mathbf k_1}\right\rangle  \left\langle u_{h,\mathbf k_2-\mathbf q}|u_{h,\mathbf  k_2} \right\rangle c_{e,\mathbf k_1+\mathbf q}^{\dagger} c_{h,\mathbf k_2} c_{e,\mathbf k_2-\mathbf q}^{\dagger }  c_{e,\mathbf k_1}.
\end{split}
\end{equation}
We next expand the exciton operator $X_{\mathbf Q}$ and interaction operator $\mathcal V$ in terms of electron hole creation operators and use the Wick's theorem to simplify the interaction between exciton.
The interaction between electrons $\mathcal V^{ee}$ leads to
\begin{equation}
\begin{split}
    & \left\langle \text{FS}\left| X_{\mathbf Q+\mathbf Q_1} X_{ \mathbf Q_2-\mathbf Q} \mathcal{V}^{ee} X_{\mathbf Q_2}^{\dagger } X_{\mathbf Q_1}^{\dagger }\right|\text{FS}\right\rangle \\
    & = \sum_{\mathbf k_1, \mathbf k_2}V(\mathbf Q) \left\langle u_{e,\mathbf k_1+\mathbf Q+\mathbf Q_1}|u_{e,\mathbf k_1+\mathbf  Q_1}\right\rangle  \left\langle u_{e,\mathbf k_2-\mathbf Q+\mathbf Q_2} |u_{e,\mathbf k_2+\mathbf Q_2}\right\rangle \\
    & \times \Psi\left(e,\mathbf k_1+\mathbf Q+\mathbf Q_1, h,\mathbf k_1\right)^* \Psi\left(e, \mathbf k_2-\mathbf Q+\mathbf Q_2,h,\mathbf k_2\right)^*  \Psi\left(e, \mathbf k_1+\mathbf Q_1,h, \mathbf k_1\right) \Psi\left(e, \mathbf k_2+\mathbf Q_2, h, \mathbf k_2\right)  \\
    & -V \left(\mathbf k_2- \mathbf k_1-\mathbf Q-\mathbf Q_1+\mathbf Q_2\right)  \left\langle u_{e,\mathbf k_2-\mathbf Q+\mathbf Q_2}|u_{e,\mathbf k_1+\mathbf Q_1}\right\rangle \left\langle u_{e,\mathbf k_1+\mathbf Q+\mathbf Q_1}|u_{e,\mathbf k_2+\mathbf Q_2}\right\rangle \\
    & \times \Psi \left(e, \mathbf k_1+\mathbf Q+\mathbf Q_1,h,\mathbf k_1\right)^* \Psi\left(e, \mathbf k_2-\mathbf Q+\mathbf Q_2,h,\mathbf k_2\right)^* \Psi \left(e, \mathbf k_1+\mathbf Q_1,h,\mathbf k_1\right) \Psi \left(e, \mathbf k_2+\mathbf Q_2,h, \mathbf k_2\right)   \\
    & + V \left(-\mathbf Q-\mathbf Q_1+\mathbf Q_2\right) \left\langle u_{e,\mathbf k_2+\mathbf Q+\mathbf Q_1}|u_{e,\mathbf k_2+\mathbf Q_2}\right\rangle  \left\langle u_{e,\mathbf k_1-\mathbf Q+\mathbf Q_2}|u_{e,\mathbf k_1+\mathbf Q_1}\right\rangle \\
    &\times \Psi\left(e, \mathbf k_2+\mathbf Q + \mathbf Q_1,h,\mathbf k_2\right)^* \Psi\left(e,\mathbf k_1-\mathbf Q+\mathbf Q_2,h,\mathbf k_1\right)^* \Psi\left(e, \mathbf k_1+\mathbf Q_1,h, \mathbf k_1\right) \Psi\left(e, \mathbf k_2+\mathbf Q_2,h,\mathbf k_2\right)  \\
    & -V\left(-\mathbf k_1+\mathbf k_2+\mathbf Q\right) \left\langle u_{e,\mathbf k_2+\mathbf Q+\mathbf Q_1}|u_{e,\mathbf k_1+\mathbf Q_1}\right\rangle  \left\langle u_{e,\mathbf k_1-\mathbf Q+\mathbf Q_2}|u_{e,\mathbf k_2+\mathbf Q_2}\right\rangle \\
    & \times \Psi\left(e, \mathbf k_2+\mathbf Q+\mathbf Q_1, h, \mathbf k_2\right)^* \Psi \left(e, \mathbf k_1-\mathbf Q+\mathbf Q_2,h, \mathbf k_1\right)^* \Psi\left(e, \mathbf k_1+\mathbf Q_1,h,\mathbf k_1\right) \Psi\left(e,\mathbf k_2+\mathbf Q_2,h,\mathbf k_2\right).
\end{split}
\label{eq: Vee}
\end{equation}
The interaction between holes $\mathcal V^{hh}$ gives
\begin{equation}
\begin{split}
    &\left\langle \text{FS}\left| X_{\mathbf Q+\mathbf Q_1} X_{ \mathbf Q_2-\mathbf Q} \mathcal{V}^{hh} X_{\mathbf Q_2}^{\dagger } X_{\mathbf Q_1}^{\dagger}\right|\text{FS}\right\rangle  \\
    & = \sum_{\mathbf k_1, \mathbf k_2} V(\mathbf Q)\left\langle u_{h,\mathbf k_1}|u_{h,\mathbf k_1-\mathbf Q}\right\rangle \left\langle u_{h,\mathbf k_2}|u_{h,\mathbf k_2+\mathbf Q}\right\rangle \\
    &\times \Psi \left(e,\mathbf k_1+\mathbf Q_1,h, \mathbf k_1-\mathbf Q\right)^* \Psi \left(e,\mathbf k_2+\mathbf Q_2,h,\mathbf k_2+\mathbf Q\right)^* \Psi \left(e,\mathbf k_1+\mathbf Q_1,h,\mathbf k_1\right) \Psi \left(e,\mathbf k_2+\mathbf Q_2,h,\mathbf k_2\right)  \\
    &-V\left(-\mathbf k_1+\mathbf k_2+\mathbf Q\right) \left \langle u_{h,\mathbf k_1}|u_{h,\mathbf k_2+\mathbf Q} \right\rangle  \left\langle u_{h,\mathbf  k_2}|u_{h,\mathbf k_1-\mathbf Q}\right\rangle \\
    & \times \Psi \left(e,\mathbf k_1+\mathbf Q_1,h,\mathbf k_1-\mathbf Q\right)^* \Psi \left(e,\mathbf k_2+\mathbf Q_2,h,\mathbf k_2+\mathbf Q\right)^* \Psi \left(e,\mathbf k_1+\mathbf Q_1,h,\mathbf k_1\right) \Psi \left(e,\mathbf k_2+\mathbf Q_2,h,\mathbf k_2\right)  \\
    & + V\left(\mathbf Q+\mathbf Q_1-\mathbf Q_2\right) \left\langle u_{h,\mathbf k_1}|u_{h,\mathbf k_1+\mathbf Q+\mathbf Q_1-\mathbf Q_2}\right\rangle  \left\langle u_{h,\mathbf k_2}|u_{h,\mathbf k_2-\mathbf Q-\mathbf Q_1+\mathbf Q_2}\right\rangle \Psi \left(e,\mathbf k_1+\mathbf Q_1,h,\mathbf k_1+\mathbf Q+\mathbf Q_1-\mathbf Q_2\right)^*\\
    & \times  \Psi \left(e,\mathbf k_2+\mathbf Q_2,h,\mathbf k_2-\mathbf Q-\mathbf Q_1+\mathbf Q_2\right)^* \Psi \left(e,\mathbf k_1+\mathbf Q_1,h,\mathbf k_1\right) \Psi \left(e,\mathbf k_2+\mathbf Q_2,h,\mathbf k_2\right)  \\
    &-V \left(\mathbf k_1-\mathbf k_2+\mathbf Q+\mathbf Q_1-\mathbf Q_2\right) \left\langle u_{h,\mathbf k_1}|u_{h,\mathbf k_2-\mathbf Q-\mathbf Q_1+\mathbf Q_2}\right\rangle  \left\langle u_{h,\mathbf k_2}|u_{h,\mathbf k_1+\mathbf Q+\mathbf Q_1-\mathbf Q_2}\right\rangle \Psi \left(e,\mathbf k_2+\mathbf Q_2,h,\mathbf k_2\right)  \\
    &\times \Psi \left(e,\mathbf k_1+\mathbf Q_1,h,\mathbf k_1+\mathbf Q+\mathbf Q_1-\mathbf Q_2\right)^* \Psi\left(e,\mathbf k_2+\mathbf Q_2,h,\mathbf k_2-\mathbf Q-\mathbf Q_1+\mathbf Q_2\right)^* \Psi \left(e,\mathbf k_1+\mathbf Q_1,h,\mathbf k_1\right) 
\end{split}
\label{eq: Vhh}
\end{equation}
The interaction between the electron and hole $\mathcal V^{eh}$ gives
\begin{equation}
\begin{split}
    &\left\langle \text{FS}\left| X_{\mathbf Q+\mathbf Q_1} X_{ \mathbf Q_2-\mathbf Q} \mathcal{V}^{eh} X_{\mathbf Q_2}^{\dagger } X_{\mathbf Q_1}^{\dagger}\right|\text{FS}\right\rangle \\
    & =-2 V(\mathbf Q) \left\langle u_{e,\mathbf Q+\mathbf k_1+\mathbf Q_1}|u_{e,\mathbf k_1+\mathbf Q_1}\right\rangle  \left\langle u_{h,\mathbf k_2}|u_{h,\mathbf Q+\mathbf k_2}\right\rangle \\
    &\times \Psi \left(e, \mathbf Q+\mathbf k_1+\mathbf Q_1,h, \mathbf k_1\right)^* \Psi \left(e, \mathbf k_2+\mathbf Q_2,h,\mathbf Q+\mathbf k_2\right)^* \Psi \left(e,\mathbf k_1+\mathbf Q_1,h,\mathbf k_1\right) \Psi \left(e,\mathbf k_2+\mathbf Q_2,h,\mathbf k_2\right) \\
    &-2 V\left(-\mathbf Q-\mathbf Q_1+\mathbf Q_2\right) 
    \left\langle u_{e,-\mathbf Q+\mathbf k_1+\mathbf Q_2}|u_{e,\mathbf k_1+\mathbf Q_1}\right\rangle  \left\langle u_{h,\mathbf  k_2}|u_{h,-\mathbf Q+\mathbf k_2-\mathbf Q_1+\mathbf Q_2}\right\rangle  \\
    & \times \Psi \left(e, -\mathbf Q+\mathbf k_1+\mathbf Q_2, h, \mathbf k_1\right)^* \Psi \left(e, \mathbf k_2+\mathbf Q_2, h,-\mathbf Q+\mathbf k_2-\mathbf Q_1+\mathbf Q_2\right)^* \Psi \left(e, \mathbf k_1+\mathbf Q_1,h,\mathbf k_1\right) \Psi \left(e,\mathbf k_2+\mathbf Q_2,h,\mathbf k_2\right) \\
    &+2 V\left(-\mathbf k_1+\mathbf k_2+\mathbf Q\right) 
    \left\langle u_{e,\mathbf Q+\mathbf k_2+\mathbf Q_1}|u_{e,\mathbf k_1+\mathbf Q_1}\right\rangle  \left\langle u_{h,\mathbf k_1}|u_{h,\mathbf Q+\mathbf k_2}\right\rangle \\
    &\times \Psi \left(e,\mathbf Q+\mathbf k_2+\mathbf Q_1,h,\mathbf k_2\right)^* \Psi \left(e,\mathbf k_2+\mathbf Q_2,h,\mathbf Q+\mathbf k_2\right)^* \Psi \left(e,\mathbf k_1+\mathbf Q_1,h,\mathbf k_1\right) \Psi \left(e,\mathbf k_2+\mathbf Q_2,h,\mathbf k_2\right) \\
    &+2 V\left(-\mathbf k_1+\mathbf k_2-\mathbf Q-\mathbf Q_1+\mathbf Q_2\right) 
    \left\langle u_{e,-\mathbf Q+\mathbf k_2+\mathbf Q_2}|u_{e,\mathbf k_1+\mathbf Q_1}\right\rangle  \left\langle u_{h,\mathbf k_1}|u_{h,-\mathbf Q+\mathbf k_2-Q_1+\mathbf Q_2}\right\rangle \\
    &\times \Psi \left(e, \mathbf k_2+\mathbf Q_2,h,-\mathbf Q+\mathbf k_2-\mathbf Q_1+\mathbf Q_2\right)^* \Psi \left(e,-\mathbf Q+\mathbf k_2+\mathbf Q_2,h,\mathbf k_2\right)^* \Psi \left(e,\mathbf k_1+\mathbf Q_1,h,\mathbf k_1\right) \Psi \left(e,\mathbf k_2+\mathbf Q_2,h,\mathbf k_2\right) \\
    &+2 V\left(\mathbf k_2-\mathbf k_1\right)\left\langle u_{h,\mathbf k_1}|u_{h,\mathbf k_2}\right\rangle  \left\langle u_{e,\mathbf  k_2-\mathbf Q+\mathbf Q_2}|u_{e,\mathbf k_1-\mathbf Q+\mathbf Q_2}\right\rangle \\
    &\times \Psi \left(e, \mathbf k_1+\mathbf Q_1, h,\mathbf k_1-\mathbf Q\right)^* \Psi \left(e, \mathbf k_2-\mathbf Q+\mathbf Q_2, h,\mathbf k_2\right)^* \Psi \left(e, \mathbf k_1+\mathbf Q_1, h, \mathbf k_1\right) \Psi \left(e, \mathbf k_1-\mathbf Q+\mathbf Q_2, h, \mathbf k_1-\mathbf Q\right)   \\
    &+2 V\left(\mathbf k_2-\mathbf k_1\right)\left\langle u_{e,\mathbf k_2+\mathbf Q+\mathbf Q_1}|u_{e,\mathbf k_1+\mathbf Q+\mathbf Q_1}\right\rangle \left\langle u_{h,\mathbf k_1}|u_{h,\mathbf k_2}\right\rangle   \Psi \left(e,\mathbf k_1+\mathbf Q+\mathbf Q_1,h,\mathbf k_1+\mathbf Q+\mathbf Q_1-\mathbf Q_2\right)  \\
    &\times \Psi \left(e, \mathbf k_1+\mathbf Q_1,h,\mathbf k_1+\mathbf Q+\mathbf Q_1-\mathbf Q_2\right)^* \Psi \left(e,\mathbf k_2+\mathbf Q+\mathbf Q_1,h,\mathbf k_2\right){}^* \Psi \left(e,\mathbf k_1+\mathbf Q_1,h,\mathbf k_1\right)
\end{split}
\label{eq: Veh}
\end{equation}

\begin{figure}
    \centering
\includegraphics[width=0.5\columnwidth]{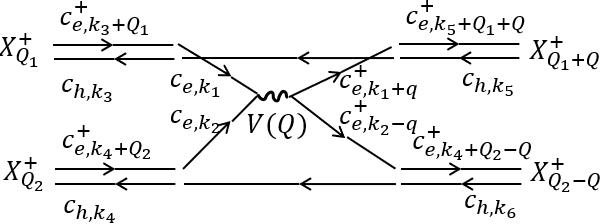}
\caption{
The Feynman diagram for the direct interaction of electrons in two exciton.
}
\label{fig: feynman diagram direct} 
\end{figure}

In the dilute limit,  the exciton interaction is dominated by the long range direct interaction term due to the much larger distance between exciton than the size of exciton.
The direct interaction between two exciton is the annihilation of electron/hole of the incoming exciton $X_{\mathbf Q_1}^{\dagger }(X_{\mathbf Q_2}^{\dagger})$ and the creation of the corresponding electron/hole of the outcoming exciton in $X_{\mathbf Q_1 + \mathbf Q}(X_{\mathbf Q_2 - \mathbf Q})$ with the Feynman diagram of the direct part of $\mathcal V^{ee}$ shown in Fig.~\ref{fig: feynman diagram direct}.
Collecting the $V(\mathbf Q)$ terms in \eqnref{eq: Vee}, \eqnref{eq: Vhh}, and \eqnref{eq: Veh} results in 
\begin{equation}
    \mathcal V^X = V(\mathbf Q) \left(F_e\left(\mathbf Q_1,\mathbf Q\right)-F_h\left(\mathbf Q_1,\mathbf Q\right)\right) \left(F_e\left(\mathbf Q_2,-\mathbf Q\right)-F_h\left(\mathbf Q_2,-\mathbf Q\right)\right)
\end{equation}
with 
\begin{equation}
\begin{split}
    F_e\left(\mathbf Q_1,\mathbf Q\right) &=\sum _{\mathbf k_1} \Psi \left(e,\mathbf k_1+\mathbf Q+\mathbf Q_1,h,\mathbf k_1\right)^* \Psi \left(e,\mathbf k_1+\mathbf Q_1,h,\mathbf k_1\right) \left\langle u_{e, \mathbf k_1+\mathbf Q+\mathbf Q_1}|u_{e,\mathbf k_1+\mathbf Q_1}\right\rangle\\
    F_e\left(\mathbf Q_2,-\mathbf Q\right) &=\sum_{\mathbf k_2} \Psi \left(e,\mathbf k_2-\mathbf Q+\mathbf Q_2,h,\mathbf k_2\right)^* \Psi \left(e,\mathbf k_2+\mathbf Q_2,h,\mathbf k_2\right) \left\langle u_{e,\mathbf k_2-\mathbf Q+\mathbf Q_2}|u_{e,\mathbf k_2+\mathbf Q_2}\right\rangle \\
    F_h\left(\mathbf Q_1,\mathbf Q\right) &=\sum_{\mathbf k_1} \Psi \left(e,\mathbf k_1+\mathbf Q_1,h,\mathbf k_1\right) \Psi \left(e,\mathbf k_1+\mathbf Q_1,h,\mathbf k_1-\mathbf Q\right)^* \left\langle u_{h,\mathbf k_1}|u_{h,\mathbf k_1-\mathbf Q}\right\rangle\\
    F_h\left(\mathbf Q_2,-\mathbf Q\right) &=\sum_{\mathbf k_2} \Psi \left(e,\mathbf k_2+\mathbf Q_2,h,\mathbf k_2\right) \Psi \left(e,\mathbf k_2+\mathbf Q_2,h,\mathbf k_2+\mathbf Q\right)^* \left\langle u_{h,\mathbf  k_2}|u_{h,\mathbf k_2+\mathbf Q}\right\rangle.
\end{split}
\end{equation}
Next, we extend the direct Coulomb interaction to the Bloch states in MBZ by replacing the $\delta_{\mathbf k_1,\mathbf k_2}$ with $\tilde \delta_{\mathbf k_1,\mathbf k_2}$ up to moir\'e reciprocal lattice vectors in the Wick's theorem.
The direct Coulomb interaction is
\begin{equation}
    \mathcal V^X = \sum_{\mathbf G} V(\mathbf Q + \mathbf G) \left(F_{e, \mathbf G}\left(\mathbf Q_1,\mathbf Q\right)-F_{h,\mathbf G}\left(\mathbf Q_1,\mathbf Q\right)\right) \left(F_{e,-\mathbf G}\left(\mathbf Q_2,-\mathbf Q\right)-F_{h,-\mathbf G}\left(\mathbf Q_2,-\mathbf Q\right)\right)
\label{eq: direct exciton interaction G}
\end{equation}
with 
\begin{equation}
\begin{split}
    F_{e,\mathbf G} \left(\mathbf Q_1,\mathbf Q\right) =\sum_{\mathbf k_1} \Psi \left(e,\mathbf k_1+\mathbf Q+\mathbf Q_1,h,\mathbf k_1\right)^* \Psi \left(e,\mathbf k_1+\mathbf Q_1,h,\mathbf k_1\right) \left\langle u_{e,\mathbf G+\mathbf k_1+\mathbf Q+\mathbf Q_1}|u_{e,\mathbf k_1+\mathbf Q_1}\right\rangle \\
    F_{e,-\mathbf G}\left(\mathbf Q_2,-\mathbf Q\right)=\sum_{\mathbf k_2} \Psi \left(e,\mathbf k_2-\mathbf Q+\mathbf Q_2,h,\mathbf k_2\right)^* \Psi \left(e,\mathbf k_2+\mathbf Q_2,h,\mathbf k_2\right) \left\langle u_{e, -\mathbf G+\mathbf k_2-\mathbf Q+\mathbf Q_2}|u_{e,\mathbf k_2+\mathbf Q_2}\right\rangle \\
    F_{h,\mathbf G}\left(\mathbf Q_1,\mathbf Q\right)=\sum_{\mathbf k_1} \Psi \left(e,\mathbf k_1+\mathbf Q_1,h,\mathbf k_1\right) \Psi \left(e,\mathbf k_1+\mathbf Q_1,h,\mathbf k_1-\mathbf Q\right)^* \left\langle u_{h,\mathbf k_1}|u_{h,-\mathbf G+\mathbf k_1-\mathbf Q}\right\rangle \\
    F_{h,-\mathbf G}\left(\mathbf Q_2,-\mathbf Q\right)=\sum _{\mathbf k_2} \Psi \left(e,\mathbf k_2+\mathbf Q_2,h,\mathbf k_2\right) \Psi \left(e, \mathbf k_2+\mathbf Q_2,h,\mathbf k_2+\mathbf Q\right)^* \left\langle u_{h,\mathbf k_2}|u_{h,\mathbf G+\mathbf k_2+\mathbf Q}\right\rangle.
\end{split}
\label{eq: form factor V}
\end{equation}

In the small $\mathbf Q$ limit, the direct exciton interaction reproduces the classical dipole-dipole interaction.
The classical dipole-dipole interaction \cite{lange2016physical} between two dipole displacement vector $\mathbf d_1, \mathbf d_2$ in the momentum space is
\begin{equation}
    V^d(\mathbf Q) = -\pi \frac{\mathbf d_1 \cdot \mathbf d_2}{r_0} + \vert \mathbf Q\vert  \frac{(\mathbf d_1 \cdot \mathbf Q)(\mathbf d_2 \cdot \mathbf Q)}{\vert \mathbf Q\vert^2},
\label{eq: classic dipole dipole interaction}
\end{equation}
where $r_0$ is the short-range cutoff in the real space for regularization.
It is long-range by $V^d(\mathbf Q) \propto \vert \mathbf Q\vert $ and anisotropic for different direction of $\mathbf Q$.
Remarkably, when $\mathbf d_1 = - \mathbf d_2$, the long-range interaction can be attractive.
In the small $\mathbf Q$ limit and $\mathbf G=\mathbf 0$, the form factors in the $\mathcal V^X$ of \eqnref{eq: form factor V} is approximated by
\begin{equation}
\begin{split}
    F_{e,\mathbf G =0}\left(\mathbf Q_1,\mathbf Q\right) & \approx -1 + \sum_{\mathbf k_1} \Psi \left(e,\mathbf k_1+\mathbf Q+\mathbf Q_1,h,\mathbf k_1\right)^* \Psi \left(e,\mathbf k_1+\mathbf Q_1,h,\mathbf k_1\right) \\
    & + \sum_{\mathbf k_1} \Psi \left(e,\mathbf k_1+\mathbf Q_1,h,\mathbf k_1\right)^* \Psi \left(e,\mathbf k_1+\mathbf Q_1,h,\mathbf k_1\right) \left\langle u_{e,\mathbf k_1+\mathbf Q+\mathbf Q_1}|u_{e,\mathbf k_1+\mathbf Q_1}\right\rangle \\
    F_{h,\mathbf G =0}\left(\mathbf Q_1,\mathbf Q\right) & \approx -1 + \sum_{\mathbf k_1} \Psi \left(e,\mathbf k_1+\mathbf Q_1,h,\mathbf k_1\right) \Psi \left(e,\mathbf k_1+\mathbf Q_1,h,\mathbf k_1-\mathbf Q\right)^* \\
    & +\sum_{\mathbf k_1} \Psi \left(e,\mathbf k_1+\mathbf Q_1,h,\mathbf k_1\right) \Psi \left(e,\mathbf k_1+\mathbf Q_1,h,\mathbf k_1\right)^* \left\langle u_{h,\mathbf k_1}|u_{h,\mathbf k_1 - \mathbf Q}\right\rangle 
\end{split}
\end{equation}
Comparing with the dipole displacement vector definition in \eqnref{eq: dipole definition SM momentum}
\begin{equation}
\begin{split}
    &F_{e,\mathbf G =0}\left(\mathbf Q_1,\mathbf Q\right)- F_{h,\mathbf G=0}\left(\mathbf Q_1,\mathbf Q\right) \approx 
    \sum_{\mathbf k_1} (\mathbf Q \cdot \partial_\mathbf k \Psi \left(e,\mathbf k_1+\mathbf Q_1 +\mathbf Q,h,\mathbf k_1\right)^*) \Psi \left(e,\mathbf k_1+\mathbf Q_1,h,\mathbf k_1\right) \\
    & +\sum_{\mathbf k_1} \Psi \left(e,\mathbf k_1+\mathbf Q_1,h,\mathbf k_1\right) \Psi \left(e,\mathbf k_1+\mathbf Q_1,h,\mathbf k_1\right)^* \mathbf Q \cdot (\left\langle \partial_{\mathbf k_1} u_{e,\mathbf k_1+\mathbf Q_1}|u_{e,\mathbf k_1+\mathbf Q_1}\right\rangle -\left\langle u_{h,\mathbf k_1}|\partial_{\mathbf k_1} u_{h,\mathbf k_1}\right\rangle ) \\
    & \approx - i \mathbf Q \cdot \mathbf d_{\mathbf Q_1},
\end{split}
\label{eq: exciton interaction form factor}
\end{equation}
where the last line uses $\mathbf Q \cdot \partial_\mathbf k \Psi \left(e,\mathbf k_1+\mathbf Q_1 +\mathbf Q,h,\mathbf k_1\right)^* \approx \mathbf Q \cdot \partial_\mathbf k \Psi \left(e,\mathbf k_1+\mathbf Q_1,h,\mathbf k_1\right)^*$ to the linear order in $\mathbf Q$.
Similarly, 
\begin{equation}
    F_{e,-\mathbf G=0}\left(\mathbf Q_2,-\mathbf Q\right)-F_{h,-\mathbf G=0}\left(\mathbf Q_2,-\mathbf Q\right) \approx i \mathbf Q \cdot \mathbf d_{\mathbf Q_2}.
\end{equation}
Summarizing the form factors leads to the dipole-dipole interaction for the direct exciton interaction
\begin{equation}
    \mathcal V^X \approx V(\mathbf Q) (\mathbf Q \cdot \mathbf d_{\mathbf Q_1})(\mathbf Q \cdot \mathbf d_{\mathbf Q_2}) = U_0 \frac{\tanh \vert \mathbf Q\vert d}{\vert \mathbf Q\vert a_{\text{m},0}}(\mathbf Q \cdot \mathbf d_{\mathbf Q_1})(\mathbf Q \cdot \mathbf d_{\mathbf Q_2}).
\label{eq: exciton interaction small Q}
\end{equation}
Here, we neglect the repulsion from the out-of-plane dipole of exciton at the angstrom scale of the interlayer distance \cite{perea2025exciton, gotting2022moire, batsch1993dipole, schindler2008analysis} induced by the displacement field, whose moment is an order smaller than the nanometer in-plane dipole moment discussed here.

\section{Exciton in gapped Dirac cones}
\label{sec: gapped Dirac cone}
\subsection{Model and quantum geometry}

In this section, we use the gapped Dirac cones with asymmetrical mass for the electron and hole as the $\mathbf k\cdot \mathbf p$ model at $\kappa$ for the tMoTe$_2$ at the large displacement field.

\begin{figure}
    \centering
\includegraphics[width=0.3\columnwidth]{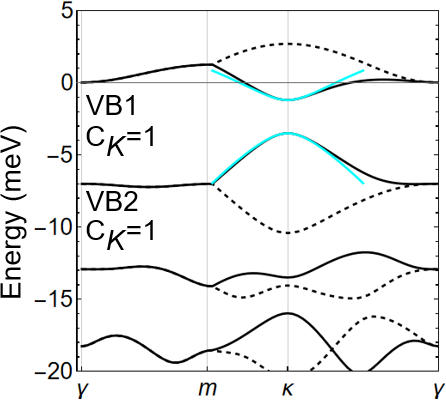}
\caption{
The spectrum of continuum model with $\Delta_D=9$ meV for the valley $K$ (solid black) and the valley $K'$ (dashed black).
The cyan solid line is the spectrum of the fitted gapped Dirac cone.
}
\label{fig: gapped Dirac cone fit} 
\end{figure}

We fit the gapped Dirac model to the energy at valley $K$ of VB1 and VB2 near $\kappa$ at $\Delta_D = 9$ meV.
The angular momentum of state at $\kappa$ under $\mathcal C_{3z}$ is $J_z = +1/2$ for VB1 and $J_z = 3/2$ for VB2.
The low energy effective model can be described by the gapped Dirac cone with asymmetrical mass as
\begin{equation}
    h^\text{D}(\mathbf k) = (\epsilon_0 + \Delta_0 k^2) \sigma_0 + v_\text{F} (k_x \sigma_x + k_y \sigma_y) + \Delta_z \sigma_z,
\end{equation}
where $\sigma_0$ is the identity matrix, $\sigma_{x,y,z}$ are Pauli matrices, $v_\text{F}$ is the Fermi velocity, and $\epsilon_0$ is a constant energy shift.
The energy spectrum is
\begin{equation}
    E^\text{D}_\pm (\mathbf k) = (\epsilon_0 + \Delta_m k^2) \pm \sqrt{\Delta_z^2 + v_\text{F}^2 k^2} 
\end{equation}
With $+$ for the electron band and $-$ for the hole band.
Near $\mathbf k = \gamma$, the energy can be approximated by
\begin{equation}
    E^\text{D}_\pm (\mathbf k) = \epsilon_0 \pm \vert\Delta_z\vert + (\Delta_0 \pm \frac{v_\text{F}^2}{2 \vert\Delta_z\vert})  k^2
    \label{eq: dirac energy approximation}
\end{equation}
The mass for electrons and holes are
\begin{equation}
    \frac{1}{2 m_e} = \frac{v_\text{F}^2}{2 \vert\Delta_z\vert} + \Delta_0, \quad \frac{1}{2 m_h} = \frac{v_\text{F}^2}{2 \vert\Delta_z\vert} - \Delta_0,
\end{equation}
which is different by $\Delta_0$.
The fitted parameters are
\begin{equation}
    \Delta_z = -1.15\text{meV}, \quad \epsilon_0 = -2.35\text{meV}, \quad 
    \Delta_0 = -0.16\text{meV} a_{\text{m},0}^{2}, \quad v_\text{F} = 1.77\text{meV} a_{\text{m},0}.
\end{equation}
The fitted spectrum is shown in Fig.~\ref{fig: gapped Dirac cone fit}, which agrees well with the spectrum of the continuum model.
The eigenstates can be taken as
\begin{equation}
    \ket{u_+(\mathbf k)} = 
    \begin{pmatrix}
    e^{-i \varphi_\mathbf k} \cos \theta_\mathbf k/2 \\
    \sin \theta_\mathbf k/2
    \end{pmatrix}, \quad
        \ket{u_-(\mathbf k)} = 
    \begin{pmatrix}
     \sin \theta_\mathbf k/2 \\
    -e^{i \varphi_\mathbf k}\cos \theta_\mathbf k/2
    \end{pmatrix},
\end{equation}
where $\cos \theta_\mathbf k = \Delta_z / \sqrt{\Delta_z^2 + v_\text{F}^2 k^2}$ and $\tan \varphi_\mathbf k = k_y /k_x$.
The gauge is chosen to avoid singularity at $\mathbf k=0$.
The Berry connection is
\begin{equation}
    A_\pm(\mathbf k) = \pm \frac{1}{k^2}(1+\frac{\Delta_z}{\sqrt{\Delta_z^2 + v_\text{F}^2 k^2}})\frac{1}{2}(\hat z \times \mathbf k).
\end{equation}
The Berry curvature and the quantum metric of the gapped Dirac cones is 
\begin{equation}
    \Omega_\pm (\mathbf k) = \mp \frac{\Delta_z v_\text{F}^2}{2 (\Delta_z^2 + v_{\text{F}}^2k^2)^{3/2}}, \quad 
    g_{\pm,ij} (\mathbf k) = \frac{1}{4} (\frac{\Delta_z v_\text{F}}{\Delta_z^2 + v_{\text{F}}^2k^2})^2\frac{k_i k_j}{k^2} +\frac{1}{4} \frac{\epsilon_{im}k_m \epsilon_{jn} k_n}{k^2} \frac{v_\text{F}^2 }{\Delta_z^2 + v_{\text{F}}^2k^2}.
\end{equation}
Near $\mathbf k = \gamma$, the Berry curvature and the diagonal quantum metric can be approximated by
\begin{equation}
    A_\pm (\mathbf k) = \mp \frac{\text{sgn}(\Delta_z) v_\text{F}^2}{2 \Delta_z^2} \frac{1}{2}(\hat z \times \mathbf k), \quad
    \Omega_\pm(\mathbf k) =  \mp \frac{\text{sgn}(\Delta_z) v_\text{F}^2}{2 \Delta_z^2}, \quad
    g_{\pm,ii} (\mathbf k) = \frac{1}{4} \frac{v_{\text{F}}^2}{\Delta_z^2}, \quad
    g_{\pm,i\neq j} (\mathbf k) =0.
\label{eq: dirac quantum geometry approximation}
\end{equation}
The sign of the Berry curvature is positive for the electron band, complying with that of the continuum model at $\kappa$ shown in Fig.~\ref{fig: quantum geometry D6}(b)(f).
The Berry connection has the form of $A_\pm(\mathbf k)= \Omega_\pm (\mathbf k=0)\hat z \times \mathbf k/2$ as those in the continuum model fixed by the optimal Chern gauge in Sec.~\ref{sec: quantum geometry of the top two valence minibands}.

\subsection{Numerical calculation of exciton dipole}
In this section, we discuss the exciton in gapped Dirac cones \cite{wu2015exciton} and its in-plane exciton dipole.

\begin{figure}
    \centering
\includegraphics[width=\columnwidth]{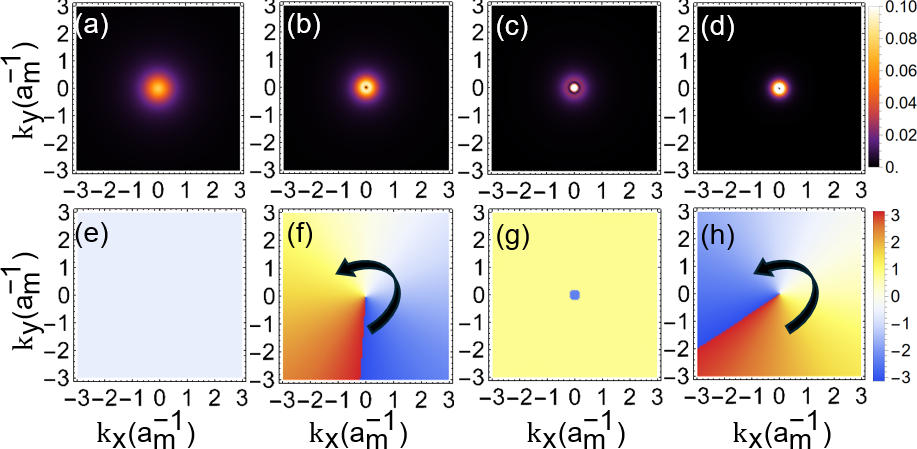}
\caption{
(a)-(d) The amplitude of the exciton envelope function in momentum space $\vert \Psi_N(e,\mathbf k,h,\mathbf k) \vert$ for $N=1s, 2p_+, 2s, 2p_-$, respectively.
(e)-(h) The phase of the exciton envelope function in momentum space $\arg \Psi_N(e,\mathbf k,h,\mathbf k)$ for $N=1s, 2p_+, 2s, 2p_-$, respectively.
The black arrows in (f)(h) label the counterclockwise direction.
}
\label{fig: exciton wavefunction in dirac cone} 
\end{figure}

The exciton in gapped Dirac cones shows hydrogenic behavior.
Here, we focus on the exciton near $\mathbf Q = \gamma$ and consider only the direct term in BSE of \eqnref{eq: BSE}.
The reduced mass, Bohr radius, and the Rydberg energy are estimated as
\begin{equation}
    \frac{1}{2m_r} = \frac{1}{2m_e} + \frac{1}{2m_h} = \frac{v_\text{F}^2}{\vert\Delta_z\vert}, \quad a_\text{B} = \frac{2 \epsilon_0\epsilon_r \hbar^2}{e^2 m_r} \approx 3.67 a_{\text{m},0}, \quad \text{Ry} = \frac{e^2}{  2\epsilon_0\epsilon_r a_\text{B} } \approx 0.20\text{meV}.
\end{equation}
Fig.~\ref{fig: exciton wavefunction in dirac cone} shows the exciton envelope function in the momentum space for the four lowest exciton energy state.
Since the angular momentum $l_z$ about $z$ axis is a good quantum number, we can take $N=nl_z$.
Fig.~\ref{fig: exciton wavefunction in dirac cone} (a)(e) is the $1s$ exciton with real exciton envelope function up to a constant phase.
Fig.~\ref{fig: exciton wavefunction in dirac cone} (b)(f) is the $2p_+$ exciton envelope function with a node at $\gamma$ and its phase increasing following the counterclockwise direction (black arrow) as $e^{i\varphi_\mathbf k}$, where $\varphi_\mathbf k$ is the azimuthal angle of momentum $\mathbf k$.
Fig.~\ref{fig: exciton wavefunction in dirac cone} (c)(g) is the $2s$ exciton with a line node near $\gamma$ and the real envelope function changing its sign shown by the $\pi$ phase different at $\mathbf k=\gamma$ and $\mathbf k\rightarrow \infty$.
Fig.~\ref{fig: exciton wavefunction in dirac cone} (d)(h) is the $2p_-$ exciton envelope function with a node at $\gamma$ and its phase winding as $e^{-i\varphi_\mathbf k}$.
The binding energies for these exciton are $E^\text{b}_{1s} = 0.734$meV, $E^\text{b}_{2p_+} = 0.166$meV, $E^\text{b}_{2s} = 0.110$meV, and $E^\text{b}_{2p_-} = 0.106$meV.
Qualitatively, the exciton envelope functions in gapped Dirac cone are similar to those hydrogenic wavefunctions while they differ by the splitting of $2s$ and $2p_\pm$ binding energy.

\begin{figure}
    \centering
\includegraphics[width=0.8\columnwidth]{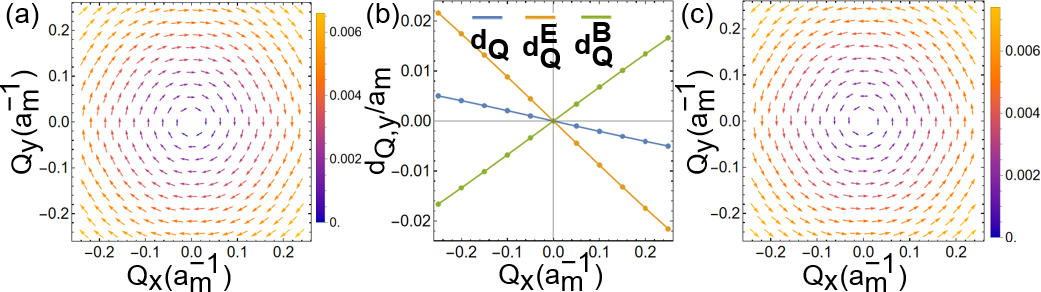}
\caption{
(a) The helical exciton dipole structure near $\mathbf Q$ for the lowest energy 1s exciton.
The dipole direction is represented by the arrow direction and the dipole magnitude is the length and color of the arrowss.
(b) The decomposition of exciton dipole $\mathbf d_\mathbf Q = (0,\mathbf d_{\mathbf Q,y})$ for $\mathbf Q=(Q_x,0)$ into its envelope function contribution $\mathbf d_\mathbf Q^\text{E}$ and its Bloch function contribution $\mathbf d_\mathbf Q^\text{B}$.
(c) The helical exciton dipole structure near $\mathbf Q$ of 1s exciton for opposite $\Delta_z$.
%(c) The amplitude of the exciton envelope function $\vert \Psi_{N=1s}(e,\mathbf k+\alpha_e \mathbf Q,h, \mathbf k-\alpha_h \mathbf Q)$ along $\mathbf k=(k_x,0)$ for $\mathbf Q=\gamma$ (gray solid line) and $\mathbf Q=(Q_x,0))$ (black dashed line).
}
\label{fig: exciton dipole in dirac cone} 
\end{figure}

We next show the helical exciton dipole structure of gapped Dirac cones and the part cancellation between the envelope function and Bloch contribution to dipoles.
The electron hole basis for exciton is chosen as 
\begin{equation}
    c^\dagger_{e, \mathbf k + \alpha_e \mathbf Q}c_{h, \mathbf k - \alpha_h \mathbf Q} \ket{\text{FS}},
\end{equation}
where $\alpha_e = m_e / (m_e + m_h) \approx 0.559, \alpha_h = m_h / (m_e + m_h)\approx 0.441$ and $\alpha_e + \alpha_h =1$.
The exciton dipole shows a helical structure $\mathbf d_{1s,\mathbf Q} = v_{1s} \hat z \times \mathbf Q$ with $v_{1s}=-0.021 a_{\text{m},0}^2$, consistent with the sign of $v$ at the large displacement field in Fig.~\ref{fig: gappedDiraccone}(a) of the main text.
Fig.~\ref{fig: exciton dipole in dirac cone}(b) decomposes the dipole into the envelope function and the Bloch contribution and shows that $\mathbf d_\mathbf Q^\text{E}$ is opposite to and dominates over $\mathbf d_\mathbf Q^\text{B}$, qualitatively complying with the exciton dipole at the large displacement field in Fig.~\ref{fig: gappedDiraccone}(a).
When the Berry curvature is reversed by taking the opposite sign of $\Delta_z$, the helical exciton dipole reverses its direction while largely keeping its magnitude as shown in Fig.~\ref{fig: exciton dipole in dirac cone}(c), qualitatively showing its origin in Berry curvature.

\subsection{Perturbative analysis of exciton dipole}
\label{sec: exciton dipole perturbation}
In this section, we discuss the qualitatively physical origin for the envelope function contribution and the opposite sign to the Bloch contribution in the limit of $v_\text{F}/ a_\text{B} \ll \vert\Delta_z\vert$ following Ref.~\cite{zhou2015berry, srivastava2015signatures}, $\vert\mathbf Q\vert\ll 1/a_\text{B}$, and the gate distance $d\rightarrow\infty$ to reproduce the $1/q$ in the Coulomb interaction.

We first identify the 2DEG/2DHG with the bare Coulomb interaction is the unperturbed BSE.
For $\mathcal H^{\text{K}}(e,\mathbf k_1+\alpha_e \mathbf Q, h, \mathbf k_1- \alpha_h \mathbf Q)$, we first expand $E_\pm^\text{D}(\mathbf k)$ to the $k^4$ order and then expand $\mathcal H^{\text{K}}(e,\mathbf k_1+\alpha_e \mathbf Q, h, \mathbf k_1- \alpha_h \mathbf Q)$ up to second order in $\mathbf Q$.
$k^4 Q^2$ is dropped while $k^2 Q^2$ term is kept.
The zeroth order of the kinetic term is 
\begin{equation}
\begin{split}
    \mathcal H^{\text{K},0}(e,\mathbf k_1+\alpha_e \mathbf Q, h, \mathbf k_1- \alpha_h \mathbf Q) = 2 \Delta_z + \frac{k_1^2}{2 m_r} + \frac{Q^2}{2(m_e + m_h)} \\
\end{split}
\end{equation}
The first order perturbation of the kinetic term is
\begin{equation}
\begin{split}
    &\mathcal H^{\text{K},1}(e,\mathbf k_1+\alpha_e \mathbf Q, h, \mathbf k_1- \alpha_h \mathbf Q) =
    \Delta _0\mathbf k_1\cdot\mathbf Q \frac{ v_{\text{F}}^2 k_1^2} {\left| \Delta _z\right|^2} 
    -\frac{k_1^2}{2 m_r}\frac{ v_\text{F}^2 k_1^2}{4 \left| \Delta _z\right| ^2} 
    -\frac{3 v^4_\text{F} (\mathbf k_1\cdot\mathbf Q)^2 \left(\alpha _e^2+\alpha _h^2\right)}{4 \left| \Delta _z\right|^3}
\end{split}
\label{eq: kinetic term 1st order}
\end{equation}
For the direct term $\mathcal H^\text{D}(e,\mathbf k_1+\alpha_e\mathbf Q, h, \mathbf k_1 -\alpha_h \mathbf Q; e,\mathbf k_2+\alpha_e\mathbf Q, h, \mathbf k_2-\alpha_h \mathbf Q)$ in BSE, we expand the form factor up to second order as
\begin{equation}
\begin{split}
    &\braket{u_{e,\mathbf k_1 +\alpha_e \mathbf Q}}{u_{e,\mathbf k_2 +\alpha_e \mathbf Q}} \approx 
    1 - i \sum_{j,j'=x,y}\frac{1}{4} \Omega_+ \epsilon_{jj'}(\mathbf k_2 - \mathbf k_1)_j (\mathbf k_1 + \mathbf k_2 + 2 \alpha_e \mathbf Q)_{j'}
    -\sum_{j,j'=x,y}\frac{1}{2}(\mathbf k_2 - \mathbf k_1)_{j}(\mathbf k_2 - \mathbf k_1)_{j} g_{+,jj} \\
    &\braket{u_{h,\mathbf k_1 - \alpha_h \mathbf Q}}{u_{e,\mathbf k_2 - \alpha_h \mathbf Q}} \approx 
    1 - i \sum_{j,j'=x,y}\frac{1}{4} \Omega_- \epsilon_{jj'}(\mathbf k_2 - \mathbf k_1)_j (\mathbf k_1 + \mathbf k_2 - 2 \alpha_h \mathbf Q)_{j'}
    -\sum_{j,j'=x,y}\frac{1}{2}(\mathbf k_2 - \mathbf k_1)_{j}(\mathbf k_2 - \mathbf k_1)_{j} g_{-,jj}.
\end{split}
\end{equation}
In the limit of $v_\text{F}/ a_\text{B} \ll \vert\Delta_z\vert$, the approximation for the Berry connection and the quantum geometry in \eqnref{eq: dirac quantum geometry approximation} is valid and $\Omega_\pm (\mathbf k), g_{\pm, ii}(\mathbf k)$ are taken as constant.
The zero order of the direct term is 
\begin{equation}
    \mathcal H^\text{D,0}(e,\mathbf k_1+\alpha_e\mathbf Q, h, \mathbf k_1 -\alpha_h \mathbf Q; e,\mathbf k_2+\alpha_e\mathbf Q, h, \mathbf k_2-\alpha_h \mathbf Q) = V(\mathbf k_1-\mathbf k_2).
\end{equation}
The perturbation of the direct term is
\begin{equation}
\begin{split}
    & \mathcal H^\text{D,1}(e,\mathbf k_1+\alpha_e\mathbf Q, h, \mathbf k_1 -\alpha_h \mathbf Q; e,\mathbf k_2+\alpha_e\mathbf Q, h, \mathbf k_2-\alpha_h \mathbf Q) = \sum_{j,j'=x,y} V(\mathbf k_1-\mathbf k_2)\Big(\\
    &-i \frac{1}{2}(\Omega_+ - \Omega_-)\epsilon_{jj'} \mathbf k_{2,j} \mathbf k_{1,j'}
    -i \frac{1}{2} (\alpha_e \Omega_+ + \alpha_h \Omega_-)\epsilon_{jj'} (\mathbf k_2 -\mathbf k_1)_{j}\mathbf Q_{j'} 
    - \frac{1}{2}(\mathbf k_2 - \mathbf k_1)_{j}(\mathbf k_2 - \mathbf k_1)_{j} (g_{+,jj}+g_{-,jj})\Big).
\end{split}
\label{eq: direct term 1st order}
\end{equation}
So the zero order BSE is 
\begin{equation}
\begin{split}
    &(2 \Delta_z + \frac{k_1^2}{2 m_r} + \frac{Q^2}{2(m_e + m_h)})\Psi_N^0 (e,\mathbf k_1+\alpha_e \mathbf Q, h, \mathbf k_1- \alpha_h \mathbf Q) 
    +\sum_{\mathbf k_2} V(\mathbf k_1-\mathbf k_2) (e,\mathbf k_2+\alpha_e \mathbf Q, h, \mathbf k_2- \alpha_h \mathbf Q) \\
    &= E^0_{N,\mathbf Q} \Psi_N^0 (e,\mathbf k_1+\alpha_e \mathbf Q, h, \mathbf k_1- \alpha_h \mathbf Q),
\end{split}
\end{equation}
which is the BSE for 2DEG/2DHG with the Coulomb interaction.
The exciton envelope function at finite $\mathbf Q$ is
\begin{equation}
    \Psi_N^0 (e,\mathbf k_1+\alpha_e \mathbf Q, h, \mathbf k_1- \alpha_h \mathbf Q) =\Psi_N^0 (e,\mathbf k_1, h, \mathbf k_1)
\end{equation}
with the energy 
\begin{equation}
    E^0_{N,\mathbf Q} = E^0_{N,\mathbf Q=0}+ \frac{Q^2}{2(m_e + m_h)},
\end{equation}
which serves as the unperturbed basis.
The unperturbed exciton wavefunction in 2D \cite{hage2008two} is 
\begin{equation}
    \Psi_{nl}^0(\mathbf k)=\frac{1}{2} \left(\frac{\frac{2}{n-1/2}}{k^2+(\frac{1}{n-1/2})^2}\right)^{3/2} i^{| l| } e^{i l \varphi_\mathbf k} \sqrt{\frac{2 (-| l| +n-1)!}{\pi  (|l|+n-1)!}} L_{n-1}^{| l| }\left(\frac{k^2-(\frac{1}{n-1/2})^2}{k^2+(\frac{1}{n-1/2})^2}\right),
\label{eq: 2d hydrogen wavefunction}
\end{equation}
$ L_{n-1}^{| l| }(x)$ is the Legendre functions.

\renewcommand{\arraystretch}{2}
\begin{table}[t] % optional floating environment
\centering
\begin{tabular}{|c|c|c|}
\hline
$\mathcal H^{1} (\mathbf k_1, \mathbf k_2, \mathbf Q)$ & $l_\mathcal H$ & $(l_1, l_2)$ \\ \hline
$\Delta _0\mathbf k_1\cdot\mathbf Q \frac{ v_{\text{F}}^2 k_1^2} {\left| \Delta _z\right|^2} $        & $\pm 1$        & $(l\pm1,l)$        \\ \hline
$-\frac{k_1^2}{2 m_r}\frac{ v_\text{F}^2 k_1^2}{4 \left| \Delta _z\right| ^2}$        & $0$         & $(l,l)$       \\ \hline
$-\frac{3 v^4_\text{F} (\mathbf k_1\cdot\mathbf Q)^2 \left(\alpha _e^2+\alpha _h^2\right)}{4 \left| \Delta _z\right|^3}$       & $\pm2,0$         & $(l\pm2 \text{ or } l,l)$       \\ \hline
$-i \frac{1}{2}(\Omega_+ - \Omega_-)\epsilon_{jj'} \mathbf k_{2,j} \mathbf k_{1,j'}$ & $0$         & $(l,l)=\pm(1,1)$       \\ \hline
$-i \frac{1}{2} (\alpha_e \Omega_+ + \alpha_h \Omega_-)\epsilon_{jj'} (\mathbf k_2 -\mathbf k_1)_{j}\mathbf Q_{j'} $ & $\pm1$         & $(l \pm 1,l)$       \\ \hline
$- \frac{1}{2}(\mathbf k_2 - \mathbf k_1)_{j}(\mathbf k_2 - \mathbf k_1)_{j} (g_{+,jj}+g_{-,jj})$ & $0$         & $(l,l)$ \\ \hline
\end{tabular}
\caption{$l_\mathcal H$ contained in each perturbation of the kinetic and direct term of BSE. $\mathcal H^{X,1}_{n_1 l_1,n_2 l_2}$ is nonzero for $(l_1,l_2)$ listed in the third column.}
\label{tab: lh}
\end{table}

We apply the first order perturbation to the exciton and analyze the perturbation to BSE under the rotation symmetry.
The first order perturbation in energy is 
\begin{equation}
    \mathcal H^{X,1}_{n_1 l_1, n_2 l_2}(\mathbf Q) = \sum_{\mathbf k_1,\mathbf k_2} \Psi_{{n_1 l_{1}}}^0 (e,\mathbf k_1+\alpha_e \mathbf Q, h, \mathbf k_1- \alpha_h \mathbf Q)^* \mathcal H^{1} (\mathbf k_1, \mathbf k_2, \mathbf Q) \Psi_{n_2l_{2}}^0 (e,\mathbf k_2+\alpha_e \mathbf Q, h, \mathbf k_2- \alpha_h \mathbf Q).
\end{equation}
$\mathcal H^{1} (\mathbf k_1, \mathbf k_2, \mathbf Q)$ represents one of the perturbation term to BSE.
We make use of the rotation symmetry to constrain the perturbation $\mathcal H^{1} (\mathbf k_1, \mathbf k_2, \mathbf Q)$ to be nonzero only for certain $l_1,l_2$ by 
\begin{equation}
\begin{split}
    \mathcal H^{X,1}_{n_1 l_1, n_2 l_2}(\mathbf Q) & = \sum_{\mathbf k_1,\mathbf k_2} \Psi_{{n_1 l_{1}}}^0 (e,\mathbf k_1, h, \mathbf k_1)^* \mathcal H^{1} (\mathbf k_1, \mathbf k_2, \mathbf Q) \Psi_{n_2l_{2}}^0 (e,\mathbf k_2, h, \mathbf k_2) \\
    &=\sum_{\mathbf k_1,\mathbf k_2} \Psi_{{n_1 l_{1}}}^0 (e,\mathcal R\mathbf k_1, h, \mathcal R\mathbf k_1)^* \mathcal H^{1} (\mathcal R\mathbf k_1, \mathcal R\mathbf k_2, \mathbf Q) \Psi_{n_2l_{2}}^0 (e,\mathcal R\mathbf k_2, h, \mathcal R\mathbf k_2) \\
    &=\sum_{\mathbf k_1,\mathbf k_2} e^{-i l_1 \phi_\mathcal R}\Psi_{{n_1 l_{1}}}^0 (e,\mathbf k_1, h, \mathbf k_1)^* e^{i l_\mathcal H \phi_\mathcal R} \mathcal H^{1} (\mathbf k_1,\mathbf k_2, \mathbf Q)e^{i l_2 \phi_\mathcal R} \Psi_{n_2l_{2}}^0 (e,\mathbf k_2, h,\mathbf k_2)
\end{split}
\end{equation}
with $\varphi_\mathcal R$ is the rotation angle of the rotation $\mathcal R$,
which leads to 
\begin{equation}
    l_1=l_\mathcal H + l_2.
\end{equation}
We list $l_\mathcal H$ for each perturbation in kinetic and direct term as in Tab.~\ref{tab: lh} by making use of 
\begin{equation}
    \mathbf k_1 \cdot \mathbf k_2 = \frac{1}{2} (k_{1+}k_{2-}+k_{1-}k_{2+}), \quad 
    \hat z \cdot \mathbf k_1 \times \mathbf k_2 = \frac{1}{2 i } (-k_{1+}k_{2-}+k_{1-}k_{2+}),
\end{equation}
with $k_\pm = k_x \pm i k_y$.
We consider for example the degeneracy between $2s$ and $2p_\pm$ under the perturbation at $\mathbf Q = 0$.
The non parabolic single particle kinetic terms $-\frac{k_1^2}{2 m_r}\frac{ v_\text{F}^2 k_1^2}{4 \left| \Delta _z\right| ^2}$ and the quantum metric term $- \frac{1}{2}(\mathbf k_2 - \mathbf k_1)_{j}(\mathbf k_2 - \mathbf k_1)_{j} (g_{+,jj}+g_{-,jj})$ split the exciton $2s$ with $2p_\pm$ while $-i \frac{1}{2}(\Omega_+ - \Omega_-)\epsilon_{jj'} \mathbf k_{2,j} \mathbf k_{1,j'}$ lifts the remaining degeneracy between $2p_\pm$.
Real $\mathcal H^{1} (\mathbf k_1, \mathbf k_2, \mathbf Q)$ keeps the degeneracy between $2p_\pm$ by $\mathcal H^{X,1}_{2 p_+, 2p_+}(\mathbf Q=0)=\mathcal H^{X,1}_{2 p_-, 2 p_-}(\mathbf Q=0)$ while imaginary $\mathcal H^{1} (\mathbf k_1, \mathbf k_2, \mathbf Q)$ leads to the splitting of $2p_\pm$ by $\mathcal H^{X,1}_{2p_+,2p_+}(\mathbf Q=0)=-\mathcal H^{X,1}_{2p_-, 2p_-}(\mathbf Q=0)$, basing on $\Psi_{np_+}^0 (e,\mathbf k_1, h, \mathbf k_1)^* = \Psi_{np_-}^0 (e,\mathbf k_1, h, \mathbf k_1)$ from the time reversal symmetry of the unpertrubed BSE.
The first order perturbation wavefunction is
\begin{equation}
    \Psi_{n_1l_1}^1 (e,\mathbf k_2+\alpha_e \mathbf Q, h, \mathbf k_2- \alpha_h \mathbf Q) = \sum_{n_2 l_2,\mathcal H^{1}} \frac{\mathcal H^{X,1}_{n_2 l_2, n_1 l_1}(\mathbf Q)}{E^0_{n_1 l_1,\mathbf Q=0}-E^0_{n_2 l_2,\mathbf Q=0}}  \Psi_{n_2l_{2}}^0 (e,\mathbf k_2, h, \mathbf k_2).
\end{equation}

We next study the exciton dipole under the first order perturbation and show the envelope function contribution comes from the $sp$ hybridization and has the opposite sign to the Bloch contribution.
The exciton dipole under the first order perturbation is
\begin{equation}
    \mathbf d_{nl,\mathbf Q} =  \frac{1}{2}(\alpha_e\Omega_+ +\alpha_h\Omega_-)\hat z\times \mathbf Q + 2\Re\sum _\mathbf k \Psi _{nl}^0(e,\mathbf k,h,\mathbf k)^*  (-i \partial_\mathbf k\Psi _{nl}^1(e,\mathbf k + \alpha_e \mathbf Q,h,\mathbf k-\alpha_h \mathbf Q)) ,
\end{equation}
where the first term is the Bloch contribution $\mathbf d^\text{B}_{nl,\mathbf Q}$ and the second term is the envelope function contribution $\mathbf d^\text{E}_{nl,\mathbf Q}$.
For $\mathbf d^\text{B}_{nl,\mathbf Q}$, the physical origin is the remaining Berry curvature weighted by effective mass from the cancellation between electron and hole Bloch states.
The linear coefficient near $\gamma$
\begin{equation}
    v^\text{B}_{nl} = \alpha_e\Omega_+ +\alpha_h\Omega_->0
\label{eq: sign of Bloch contribution to dipole}
\end{equation}
with the sign following $\Omega_+$ because $\alpha_e > \alpha_h$ and $\Omega_+ = -\Omega_->0$, complying with the positive slope of $\mathbf d^\text{B}_{1s,\mathbf Q}$ in Fig.~\ref{fig: exciton dipole in dirac cone}(b).
For $\mathbf d^\text{E}_{nl,\mathbf Q}$, expanding the first order perturbation exciton envelope function leads to 
\begin{equation}
    \mathbf d^\text{E}_{nl,\mathbf Q} = 2\Re \sum_{n_2 l_2,\mathcal H^{1}} \frac{\mathcal H^{X,1}_{n_2 l_2, n l}(\mathbf Q)}{E^0_{n l,\mathbf Q=0}-E^0_{n_2 l_2,\mathbf Q=0}} \sum _\mathbf k   \Psi _{nl}^0(e,\mathbf k,h,\mathbf k)^*  (-i \partial_\mathbf k\Psi_{n_2l_{2}}^0 (e,\mathbf k, h, \mathbf k)) = \sum_{\mathcal H^{1}} \mathbf d^{\text{E}, \mathcal H^{X,1}}_{nl,\mathbf Q}.
\end{equation}
$\mathbf d^{\text{E}, \mathcal H^{X,1}}_{nl,\mathbf Q}$ labels the contribution to exciton dipole from the perturbation term $\mathcal H^1(\mathbf k_1,\mathbf k_2,\mathbf Q)$.
$-i\partial_\mathbf k$ is nonzero for $l_2 = l \pm 1$, leading to $l_\mathcal H = \pm 1$.
From Tab.~\ref{tab: lh}, $\Delta _0\mathbf k_1\cdot\mathbf Q \frac{ v_{\text{F}}^2 k_1^2} {\left| \Delta _z\right|^2} $  and $-i \frac{1}{2} (\alpha_e \Omega_+ + \alpha_h \Omega_-)\epsilon_{jj'} (\mathbf k_2 -\mathbf k_1)_{j}\mathbf Q_{j'} $ in the sum of $\mathcal H^1(\mathbf k_1,\mathbf k_2,\mathbf Q)$ can contribute to $\mathbf d^\text{E}_{nl,\mathbf Q}$.
Consider $nl=1s, n_2 l_2 = n_2 p_\pm$ and denotes
\begin{equation}
    \sum_\mathbf k \Psi_{1s}^0(e,\mathbf k,h,\mathbf k)^*  (-i \partial_\mathbf k\Psi_{n_2 p_\pm}^0 (e,\mathbf k, h, \mathbf k))= -i c_{n_2} (-i \hat x \pm \hat y)a_{\text{B}}
\end{equation}
with the constant $c_{n_2}>0$.
For $\mathcal H^1(\mathbf k_1,\mathbf k_2,\mathbf Q) = \Delta _0\mathbf k_1\cdot\mathbf Q \frac{ v_{\text{F}}^2 k_1^2} {\left| \Delta _z\right|^2} = \Delta _0 \frac{ v_{\text{F}}^2 k_1^2} {\left| \Delta _z\right|^2} \frac{1}{2}(k_{1+}Q_{-}+k_{1-}Q_{+})$,
\begin{equation}
    H^{X,1}_{n_2 p_-, 1s}(\mathbf Q) = (-i) \frac{\Delta_0 v_\text{F}^2}{\vert \Delta_z \vert^2 a_{\text{B}}^4}  c_{n_2}^{X} Q_+, \quad
    H^{X,1}_{n_2 p_+, 1s}(\mathbf Q) = (-i) \frac{\Delta_0 v_\text{F}^2}{\vert \Delta_z \vert^2 a_{\text{B}}^4} c_{n_2}^{X} Q_-,
\end{equation}
with the constant $c_{n_2}^{X} > 0$ and $-i$ comes from the $(-i)^{|l|}$ \eqnref{eq: 2d hydrogen wavefunction}.
Collecting all terms gives
\begin{equation}
\begin{split}
    \mathbf d^{\text{E}, \mathcal H^{X,1}}_{nl,\mathbf Q} & = 2 \Re \sum_{n_2}  \frac{-i\frac{\Delta_0 v_\text{F}^2}{\vert \Delta_z \vert^2 a_{\text{B}}^4} c_{n_2}^{X} Q_+ }{E^0_{1s,\mathbf Q=0}-E^0_{n_2 p_-,\mathbf Q=0}} (-i) c_{n_2} (-i \hat x - \hat y)a_{\text{B}} + \frac{-i\frac{\Delta_0 v_\text{F}^2}{\vert \Delta_z \vert^2 a_{\text{B}}^4} c_{n_2}^{X} Q_+ }{E^0_{1s,\mathbf Q=0}-E^0_{n_2 p_+,\mathbf Q=0}} (-i) c_{n_2}(-i \hat x + \hat y)a_{\text{B}} \\
    &=2 \Re \sum_{n_2}  \frac{\frac{\Delta_0 v_\text{F}^2}{\vert \Delta_z \vert^2 a_{\text{B}}^3} c_{n_2}^{X}  c_{n_2}}{E^0_{1s,\mathbf Q=0}-E^0_{n_2 p_-,\mathbf Q=0}} 2i \mathbf Q \\
    & = 0,
\end{split}
\end{equation}
complying with $\mathbf d_{N,\mathbf Q} = v_N \hat z \times \mathbf Q$ without the $\mathbf Q$ term.
For $-i \frac{1}{2} (\alpha_e \Omega_+ + \alpha_h \Omega_-)\epsilon_{jj'} (\mathbf k_2 -\mathbf k_1)_{j}\mathbf Q_{j'} =- \frac{1}{4} (\alpha_e \Omega_+ + \alpha_h \Omega_-) (-(k_{2+}-k_{1+})Q_{-}+(k_{2-}-k_{1-})Q_{+})$, we have
\begin{equation}
    H^{X,1}_{n_2 p_-, 1s}(\mathbf Q) = -i c_{n_2}^{X'} \frac{1}{4 a_\text{B}} (\alpha_e \Omega_+ + \alpha_h \Omega_-)Q_+, \quad
    H^{X,1}_{n_2 p_+, 1s}(\mathbf Q) = i c_{n_2}^{X'} \frac{1}{4 a_\text{B}} (\alpha_e \Omega_+ + \alpha_h \Omega_-)Q_-,
\end{equation}
with the constant $c_{n_2}^{X'} > 0$.
Its contribution to the dipole is
\begin{equation}
\begin{split}
    \mathbf d^{\text{E}, \mathcal H^{X,1}}_{nl,\mathbf Q} & = 2 \Re \sum_{n_2}  \frac{-i c_{n_2}^{X'} \frac{1}{4 a_\text{B}} (\alpha_e \Omega_+ + \alpha_h \Omega_-)Q_+}{E^0_{1s,\mathbf Q=0}-E^0_{n_2 p_-,\mathbf Q=0}} (-i) c_{n_2} (-i \hat x - \hat y)a_{\text{B}} + \frac{i c_{n_2}^{X'} \frac{1}{4 a_\text{B}} (\alpha_e \Omega_+ + \alpha_h \Omega_-)Q_-}{E^0_{1s,\mathbf Q=0}-E^0_{n_2 p_+,\mathbf Q=0}} (-i) c_{n_2}(-i \hat x + \hat y)a_{\text{B}} \\
    &= \sum_{n_2}  \frac{c_{n_2}^{X'}c_{n_2} (\alpha_e \Omega_+ + \alpha_h \Omega_-)}{E^0_{1s,\mathbf Q=0}-E^0_{n_2 p_-,\mathbf Q=0}} \hat z \times \mathbf Q.
\end{split}
\label{eq: envelope function contribution to dipole}
\end{equation}
Thus, for $\mathbf d^{\text{E}}_{1s,\mathbf Q}$, the physical origin is the $sp$ hybridization between the $1s$ and $n_2 p_\pm$ exciton with the linear coefficient near $\gamma$ as
\begin{equation}
    v^{\text{E}}_{1s,\mathbf Q} < 0
\end{equation}
from the negative energy denominator in \eqnref{eq: envelope function contribution to dipole}, which has the opposite sign to $v^{\text{B}}_{1s,\mathbf Q}$ in \eqnref{eq: sign of Bloch contribution to dipole} and complies with the negative slope of $\mathbf d^\text{E}_{1s,\mathbf Q}$ in Fig.~\ref{fig: exciton dipole in dirac cone}(b).

The switch of the helicity of the exciton dipole in Fig.~\ref{fig: exciton dipole in dirac cone}(c) from the $\Delta_z$ originates from the Berry curvature in both $v^{\text{E}}_{1s,\mathbf Q}, v^{\text{B}}_{1s,\mathbf Q}$.
The signs of both $v^{\text{E}}_{1s,\mathbf Q}, v^{\text{B}}_{1s,\mathbf Q}$ share a common factor  $(\alpha_e \Omega_+ + \alpha_h \Omega_-)$ as the perturbation analysis in \eqnref{eq: sign of Bloch contribution to dipole} and \eqnref{eq: envelope function contribution to dipole}.
Reversing the sign of $\Delta_z$ leads to the sign change of both $\Omega_+, \Omega_-$ and also the dipole direction.

\section{Exciton in twisted MoTe2}
\subsection{Exciton evolution under the displacement field}
\label{sec: Exciton evolution under the displacement field}
% In this section, we study the evolution of the exciton band and envelope function under the displacement field, qualitatively determined by single-particle electron and hole band.
In this section, we study the evolution of the envelope function under the displacement field, qualitatively determined by single-particle electron and hole band.

\begin{figure}
\centering
\includegraphics[width=\columnwidth]{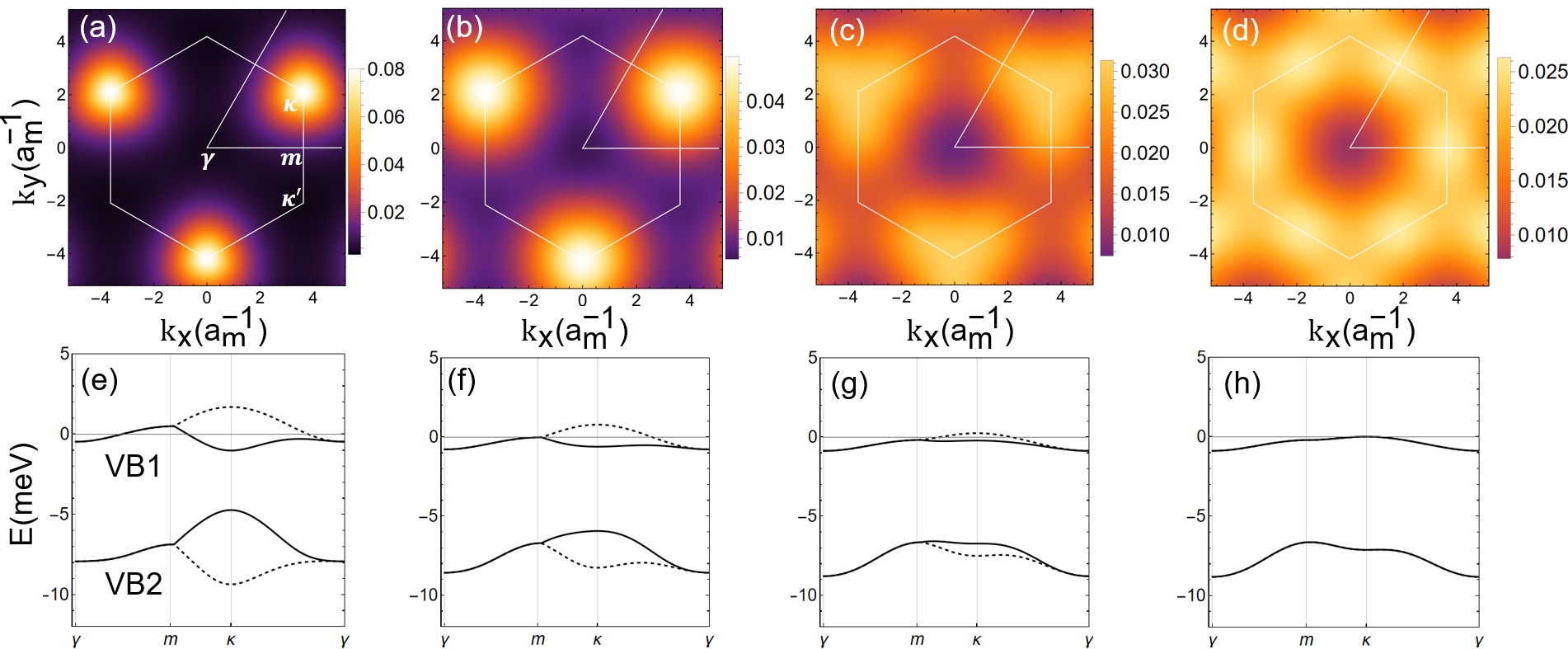}
\caption{
(a)-(d) The amplitude of the envelope function of the lowest energy exciton $\vert \Psi_{N=1} (e, \mathbf k, h, \mathbf k)\vert$ for the diplacement field induced potential difference $\Delta_D =6,3,1,0$ meV, respectively.
(e)-(h) The electron (VB1) and hole (VB2) bands $\Delta_D =6,3,1,0$ meV, respectively.
The solid line is the spectrum in valley $K$ while the dashed line is the spectrum in valley $K'$.
}
\label{fig: exciton envelop D} 
\end{figure}

The exciton envelope function shifts its peak in magnitude from $m$ to $\kappa$ when increasing the displacement field, accompanying a transition from the Frenkel type to the Wannier type.
Fig.~\ref{fig: exciton envelop D}(a)-(d) shows the exciton envelope function for the lowest energy exciton $N=1$ with decreasing displacement field.
The magnitude of the envelope function peaks at $\kappa$ at large displacement field in Fig.~\ref{fig: exciton envelop D}(a), forming Wannier type of exciton.
With decreasing $\Delta_D$, the peak value decreases and the envelope function spreads more aound $\kappa$ as shown in Fig.~\ref{fig: exciton envelop D}(b).
When $\Delta_D$ is decreased further in Fig.~\ref{fig: exciton envelop D}(c), the amplitude at $\kappa$ can be equal or smaller than that nearby and its peak shifts away from $\kappa.$
At $\Delta_D = 0$ meV in Fig.~\ref{fig: exciton envelop D}(d), the peak is located at $m$ and the envelope function spreads across the MBZ boundary (white hexagonal), forming Frenkel type exciton.

The evolution of the exciton peak is qualitatively determined by the minimum of the direct gap between the electron and hole band, accompanying a transition from the direct gap to indrect gap insulator, shown in Fig.~\ref{fig: exciton envelop D}(e)-(h).
At a large displacement field in Fig.~\ref{fig: exciton envelop D}(e), the minimum direct gap is located at $\kappa$ with the envelope function peak at the same position to minimize the energy $E^\text{K}_{N=1,\mathbf Q=\gamma}$.
When decreasing the displacement field in Fig.~\ref{fig: exciton envelop D}(f)(g), the direct gap at $\kappa$ increases and becomes comparable to the direct gap at $m$, corresponding to the fact that the peak of the envelope function shifts off $\kappa$.
At zero displacement field Fig.~\ref{fig: exciton envelop D}(h), the direct gap is minimum at $m$, corresponds to the envelope function peak at $m$.
The direct gap at $\kappa$ is a little larger than that at $m$ and is the same as that at $\kappa'$ due to the intra valley inversion, so that the envelope function is nonzero and the same at $\kappa, \kappa'$.
The direct gap at $\gamma$ is larger than those at $m, \kappa, \kappa'$, leading to the envelope function smearing over the moir\'e MBZ boundary without amplitude near $\gamma$.

\subsection{Exciton drift velocity}
\label{sec: Exciton drift velocity}
Besides the out-of-plane displacement field, the helical exciton dipole texture enables the response of the exciton to the uniform in-plane electrical field by an anomalous drift velocity \cite{cao2021quantum,tang2024inheritance} proportional to $v$.
Take the exciton at $\mathbf Q = \gamma$ as an example.
In semiclassical equation of motion, the group velocity of the center-of-mass motion is $\nabla_\mathbf Q (E_\mathbf Q + e \mathcal E \cdot\mathbf d_\mathbf Q)$, where $\mathcal E$ represents the uniform in-plane electrical field coupled to the exciton dipole.
$\nabla_\mathbf Q E_\mathbf Q$ is the usual group velocity associated with the exciton energetics and $\nabla_\mathbf Q E_\mathbf Q=0$ for the exciton at $\mathbf Q = \gamma$.
$\nabla_\mathbf Q e \mathcal E \cdot \mathbf d_\mathbf Q = e v \mathcal E \times \hat z$ from \eqnref{eq: helical dipole} of the main text is perpendicular to the $\mathcal E$ and originates in the Berry curvature of the electron and holes, termed as the anomalous drift velocity, which dominates the center-of-mass motion of the exciton with $\mathbf Q = \gamma$.
This anomalous drift velocity is analogous to those induced by the perpendicular $\mathcal E$ and external magnetic field $\mathcal B$, where the Berry curvature serves as $\mathcal B$ in the momentum space.
The $v$ can be tuned by the displacement field as shown in Fig.~\ref{fig: gappedDiraccone}(a) of the main text, enabling the electrical control of both the magnitude and the direction of the anomalous drift velocity. 
In terms of the typical in-plane electrical field $\mathcal E = 1$ V/m for the transport measurement of tMoTe$_2$ \cite{park2023observation}, the exciton drift velocity reaches $e v \vert \mathcal E \vert \sim 0.016$ m/s.

%\subsection{Higher exciton bands}
\subsection{Two-body exciton interaction}
In this section, we discuss the dipole-dipole interaction for the exciton and its tunability under the displacement field.

\begin{figure}
\centering
\includegraphics[width=\columnwidth]{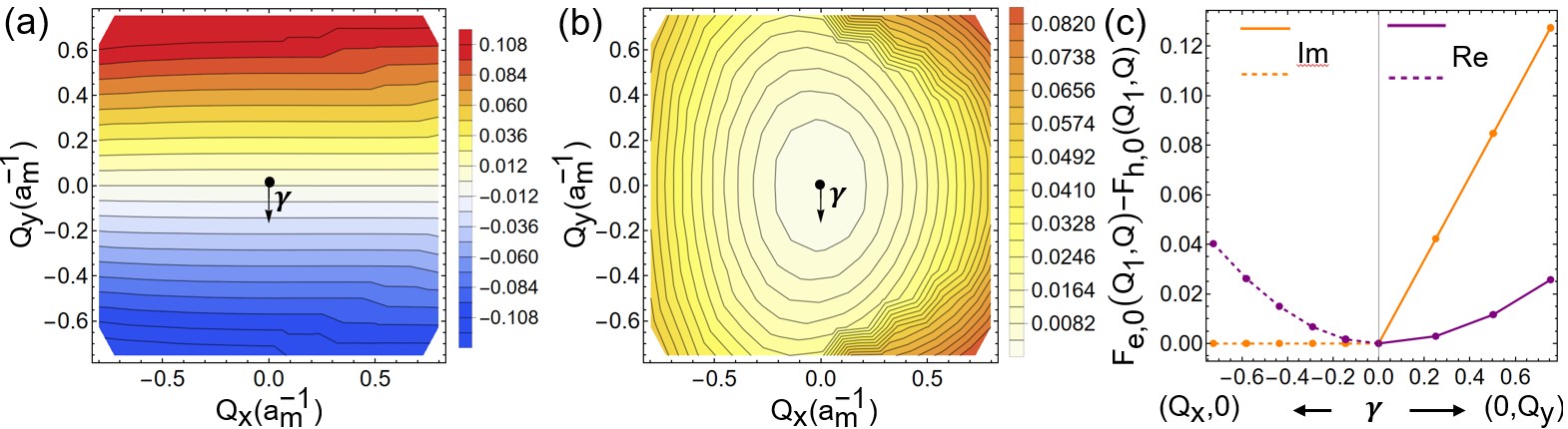}
\caption{
(a) The imaginary part of the form factor for the dipole-dipole interaction between exciton $\Im (F_{e,\mathbf G=0}(\mathbf Q_1, \mathbf Q) - F_{h,\mathbf G=0}(\mathbf Q_1, \mathbf Q))$ defined in \eqnref{eq: exciton interaction form factor}.
The arrow is the dipole $\mathbf d_{N=1, \mathbf Q_1} /a_{\text{m,0}}$ with $\mathbf Q_1 a_{\text{m,0}}= (-2\pi /5\sqrt 3,0)$.
The contour corresponds to a constant value and adjacent contours differ by the same amount.
(b) The real part of the form factor $\Re (F_{e,\mathbf G=0}(\mathbf Q_1, \mathbf Q) - F_{h,\mathbf G=0}(\mathbf Q_1, \mathbf Q))$.
(c) The line cut of (a) in the direction of positive y direction (solid orange) and the negative x direction (dashed orange).
The purple line is the cut of (b) in the same directions.
}
\label{fig: exciton interaction form factor} 
\end{figure}

We first consider the flat Chern band case with $\Delta_D = 0$ meV and check the form factor is proportional to exciton dipole as \eqnref{eq: exciton interaction form factor} in the limit of $\mathbf Q\rightarrow0$.
Fig.~\ref{fig: exciton interaction form factor} shows the real and imaginary part of the form factor for $\mathbf Q_1 = (-2\pi /5\sqrt 3,0)/a_{\text{m,0}}$.
The dipole $\mathbf d_{N=1, \mathbf Q_1}$ is in the negative y direction, consistent with $\hat z\times \mathbf Q_1$.
In Fig.~\ref{fig: exciton interaction form factor}(c), $\Im(F_{e,\mathbf G =0}\left(\mathbf Q_1,\mathbf Q\right)- F_{h,\mathbf G=0}\left(\mathbf Q_1,\mathbf Q\right))$ is zero for $\mathbf Q \perp \mathbf d_{N=1, \mathbf Q_1}$ lying on the x axis (dashed orange line) and grows linearly if $\mathbf Q \parallel  \mathbf d_{N=1, \mathbf Q_1}$ for $\mathbf Q$ lying on the y axis (solid orange line).
Fig.~\ref{fig: exciton interaction form factor}(a) shows that the contours near $\gamma$ are parallel to the x axis and equally spaced, consistent with the behavior of $\mathbf Q\cdot \mathbf d_{N=1, \mathbf Q_1}$.
The discontinuity on contours at corners comes from the phase difference of the exciton envelope function increasing with larger $\mathbf Q$, where the phase difference is $e^{-i \phi_{\mathbf Q_1 + \mathbf Q}+\phi_{\mathbf Q_1}}$ between the envelope function $\Psi\left(e,\mathbf k_1+\mathbf Q+\mathbf Q_1,h,\mathbf k_1\right)^* \Psi \left(e,\mathbf k_1+\mathbf Q_1,h,\mathbf k_1\right)$ in \eqnref{eq: form factor V} at different center of mass momenta.
The real part  $\Re(F_{e,\mathbf G =0}\left(\mathbf Q_1,\mathbf Q\right)- F_{h,\mathbf G=0}\left(\mathbf Q_1,\mathbf Q\right))$ in Fig.~\ref{fig: exciton interaction form factor}(c) depends on $\mathbf Q$ quadratically along both directions (solid and dashed purple lines) and is smaller than the imaginary part in the positive y direction (solid lines).
The contours in 2D shown in Fig.~\ref{fig: exciton interaction form factor}(b) circles around the $\gamma$ with smaller space at larger $\mathbf Q$, reflecting the behavior of $\vert\mathbf Q\vert^2$, which can be neglected compared to its imaginary part for $\mathbf Q\approx 0$.

\begin{figure}
\centering
\includegraphics[width=\columnwidth]{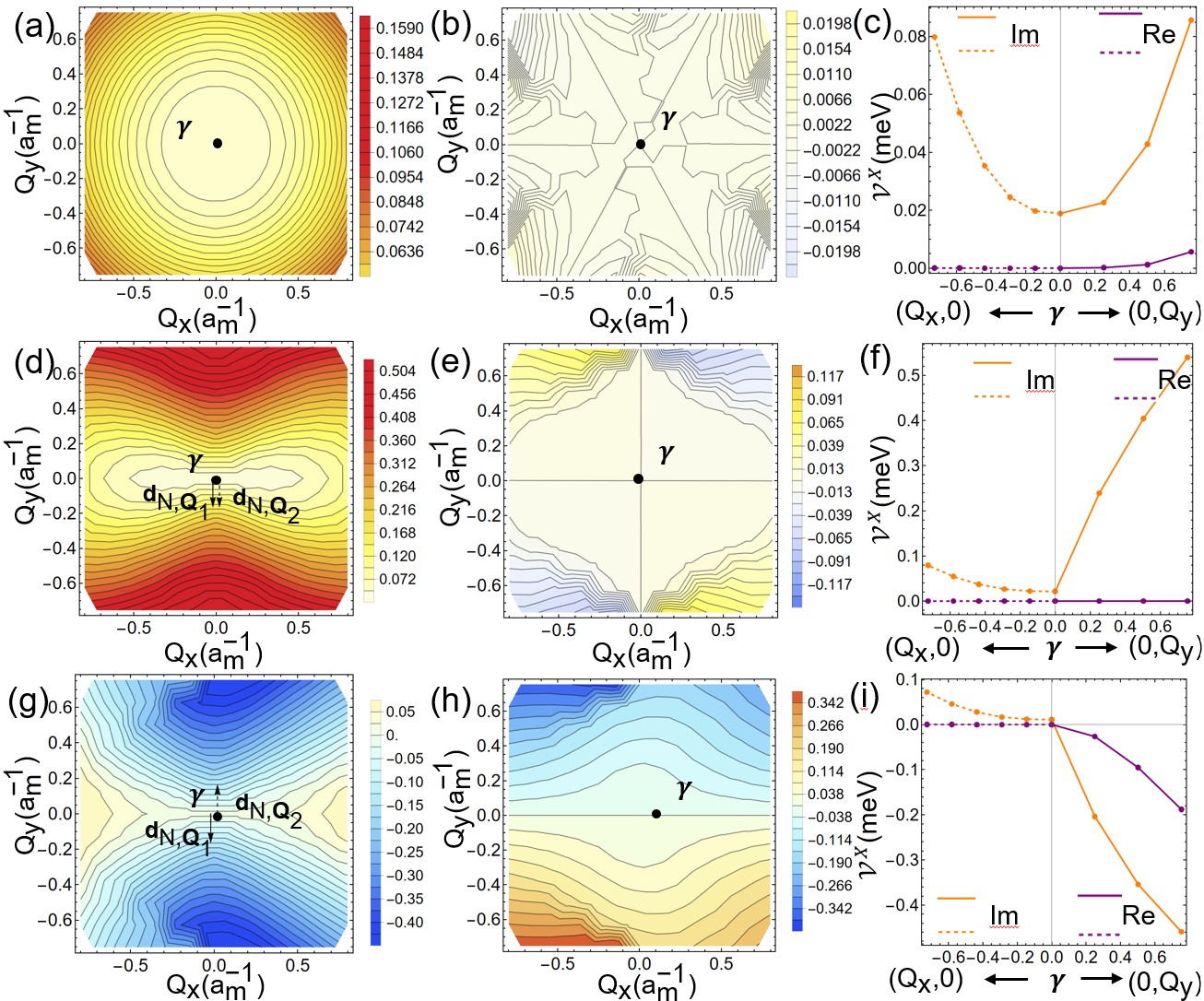}
\caption{
(a)(b) The real and imaginary part of the exciton interaction $\mathcal V^X$ versus $\mathbf Q$ for $\mathbf Q_1 = \mathbf Q_2 =0$, respectively.
One contour corresponds to a constant interaction energy and adjacent contours differ by the same amount.
(c) The line cut of (a)(b) along the positive y axis (solid lines) and negative x axis (dashed lines).
(d)(e)(f) The real, imaginary, and line cuts of the exciton interaction for $\mathbf Q_1 = \mathbf Q_2 = (-4\pi /5\sqrt 3,0)/a_{\text{m,0}}$, respectively.
The solid (dashed) arrow represents the dipole $\mathbf d_{N=1.\mathbf Q_1 }$ ($\mathbf d_{N=1.\mathbf Q_2}$).
(d)(e)(f) The real, imaginary, and line cuts of the exciton interaction for $\mathbf Q_1 = -\mathbf Q_2 = (-4\pi /5\sqrt 3,0)/a_{\text{m,0}}$, respectively.
}
\label{fig: exciton interaction flat band} 
\end{figure}

The two-body interaction between exciton is long range when finite dipoles are present.
For zero dipole at $\mathbf Q_1 = \mathbf Q_2 =\gamma$ shown in Fig.~\ref{fig: exciton interaction flat band}(c), $\mathcal V^X$ is quadratic in $\mathbf Q$ in both x and y direction, which is short range and higher order in $\mathbf Q$ than the dipole-dipole interaction in \eqnref{eq: classic dipole dipole interaction}.
The real part over the 2D $\mathbf Q$ plane is shown in Fig.~\ref{fig: exciton interaction flat band}(a) and nearly isotropic by the concentric contours around $\gamma$.
The unequal spacing between adjacent contours complies with the quadratic behavior in $\mathbf Q$.
Fig.~\ref{fig: exciton interaction flat band}(b) is the imaginary part and negligible compared to its real part.
For nonzero dipole at $\mathbf Q_1 = \mathbf Q_2 \neq \gamma$ shown in Fig.~\ref{fig: exciton interaction flat band}(f), the real part of $\mathcal V^X$ is linear in $\mathbf Q$ in the positive y direction (solid orange) for $\mathbf Q \parallel \mathbf Q_1$, which is the characteristic long range behavior of the dipole-dipole interaction.
In the 2D contour plot of $\Re\mathcal V^X$ in Fig.~\ref{fig: exciton interaction flat band}(a), the long range behavior is shown by the equal space of contours in the y direction.

The dipole-dipole interaction is anisotropic and can be attractive when the dipoles are anti-aligned.
For $\mathbf Q_1 = \mathbf Q_2 \neq \gamma$ shown in Fig.~\ref{fig: exciton interaction flat band}(d), while $\Re\mathcal V^X$ shows the linear behavior in the y direction, it is short range with quadratic $\mathbf Q$ dependence in the x direction, which reflects the anisotropy in the dipole-dipole interaction.
%The sign of $\Re\mathcal V^X$ is positive for all $\mathbf Q$, indicating the repulsion between the exciton with dipoles in the same direction.
When the direction of the $\mathbf d_{N=1, \mathbf Q_2}$ is reversed in Fig.~\ref{fig: exciton interaction flat band}(i), $\Re\mathcal V^X$ is still linear in $\mathbf Q$ but can be negative in the positive y direction (solid orange).
Comparing the 2D contour plot of $\Re\mathcal V^X$ for the opposite dipoles in Fig.~\ref{fig: exciton interaction flat band}(g) with that for the same dipoles in Fig.~\ref{fig: exciton interaction flat band}(d), there exists a region in $\mathbf Q$ that $\Re\mathcal V^X < 0$, implying the attraction between exciton with opposite dipoles.
The imaginary parts shown in Fig.~\ref{fig: exciton interaction flat band}(e)(h) is negiligible compared to their real parts near $\mathbf Q = \gamma$.

The exciton interaction interaction strength can also be tuned by the displacement field through the dipole magnitude.
Near $\gamma$, the two-body exciton interaction is proportional to $v_N^2$ by
\begin{equation}
    \mathcal V^X\approx V(\mathbf Q) v_N^2(\mathbf Q \cdot \hat z \times \mathbf Q_1)(\mathbf Q \cdot \hat z \times \mathbf Q_2).
\end{equation}
Based on Fig.~\ref{fig: gappedDiraccone}(a) of the main text, the dipole-dipole interaction first decreases, reaches the minimum at $\Delta_D = 1$ meV, and then increases again at larger $\Delta_D$.
The reverse of the helicity in exciton dipole does not lead to the sign change of the exciton interaction since $\mathcal V^X$ depends quadratically on $v_N$.

\begin{figure}
\centering
\includegraphics[width=\columnwidth]{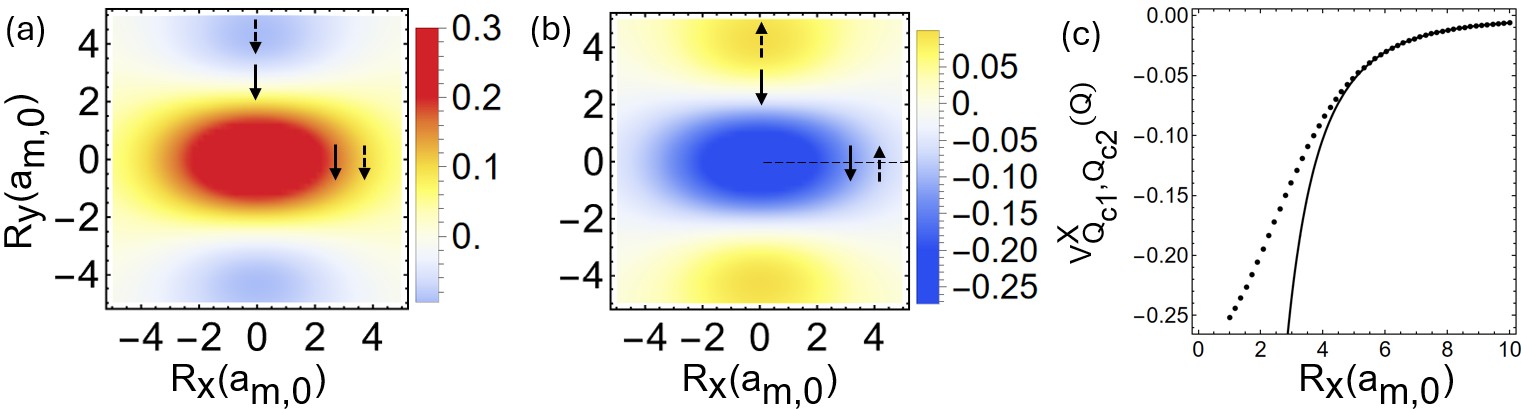}
\caption{
(a)(b) The exciton interaction potential in real space versus the relative position between two wavepackets located at $\mathbf R_{c1}, \mathbf R_{c2}$.
The solid (dashed) arrows are the exciton dipole $\mathbf d_{\mathbf Q_1}(\mathbf d_{\mathbf Q_2})$ aligned with the position $\mathbf R_{c2}- \mathbf R_{c1}$
$\mathbf Q_{c1}=\mathbf Q_{c2}$ for (a) and $\mathbf Q_{c1}=-\mathbf Q_{c2}$ for (b).
(c) The black dots are the exciton interaction potential along the dashed line in (b). The black solid line is the fitted interaction of the form $1/ \vert \mathbf R_{c2}- \mathbf R_{c1}\vert^3$.
}
\label{fig: exciton interaction real space}
\end{figure}

\subsection{Exciton interaction in real space}
\label{sec: Exciton interaction in real space}
In this section, we calculate the interaction potential of two exciton wave packets in terms of their spatial separation.

We first construct the wave packet of exciton located at $\mathbf R_c$ in the real space with its center-of-mass momentum around $\mathbf Q_c$ \cite{yu2017moire,zheng2025forster}. 
The creation operator for the exciton wave packet is
\begin{equation}
    \tilde X_{N,\mathbf Q_c}(\mathbf R_c) = \sum_\mathbf Q e^{-i \mathbf Q \cdot \mathbf R_c} \frac{1}{\sqrt \pi} w e^{-w^2(\mathbf Q-\mathbf Q_c)^2} X^\dagger_{N,\mathbf Q},
\end{equation}
where $w$ is the wave packet width in the real space and is taken as $w=2a_{\text m}$.
Considering the interaction between two exciton wave packets, the interaction by Wick's theorem is
\begin{equation}
\begin{split}
    &\tilde {\mathcal V}^X_{\mathbf Q_{c1},\mathbf Q_{c2}}(\mathbf R_{c2}-\mathbf R_{c1}) \\
    & = \langle \tilde X_{N,\mathbf Q_{c1}}(\mathbf R_{c1})  \tilde X_{N,\mathbf Q_{c2}}(\mathbf R_{c2}) \sum_{\mathbf Q_1, \mathbf Q_2, \mathbf Q} \frac{1}{2} \mathcal{V}^X_{\mathbf Q_1, \mathbf Q_2}(\mathbf Q) X_{\mathbf Q_1 + \mathbf Q}^\dagger X_{\mathbf Q_2-\mathbf Q}^{\dagger } X_{\mathbf Q_2} X_{\mathbf Q_1}  \tilde{X}_{N,\mathbf Q_{c2}}\left(\mathbf R_{c2}\right)^{\dagger } \tilde{X}_{N,\mathbf Q_{c1}}\left(\mathbf R_{c1}\right)^{\dagger } \rangle \\
    & \approx \sum_{\mathbf Q_1',\mathbf Q_1'',\mathbf Q_2', \mathbf Q_2''} \sum _{\mathbf Q_1,\mathbf Q_2, \mathbf Q} \mathcal{V}^X_{\mathbf Q_1, \mathbf Q_2}(\mathbf Q) 
    \frac{w}{\sqrt \pi} e^{i \mathbf Q_1''\cdot \mathbf R_{c1}}  e^{-w^2 \left(\mathbf Q_1''- \mathbf Q_{c1}\right)^2/2}  
    \frac{w}{\sqrt \pi} e^{i \mathbf Q_2''\cdot \mathbf R_{c2}} e^{- w^2 \left(\mathbf Q_2'' -\mathbf Q_{c2}\right)^2/2}  \\
    & \times \frac{w}{\sqrt \pi}  e^{-i \mathbf Q_2'\cdot \mathbf R_{c2}}  e^{- w^2 \left(\mathbf Q_2'- \mathbf Q_{c2}\right)^2/2} 
    \frac{w}{\sqrt \pi} e^{-i \mathbf Q_1'\cdot \mathbf R_{c1}}  e^{- w^2 \left(\mathbf Q_1'- \mathbf Q_{c1}\right)^2 / 2}   \\
    & \times (\delta_{\mathbf Q_1, \mathbf Q_2'} \delta_{\mathbf Q_2, \mathbf Q_1'} \delta _{\mathbf Q_2- \mathbf Q, \mathbf Q_1''} \delta_{\mathbf Q+ \mathbf Q_1, \mathbf Q_2''} + \delta_{\mathbf Q_1, \mathbf Q_1'} \delta_{\mathbf Q_2,\mathbf Q_2'} \delta_{\mathbf Q_2-\mathbf Q, \mathbf Q_2''} \delta _{\mathbf Q+\mathbf Q_1,\mathbf Q_1''}) \\
    &\approx \sum_\mathbf Q  \mathcal{V}^X_{\mathbf Q_{c1}, \mathbf Q_{c2}}(\mathbf Q)  e^{-i \mathbf Q\cdot \mathbf R_{c1}} e^{i \mathbf Q\cdot \mathbf R_{c2}} e^{ -w^2 \mathbf Q^2/2}.
\end{split}
\end{equation}
The approximation in the third line takes only the direct term of the Coulomb interaction.
The approximation in the fourth line approximate $\mathcal{V}_{\mathbf Q_1, \mathbf Q_2}(\mathbf Q)$ with $\mathcal{V}_{\mathbf Q_{c1}, \mathbf Q_{c2}}(\mathbf Q)$, which is \eqnref{eq: exciton interaction small Q} in the small $\mathbf Q$ limit 
\begin{equation}
    \mathcal{V}_{\mathbf Q_{c1}, \mathbf Q_{c2}}(\mathbf Q) \approx V_s + V(\mathbf Q) (\mathbf Q \cdot \mathbf d_{\mathbf Q_{c1}})(\mathbf Q \cdot \mathbf d_{\mathbf Q_{c2}}),
\end{equation}
where $V_s$ is the short range repulsion at $\mathbf Q=0$ shown in Fig.~\ref{fig: exciton interaction flat band}(f)(i).
We model the short range repulsion from other interaction terms with $w' =0.2 a_m$, an order smaller than that of the direct term.

$\tilde {\mathcal V}^X_{\mathbf Q_{c1},\mathbf Q_{c2}}(\mathbf R_{c2}-\mathbf R_{c1})$ is the dipole-dipole interaction in the real space with anisotropy and long range behavior.
Fig.~\ref{fig: exciton interaction real space}(a) shows the interaction potential for two exciton with the same dipole in the y direction with $\mathbf Q_{c1}=\mathbf Q_{c2}$.
The interaction is attractive for head-to-tail aligned dipole in the real space along the y direction while it is repulsive for parallel aligned along the x direction, consistent with the anisotropy features of the dipole-dipole interactions.
Fig.~\ref{fig: exciton interaction real space}(b) shows the interaction potential for two exciton with the opposite dipole in the y direction with $\mathbf Q_{c1}=-\mathbf Q_{c2}$.
The interaction potential is minimal when the two exciton dipole overlap with each other without the short-range repulsion from terms other than the direct Coulomb interaction.
The asymptotic behavior of the interaction at the large $\vert \mathbf R_{c2}- \mathbf R_{c1}\vert$ is proportional to $1/ \vert \mathbf R_{c2}- \mathbf R_{c1}\vert^3$ as shown in Fig.~\ref{fig: exciton interaction real space}(c), which is a long range interaction.

\end{document}